\documentclass[floatfix,aps,prd,amsmath,nofootinbib,onecolumn,10pt,superscriptaddress]{revtex4-2}

\usepackage{graphicx}
\usepackage{dcolumn}
\usepackage{bm}
\usepackage{epsfig} 
\usepackage{amsfonts}
\usepackage{amsmath}
\usepackage{amssymb}
\usepackage{appendix}
\usepackage[usenames]{color}
\usepackage[dvipsnames]{xcolor}
\usepackage{booktabs}
\usepackage{multirow}
\usepackage{longtable}
\usepackage{placeins}
\usepackage[unicode, colorlinks=true, linkcolor=linkcolor, citecolor=linkcolor, filecolor=linkcolor,urlcolor=linkcolor, pdfusetitle]{hyperref}

\hypersetup{colorlinks,citecolor=blue,linkcolor=blue,urlcolor=blue}
\hypersetup{final=true}

\begin{document}

\title{Measurements of the Angular Homogeneity Scale from DESI DR1}

\author{Mariana Lopes-Dias}
\email{marianadias@on.br}
\affiliation{Observatório Nacional, 20921-400, Rio de Janeiro, RJ, Brazil}

\author{Carlos A.P. Bengaly}
\email{}
\affiliation{Observatório Nacional, 20921-400, Rio de Janeiro, RJ, Brazil}

\author{Xiaoyun Shao}
\email{}
\affiliation{Observatório Nacional, 20921-400, Rio de Janeiro, RJ, Brazil}

\author{Rodrigo Gonçalves}
\email{}
\affiliation{Departamento de F\'{\i}sica, Universidade Federal Rural do Rio de Janeiro, Serop\'edica, RJ, 23897-000, Brazil}

\author{Paula S. Ferreira}
\email{}
\affiliation{Institute of Astronomy, School of Physics, Zhejiang University, Hangzhou, China}
\affiliation{Center for Cosmology and Computational Astrophysics, Institute for Advanced Study in Physics, Zhejiang University, Hangzhou, China.}

\author{Gabriela Coutinho de Carvalho}
\email{}
\affiliation{Faculdade de Tecnologia, Universidade do Estado do Rio de Janeiro, 27537-000, Resende, RJ, Brazil}

\author{Jailson Alcaniz}
\email{}
\affiliation{Observatório Nacional, 20921-400, Rio de Janeiro, RJ, Brazil}

\begin{abstract}
The study of the large-scale distribution of galaxies provides essential information for testing the standard cosmological model, namely the $\Lambda$CDM paradigm. This scenario is based upon two foundations: General Relativity as the theory of gravity, and  the Cosmological Principle, which states that the Universe is statistically homogeneous and isotropic on large scales – so that we can measure distances and ages in the Universe assuming the FLRW metric. 

In this work, we perform a test of the Cosmological Principle by probing the angular homogeneity scale, $\theta_H$, using the state-of-the-art observational data of Luminous Red Galaxies (LRGs) from the Dark Energy Spectroscopic Instrument Data Release 1 (DESI DR1). Our analysis is performed exclusively in two dimensions, across narrow redshift ranges inside a larger redshift sample of $0.4 < z < 1.1$, in two different surveyed regions of the sky (North and South Galactic Caps), as we want to minimize a priori dependences on an underlying cosmological model. 

We obtain that such a scale is indeed identified in all redshift ranges, and that they are consistent with mock simulations assuming the $\Lambda$CDM model. Moreover, our results are in great agreement with previous measurements using Sloan Digital Sky Survey IV extended Baryon Oscillation Spectroscopic Survey Data Release 16 (SDSS-IV eBOSS DR16), as well as between the north and south galactic caps of the DESI DR1 survey. These findings help underpinning statistical isotropy and homogeneity of the Universe as a physically valid hypothesis in light of upcoming stage-IV redshift surveys, hence are consistent with one of the fundamental pillars of the standard cosmological model.   

\end{abstract}

\maketitle

\section{Introduction}

The $\Lambda$CDM cosmological model \citep{SupernovaCosmologyProject:1998vns, SupernovaSearchTeam:1998fmf} has become established as the standard framework in Cosmology as a consequence of recent observations such as the Cosmic Microwave Background (CMB) \cite{Planck:2018yye} and Type Ia Supernovae \cite{Brout:2022vxf, Rubin:2023jdq, DES:2024jxu}, which confirm with high precision that this model successfully describes the observed data. Nevertheless, unresolved theoretical challenges remain, such as the primordial singularity and cosmic coincidence problems, in addition to significant tensions in the measurements of cosmological parameters, most notably the $\sim 5.0 \sigma$ discrepancy in the Hubble parameter $H_0$ when different observables are considered \citep{DiValentino:2021izs, Shah:2021onj, Riess:2021jrx}. Therefore, it is essential that the foundations of the standard cosmological model be continuously tested, since any statistically significant deviation from its predictions could imply modifications to our current understanding of physics.

General Relativity (GR) and the Cosmological Principle (CP) constitute the foundations of the $\Lambda$CDM model: GR as the theory of gravitation, and the CP as the postulate the homogeneity and isotropy of the Universe on large scales. Given the validity of the CP, we can describe cosmic distances and ages in the Universe by the Friedmann–Lemaître–Robertson–Walker (FLRW) metric. Still, inhomogeneities are expected on small scales. As a consequence, we observe non-uniform structures such as galaxies, galaxy clusters, voids, walls, among others. In this scenario, a transition scale from an irregular Universe to a more homogeneous one is also expected, so that different regions of the Universe become statistically indistinguishable from one another from such scale onward. This is the so-called homogeneity scale, whose identification and measurement consist on a crucial test of the CP. 

The homogeneity scale in three-dimensional space, $R_H$, has already been extensively investigated in the literature. Measurements of this transition scale using different galaxy and quasar surveys typically correspond to values around $70 - 150$ Mpc h$^{-1}$ \cite{Hogg:2004vw, Sarkar:2009iga, Scrimgeour:2012wt, Pandey:2013xz, Pandey:2015xea, Laurent:2016eqo, Sarkar:2016fir, Ntelis:2017nrj, Goncalves:2018sxa, Goncalves:2020erb, Avila:2021mbj, Kim:2021osl}, consistent with the upper limit of $\sim 260$ Mpc h$^{-1}$ reported by \cite{Yadav:2010cc}. Using different methods and datasets, this result has also been recovered in \cite{Zhang:2010fa, Valkenburg:2012td, Hoyle:2012pb, Kraljic:2014aea, Jimenez:2019cll, Camarena:2021mjr}. However, its two-dimensional counterpart, i.e., the angular homogeneity scale ($\theta_H$), has not been explored as extensively in the literature. Differently from $R_H$, it has the advantage of prior assumption of the underlying cosmological model -- which is needes to convert redshifts into distances in its computation. This test was originally proposed in \cite{Alonso:2013boa}, based on the two-point correlation function, $\omega(\theta)$, which was later applied by \cite{Alonso:2014xca, Goncalves:2017dzs, Avila:2019gdb, Camacho-Quevedo:2021bvt, Andrade:2022imy, Shao:2024qrd} using data from different surveys, where $\theta_H$ has been shown to be a powerful cosmological probe.

In this work, our main goal is to measure $\theta_H$ using Luminous Red Galaxies (LRGs) from the most recent available survey, the Dark Energy Spectroscopic Instrument Data Release 1 (DESI DR1) \cite{DESI:2025fxa}. The LRG are almost exclusively massive elliptical galaxies that dominate the center of galaxy clusters, meaning they are mostly highly coupled with dark matter \cite{Hogg2003,Eisenstein2001,Zehavi2005}. The analysis is performed in two dimensions using narrow redshift ranges of width $\Delta z = 0.01$ across the full LRG distribution of the DESI DR1 survey in both sky regions, namely the North and South Galactic Caps. The only quantities required are the celestial coordinates of each object. We extend the analyses performed in \cite{Goncalves:2017dzs, Andrade:2022imy} to broader redshift ranges, since DESI DR1 provides extensive LRG coverage over the interval $0.4 < z < 1.1$ \cite{DESI:2024aax}. This way, we are able to investigate cosmic homogeneity at epochs of the Universe that have not yet been explored through this type of approach. For completeness, we also perform a comparative analysis using data from the Sloan Digital Sky Survey IV extended Baryon Oscillation Spectroscopic Survey Data Release 16 (SDSS-IV eBOSS DR16) \cite{eBOSS:2020mzp}, in which LRGs are also used as the main tracers.

\section{Observational Data}\label{sec_observationaldata}

In this work, we use two catalogs: the Sloan Digital Sky Survey IV extended Baryon Oscillation Spectroscopic Survey Data Release 16 (SDSS-IV eBOSS DR16) \cite{eBOSS:2020mzp} and the Dark Energy Spectroscopic Instrument Data Release 1 (DESI DR1) \cite{DESI:2025fxa}. Both provide data on Luminous Red Galaxies (LRGs) (see \cite{eBOSS:2020mzp} and \cite{DESI:2024aax}), which constitute the objects of study in this work. 

\begin{itemize}
    \item eBOSS DR16

    The catalog contains 174 816 LRG's distributed over the redshift range $0.6 < z < 1.0$, covering a total area of 4 242 deg$^2$. Of this, 2 566 deg$^2$ correspond to the North Galactic Cap (NGC), while 1 676 deg$^2$ correspond to the South Galactic Cap (SGC).

    For this catalog, we used the \texttt{EZmocks} \cite{ezmocks} 1000 mock realizations. These are based on the Zel'dovich approximation, which captures large-scale coherent flows and mildly nonlinear clustering while remaining computationally efficient.

    \item DESI DR1

    The catalog contains 2 138 627 LRG's distributed over the redshift range $0.4 < z < 1.1$, covering a total area of 5 740 deg$^2$. Of this, 63.6\% corresponds to the North Galactic Cap, covering an area of 3 650.64 deg$^2$, while 36.4\% corresponds to the South Galactic Cap, covering 2 089.36 deg$^2$. 

    The DESI Collaboration provided the 1000 mock realizations named \texttt{Uchuu} \cite{uchuu}. Unlike the approximate \texttt{EZmocks}, the \texttt{Uchuu} mocks are derived from a high-resolution $N$-body simulation, accurately modeling nonlinear structure growth, halo occupation distribution, and redshift-space distortions

\end{itemize}
\begin{figure}[h!]
    \centering
    \includegraphics[width=0.85\textwidth]{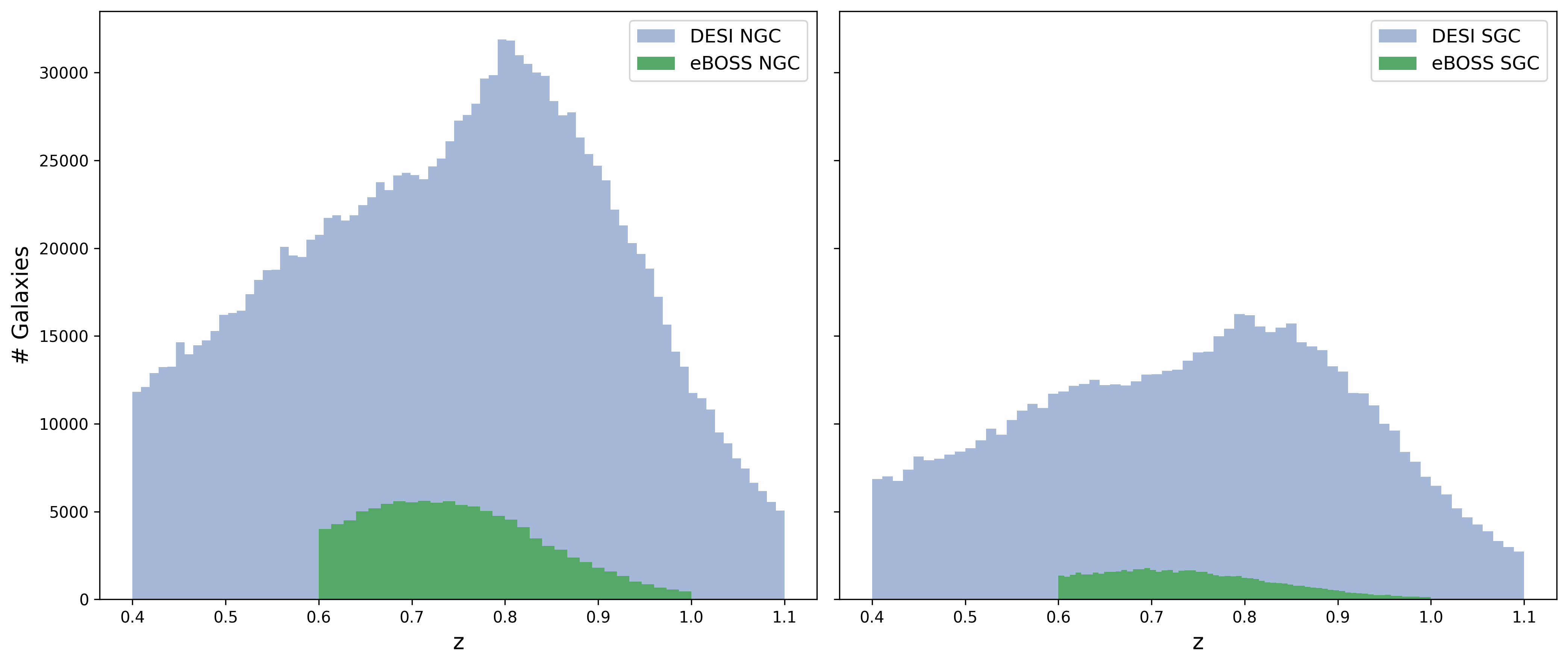}
    \caption{Histogram of the redshift distribution of LRG's in the North Galactic Cap (left panel) and South Galactic Cap (right panel). The left panel shows the eBOSS DR16 NGC catalog with 107 500 objects and DESI DR1 NGC catalog with 1 476 136 objects, while the rigth panel shows the eBOSS DR16 SGC catalog with 67 316 objects and DESI DR1 SGC catalog with 662 492 objects.}
    \label{fig:histogram_NGCandSGC}
\end{figure}
\begin{figure}[h!]
    \centering
    \includegraphics[width=0.85\textwidth]{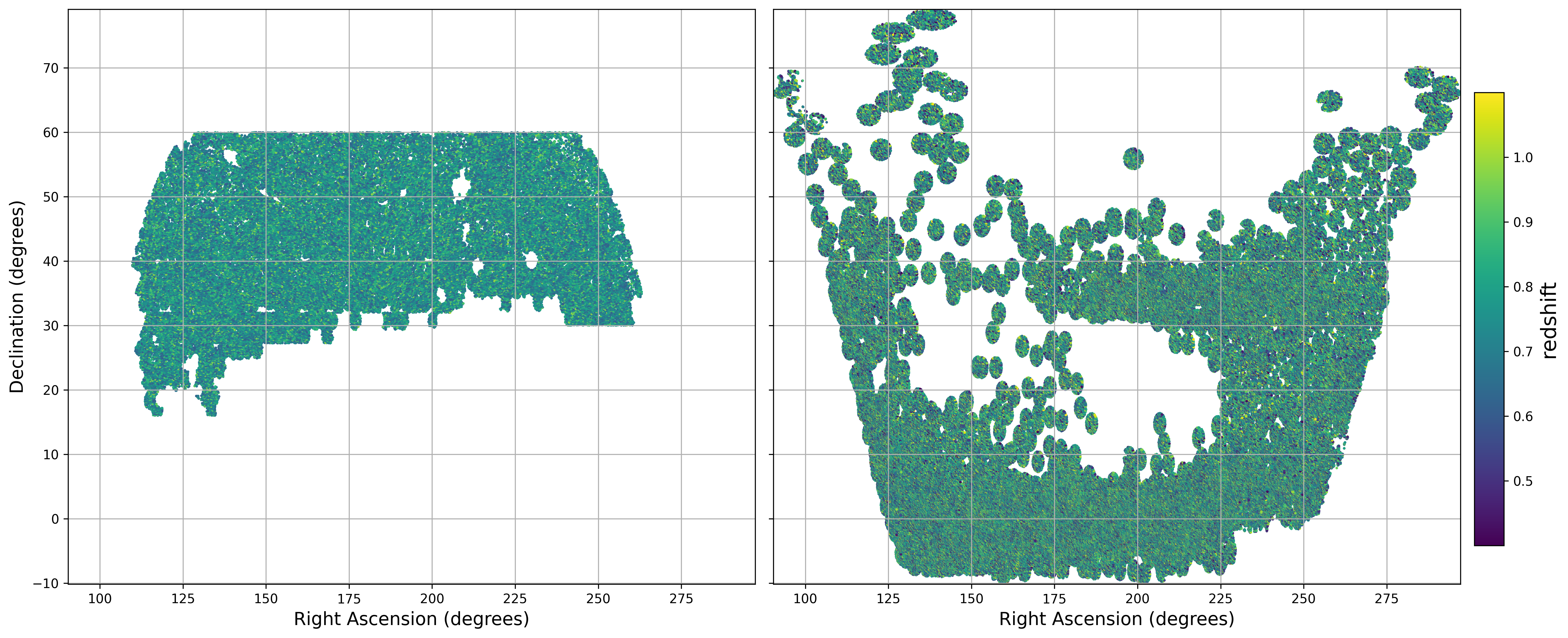}
    \vfill
    \centering
    \includegraphics[width=0.85\textwidth]{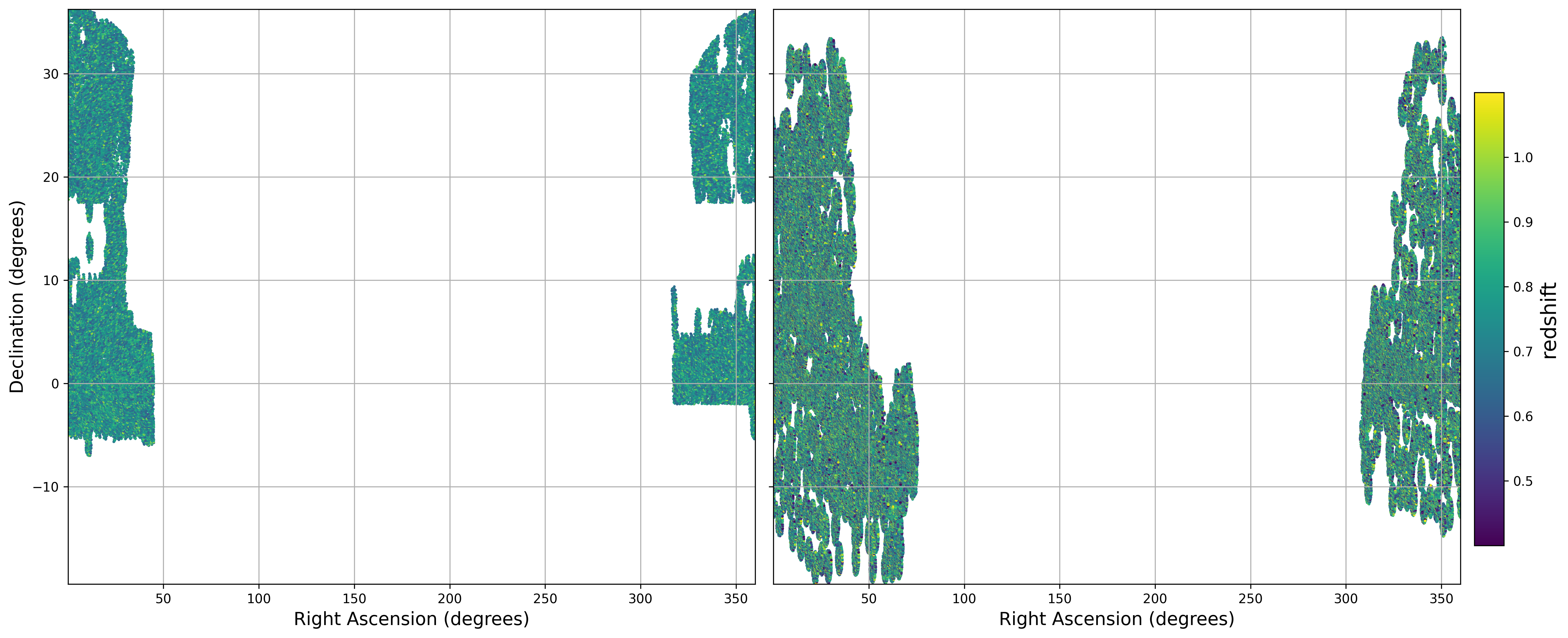}
    \caption{Footprint of Galactic Cap. Upper panels show North Hemisphere and bottom panels show South Hemisphere. Left panels correspond to the eBOSS DR16 data and the right panels correspond to the DESI DR1 data. }
\end{figure}

\newpage
\subsection{Weights}
Some systematic effects in the data are known and must be corrected. For this purpose, specific statistical weights are applied to each type of galaxy (real and random data), correcting for variations in the selection function and optimization. 

For the eBOSS DR16 data, we follow \cite{eBOSS:2020mzp}, which defines the total weight as,
\begin{equation}\label{weight_eboss}
    \omega_{total} = \omega_{sys} \, \omega_{cp} \, \omega_{noz} \, \omega_{FKP},
\end{equation}
while for DESI DR1 data, we follow \cite{DESI:2024aax}, which defines,
\begin{equation}\label{weight_desi}
    \omega_{total} = \frac{\omega_{comp} \, \omega_{zfail} \, \omega_{imsys}}{\langle \omega_{comp} \rangle  \, n_{tile}} \omega_{FKP} .
\end{equation}

For both Equations \eqref{weight_eboss} and \eqref{weight_desi}, several weights have equivalent physical roles. The weight that corrects for fluctuations in the target density due to variations in imaging conditions (such as seeing and Galactic extinction), typically derived from linear regression models, is $\omega_{sys}$, which is equivalent to $\omega_{imsys}$. The weight that corrects for incompleteness caused by fiber collisions due to their finite physical size is $\omega_{cp}$, equivalent to $\omega_{comp}$. The redshift failure weight accounts for the inability to obtain reliable redshift measurements for low signal-to-noise targets, with $\omega_{noz}$ being equivalent to $\omega_{zfail}$. They also include the Feldman–Kaiser–Peacock (FKP) weight, $\omega_{FKP}$ \cite{Feldman:1993ky}, used for optimization in regimes where the number density is low and noise is dominant. Signal-to-noise optimization is particularly relevant for the SGC data, where the density is lower than in the NGC, as we can observe in Figure \ref{fig:histogram_NGCandSGC}.

For the last weight considered, eBOSS depends only on redshift, with
\begin{equation}
    \omega_{FKP} = \frac{1}{1+n(z)P_0},
\end{equation}
whereas DESI also takes into account the number of overlapping tiles, $n_{tile}$, which corresponds to the number of observation tiles covering a given location on the sky. In this case,
\begin{equation}
    \omega_{FKP} = \frac{1}{1+[n(z)\langle C_{assign}\rangle n_{tile}] P_0},
\end{equation}
where $n(z)$ is the comoving number density, $C_{assign}$ represents the fraction of targets that were successfully assigned an optical fiber, and $P_0$ is the reference amplitude of the tracer power spectrum.

It is worth noting that $n_{tile}$ also appears explicitly in the definition of $\omega_{total}$, Equation \eqref{weight_desi}. When multiplied by the mean value  $\langle \omega_{comp} \rangle$, it ensures that the total data counts are properly normalizes, statistically correcting for the sky sampling. In the FKP weighting scheme, however, $n_{tile}$ enters through the effective galaxy density, providing a more accurate description of the survey selection function. By accounting for the varying observational coverage across the sky, the DESI implementation wields a more refined FKP weight, better suited to the survey's higher galaxy density and more complex observational strategy.

\section{Methodology}

The goal of this work is to measure the homogeneity scale of the Universe, $\theta_H$, with minimal dependence on a cosmological model. To this end, we will adopt the correlation dimension, $D_2$, as our estimator, following the approaches of previous works \cite{Scrimgeour:2012wt, Camarena:2021mjr, Andrade:2022imy}, derived from purely angular measurements. Right ascension and declination are directly observed quantities, therefore no conversion from redshift to distance is required, which is highly sensitive to the parameters of cosmological models.

The idea is to measure how the number of galaxy pairs increases as the angular separation between them grows. If the galaxies are distributed uniformly and randomly in a two-dimensional space, then $D_2 = 2$, if they are strongly clustered, then $D_2 < 2$. We aim to determine the scale at which the Universe transitions from a clumpy distribution to a homogeneous one ($D_2 = 2$). As shown in \cite{Goncalves:2017dzs} the homogeneity scale, $\theta_H$, corresponds to the angle at which $D_2 = 1.98$, i.e., 99\% of the expected value of the correlation dimension for a perfectly homogeneous distribution.

To mitigate projection effects and ensure that we are analyzing galaxies located at approximately the same distance, we perform the analysis in narrow redshift ranges (i.e. $\Delta z = 0.01$) as proposed in \cite{Alonso:2013boa} and \cite{Andrade:2022imy}.

Considering the galaxy distribution divided into these redshift ranges, we compute the two-point correlation function, $\omega(\theta)$, for each range using real data combined with random catalogs, and mock data combined with random catalogs. In this work, we adopt the Landy–Szalay estimator \cite{Landy:1993yu} to calculate $\omega(\theta)$, defined as, 
\begin{equation}\label{eq_omegatheta}
    \omega(\theta) = \frac{DD(\theta) - 2 DR(\theta) + RR(\theta)}{RR(\theta)}, 
\end{equation}
where $DD(\theta)$, $DR(\theta)$, and $RR(\theta)$ are the numbers of pairs as a function of the angular separation, corresponding to data–data, data–random, and random–random pairs, respectively. 

This pair-counting was computed by the {\sc TreeCorr} package \cite{Jarvis:2003wq}\footnote{Publicly available at \url{https://github.com/rmjarvis/treecorr}.}. Using 30 bins--rather than the standard choice of 24 commonly found in the literature--as this provides better fits to the data (see more details in Appendix \ref{appendix_bins}). Although a smaller number of bins can improve the signal-to-noise ratio, the choice of 30 bins represents a good compromise before the noise becomes significant and leads to overfitting, with a minimum separation of $0.1^\circ$ and a maximum separation of $15^\circ$. 

We can associate Equation \eqref{eq_omegatheta} with the correlation dimension, $D_2$, by means of the scaled counts-in-spheres \cite{Scrimgeour:2012wt, Laurent:2016eqo, Ntelis:2017nrj, Goncalves:2020erb}, which reads,
\begin{equation}
    \mathcal{N} (< \theta)\equiv 1 + \frac{1}{1 - \cos{\theta}} \int_0^{\theta} \omega(\theta^{\prime}) \sin{\theta^{\prime}} d\theta^{\prime} , 
\end{equation}
\begin{equation}
    D_2 (\theta) \equiv \frac{d \ln{\mathcal{N}}}{d \ln{\theta}} +2 .
\end{equation}

The value of $\theta_H$ is obtained by interpolating $\theta$ at $D_2 = 1.98$. This procedure is performed using 1000 bootstrap resamplings of the real data, in order to estimate the $D_2$ uncertainties at each angular scale, in addition to the 1000 mock catalogs described in Section \ref{sec_observationaldata}, so that we can verify whether the real data results are consistent with the simulations based on the standard model. The mean and standard deviation of the angular homogeneity scales obtained from the bootstraps and the mocks are hereafter referred as $\theta_H^{boot}$ and $\theta_H^{mock}$, respectively.

\section{Results}

\begin{figure}[h!]
    \centering
    \includegraphics[width=0.4\textwidth]{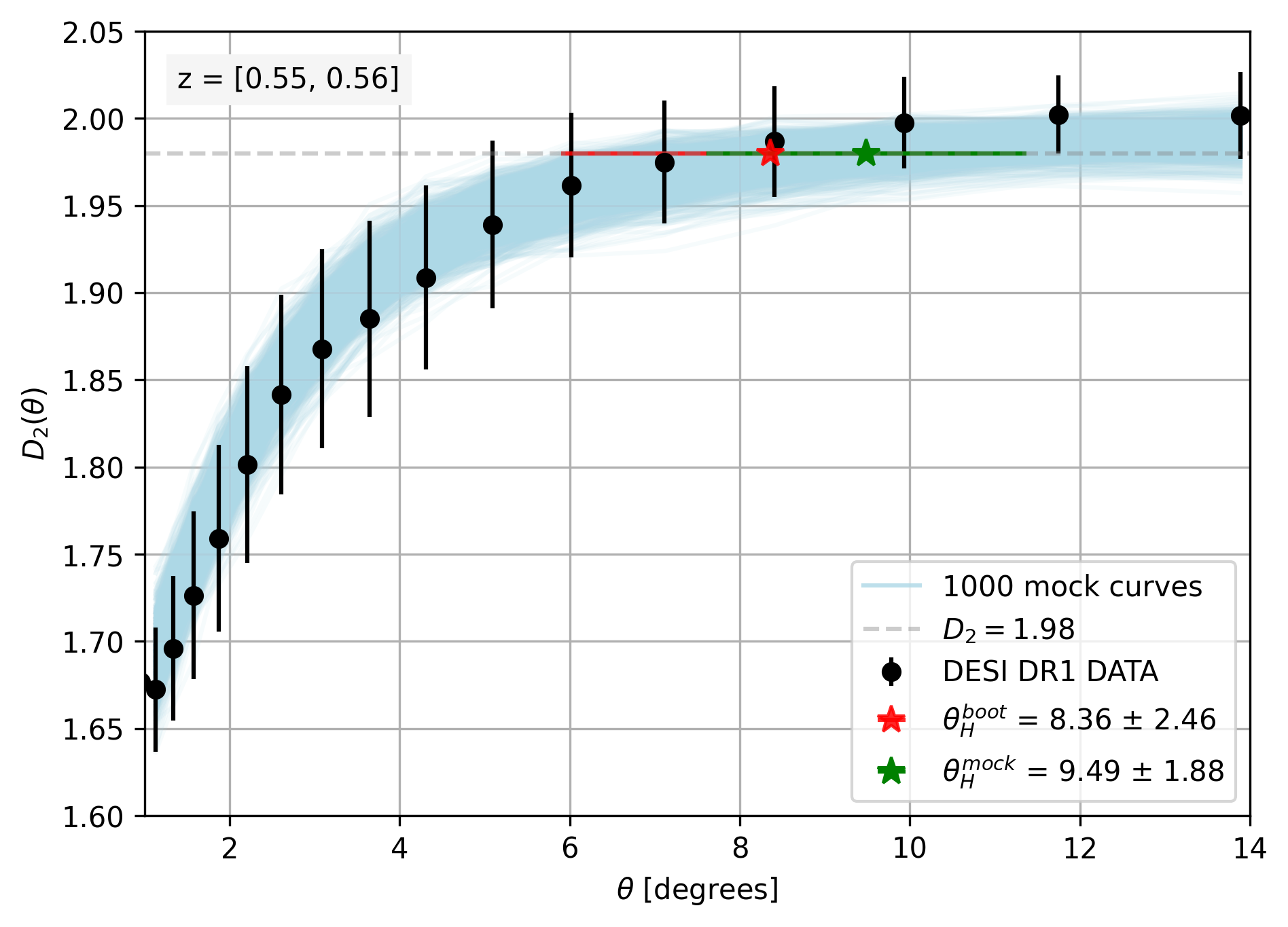}
    \hfill
    \includegraphics[width=0.4\textwidth]{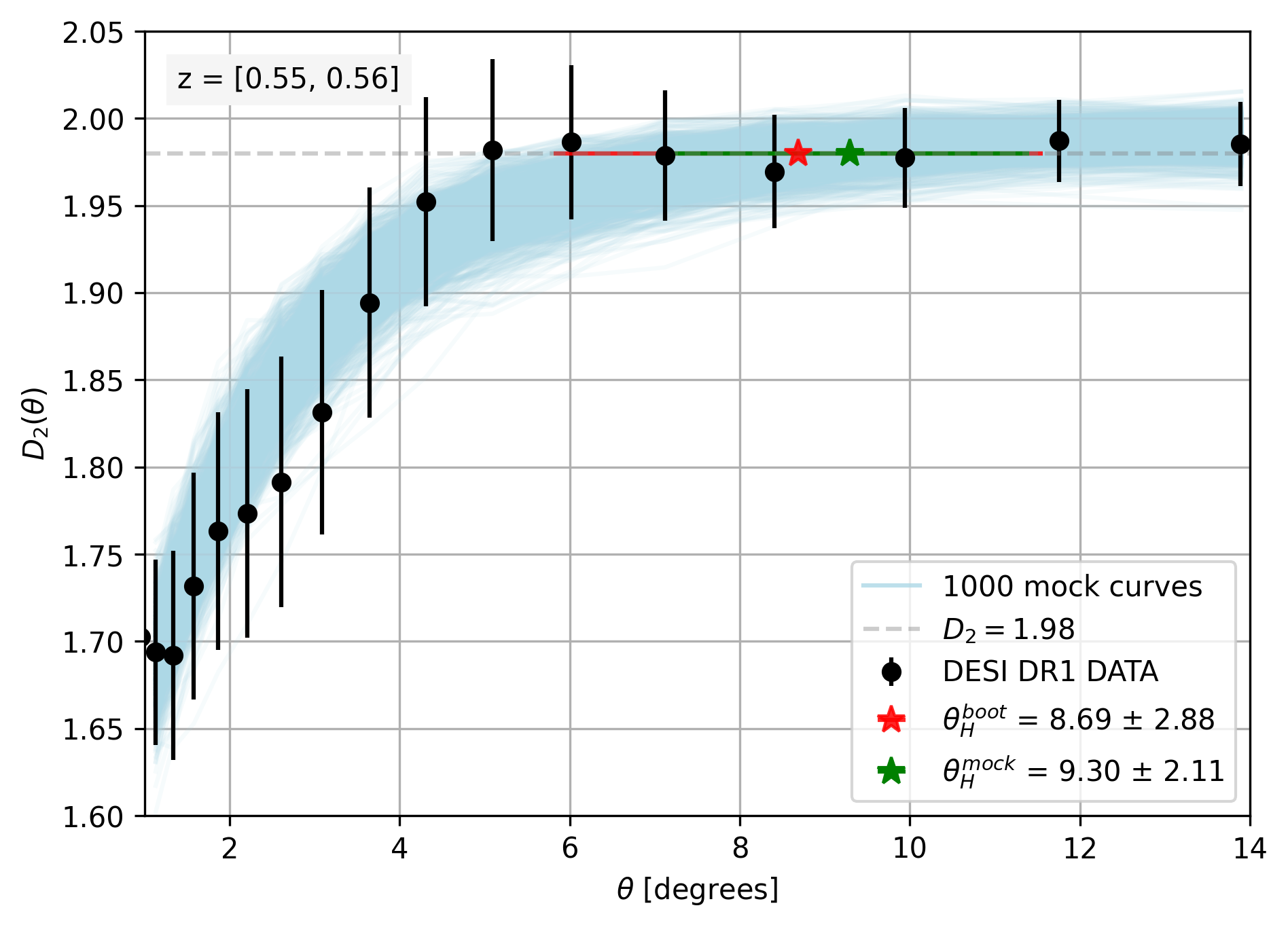}
    \caption{Curves of the correlation dimension, $D_2(\theta)$, in the same redshift range $0.55 < z < 0.56$. The left panel shows measurements using North Galactic Cap (NGC), while the right panel shows measurements using the South Galactic Cap (SGC), both shows the DESI DR1 catalog.  }
    \label{fig:d2curves_NGCxSGCdesi_045z046}
\end{figure}

\begin{figure}[h!]
    \centering
    \includegraphics[width=0.6\textwidth]{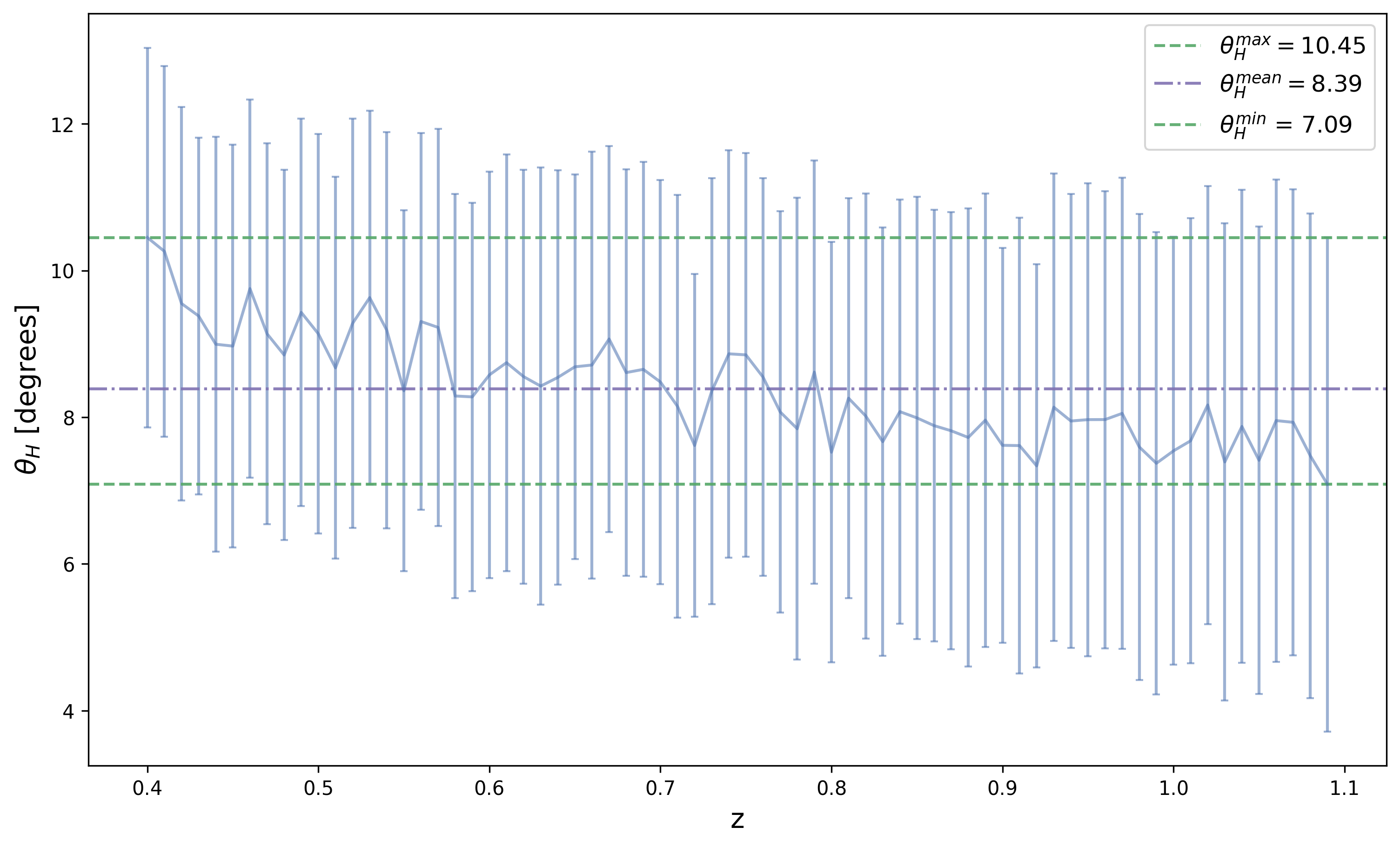}
    \caption{Measurements of the angular homogeneity scale, $\theta_H^{boot}$, for all redshift ranges of width $\Delta z = 0.01$, using the real data from the DESI DR1 NGC catalog via bootstrap. The minimum, maximum and mean values are also presented, $\theta^{min}_H$, $\theta^{max}_H$, $\theta^{mean}_H$, respectively.}
    \label{fig:thetaH_desiNGC}
\end{figure}

\begin{figure}[h!]
    \centering
    \includegraphics[width=0.6\textwidth]{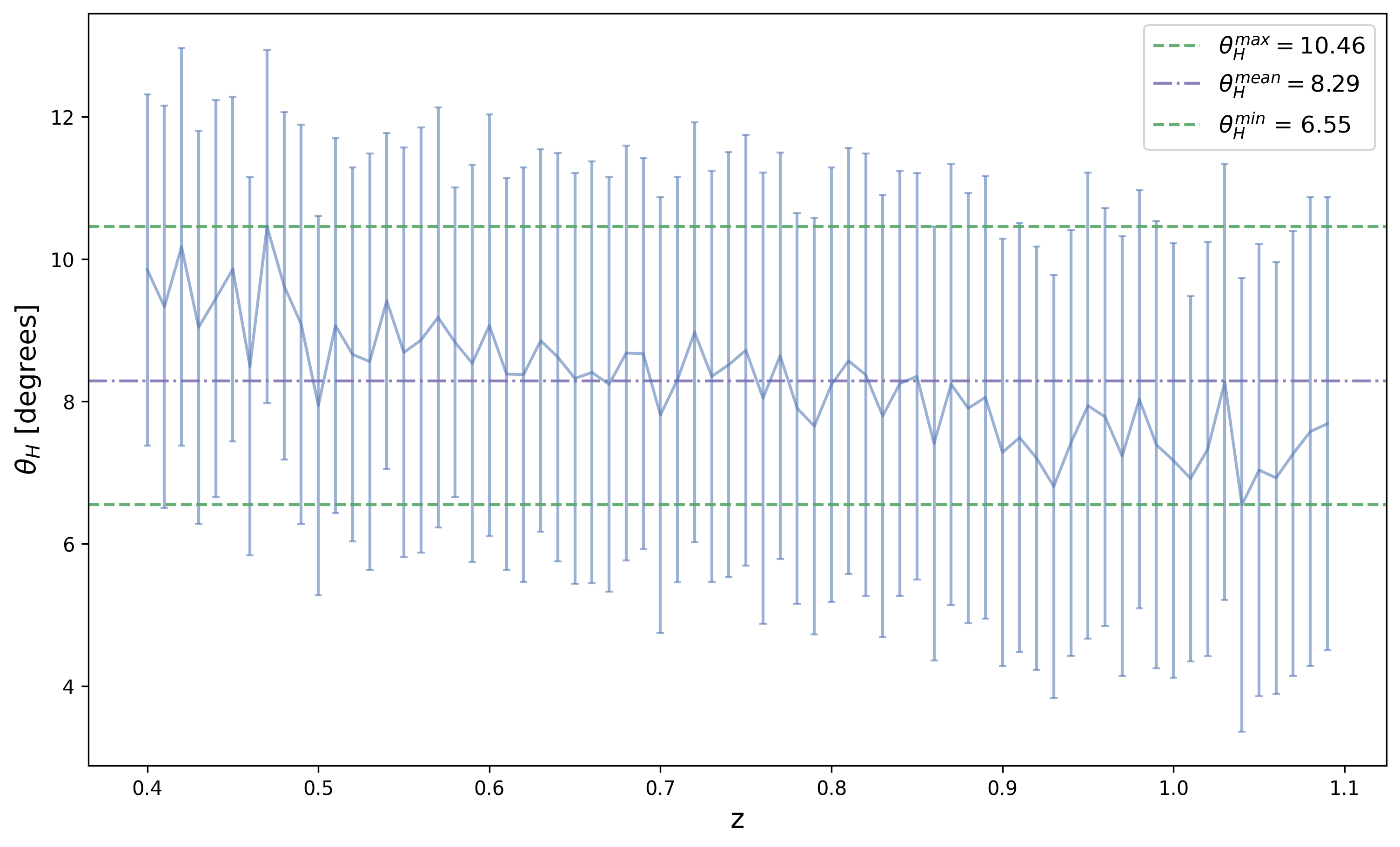}
    \caption{Measurements of the angular homogeneity scale, $\theta_H^{boot}$, for all redshift ranges of width $\Delta z = 0.01$, using the real data from the DESI DR1 SGC catalog via bootstrap. The minimum, maximum and mean values are also presented, $\theta^{min}_H$, $\theta^{max}_H$, $\theta^{mean}_H$, respectively.}
    \label{fig:thetaH_desiSGC}
\end{figure}

    As described above we perform the $D_2$ analysis for DESI DR1 LRGs in all redshift bins within $0.4 < z < 1.1$, assuming $\Delta z =0.01$, in order to identify and measure $\theta_H$. For the sake of exemplification the results for one of these bins, $z = [0.55, 0.56]$, is presented in Figure \ref{fig:d2curves_NGCxSGCdesi_045z046}. As for the results of $\theta_H$ in each of the 70 available redshift bins, they are displayed in Figure \ref{fig:thetaH_desiNGC} (for DESI NGC) and Figure \ref{fig:thetaH_desiSGC} (for DESI SGC). We find that DESI DR1 catalog indeed provides observational evidence for the angular homogeneity scale, as predicted by the standard model, due to the agreement between the mocks and observational data. From both Figures \ref{fig:thetaH_desiNGC} and \ref{fig:thetaH_desiSGC}, we observe a good agreement between the NGC and SGC results, indicating evidence not only for the homogeneity of the Universe, but also for its isotropy. This, in turn, supports the validity of the Cosmological Principle. As expected from the angular diameter distance-redshift relation, $\theta_H$ decreases with redshift progression at fixed comoving scale, it would clearly seeing in Figures \ref{fig:thetaH_desiNGC} and \ref{fig:thetaH_desiSGC} and consistent with the $\Lambda$CDM predictions from the mock simulations. For completeness, we also compute the spatial homogeneity scale, $R_H$, which can be found in Appendix \ref{appendix_values_thetah_rh}, Table \ref{tab:values_thetaH_rh}, along with all $\theta_H$ measurements. 


\clearpage
\subsection{SDSS eBOSS DR16 x DESI DR1}

We also perform a comparative analysis between the SDSS eBOSS DR16 and DESI DR1 data in the NGC. Due to the low galaxy density of the eBOSS DR16 catalog in the SGC (see Table \ref{tab:num_galaxies}, Figure \ref{fig:histogram_NGCandSGC}, and Figure \ref{fig:SGC_footprint_byz} in Appendix \ref{appendix_footprints}), we obtain noisier results in this region. For this reason, the corresponding analyses are presented in Appendix \ref{appendix_ebossSGC}, and we do not perform the comparative analysis with the DESI data for the SGC. For this comparison in the NGC we select three redshift ranges: $0.67 < z < 0.68$, $0.70 < z < 0.71$, and $0.73 < z < 0.74$, i.e., the same ones examined in \cite{Andrade:2022imy}. The number density of these galaxies can be seen in Table \ref{tab:num_galaxies}, while the galaxy distributions within these redshift ranges are shown in the Appendix \ref{appendix_footprints} in Figures \ref{fig:NGC_footprint_byz} and \ref{fig:SGC_footprint_byz}.

\begin{figure}[h!]
    \centering
    \includegraphics[width=0.45\textwidth]{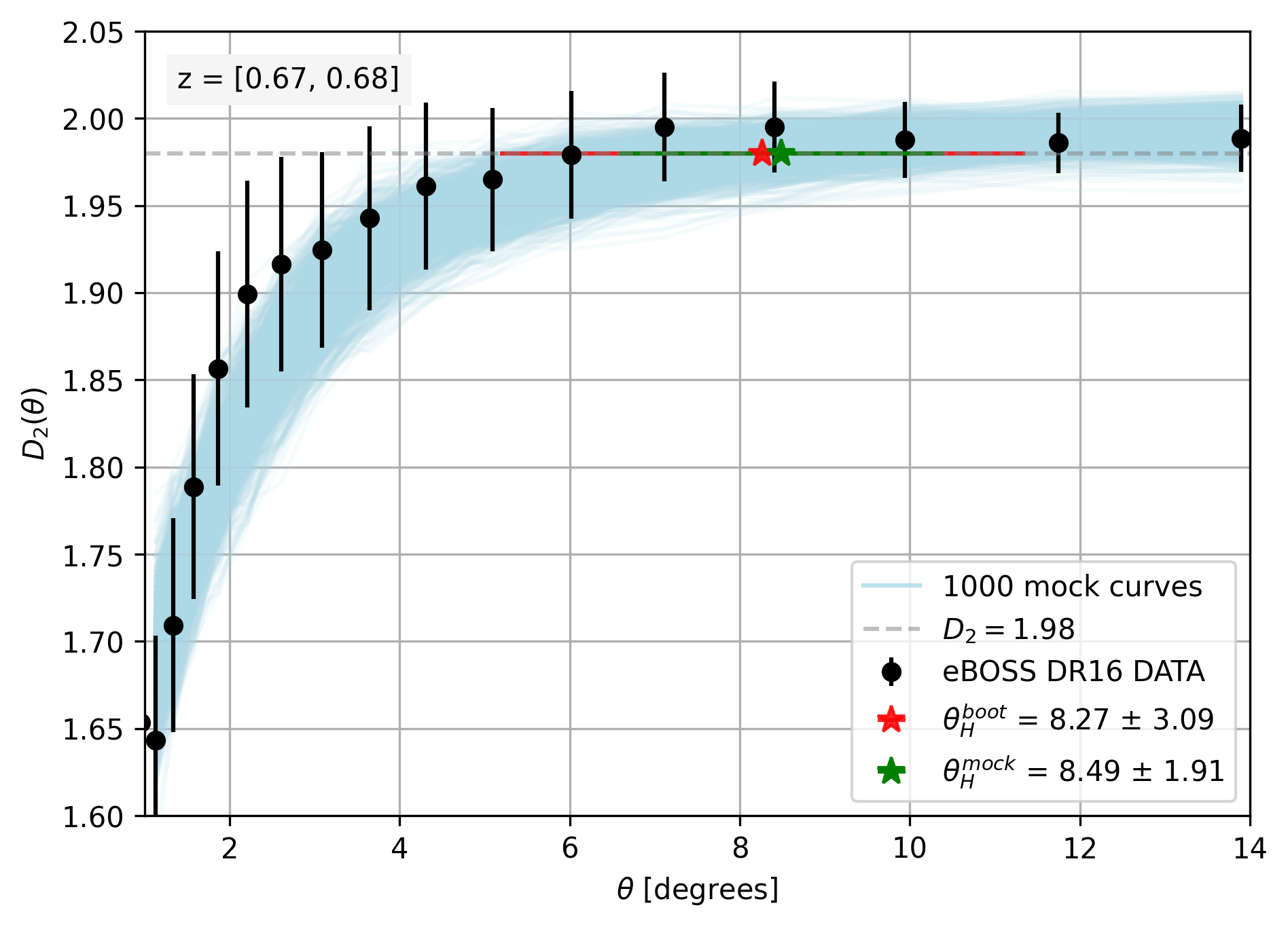}
    \hfill
    \includegraphics[width=0.45\textwidth]{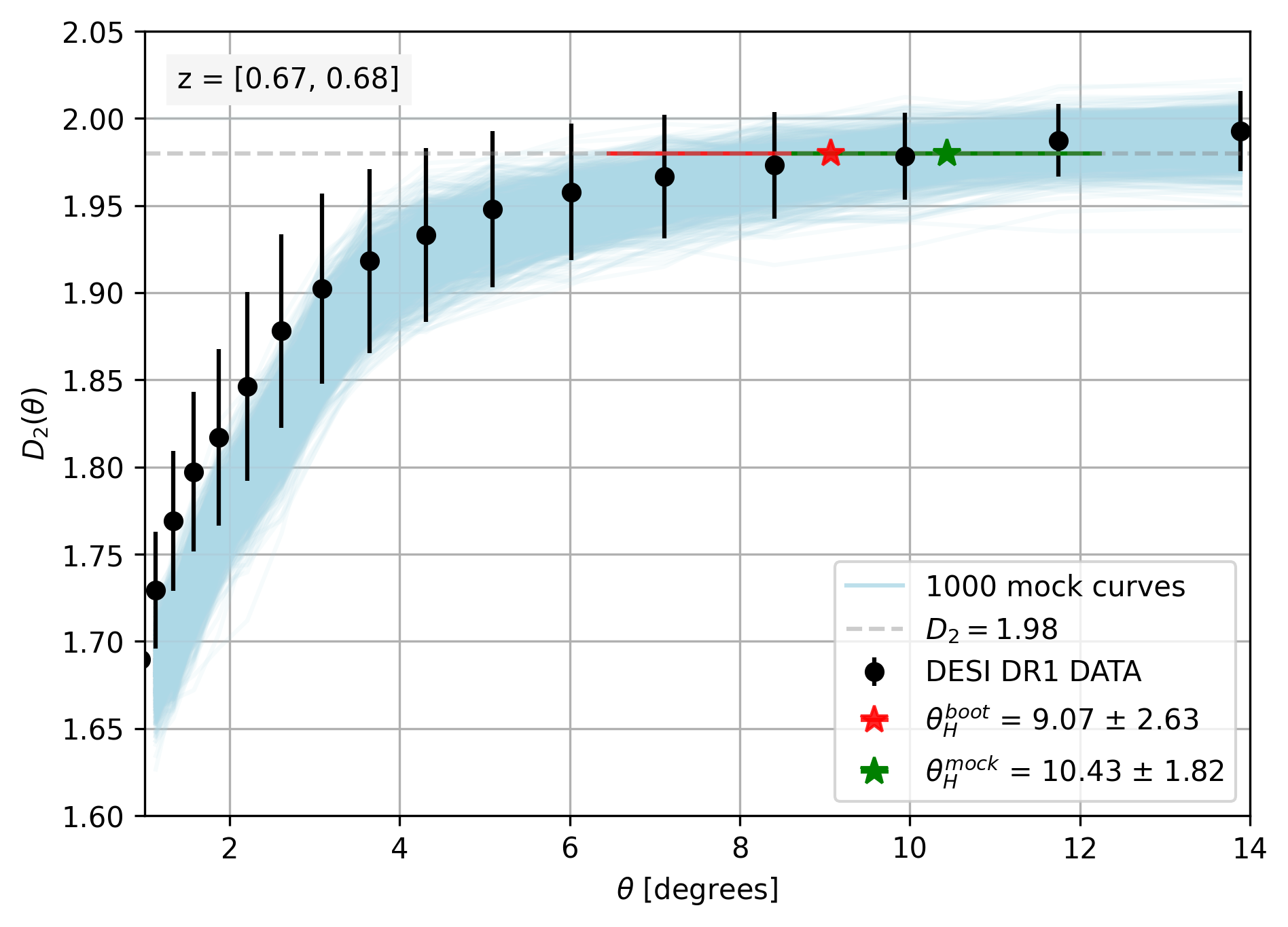}
    \caption{Measurements of the correlation dimension, $D_2(\theta)$, in the same redshift range $0.67 < z < 0.68$. The left panel shows measurements using the eBOSS DR16 catalog, while the right panel shows measurements using the DESI DR1 catalog, both shows the North Galactic Cap (NGC). }
    \label{fig:thetah_d2curves_NGC_067z068}
\end{figure}

\begin{figure}[h!]
    \centering
    \includegraphics[width=0.45\textwidth]{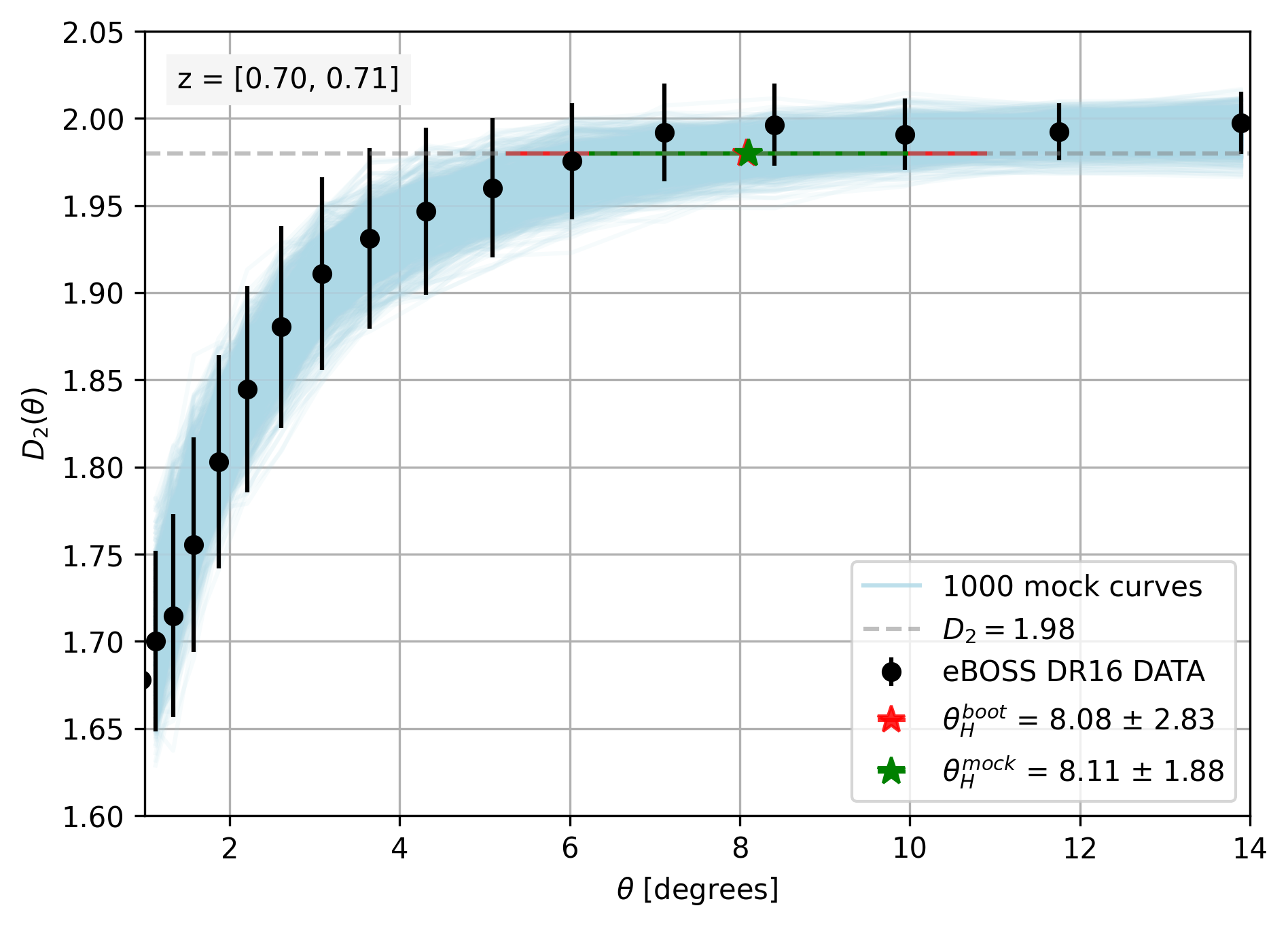}
    \hfill
    \includegraphics[width=0.45\textwidth]{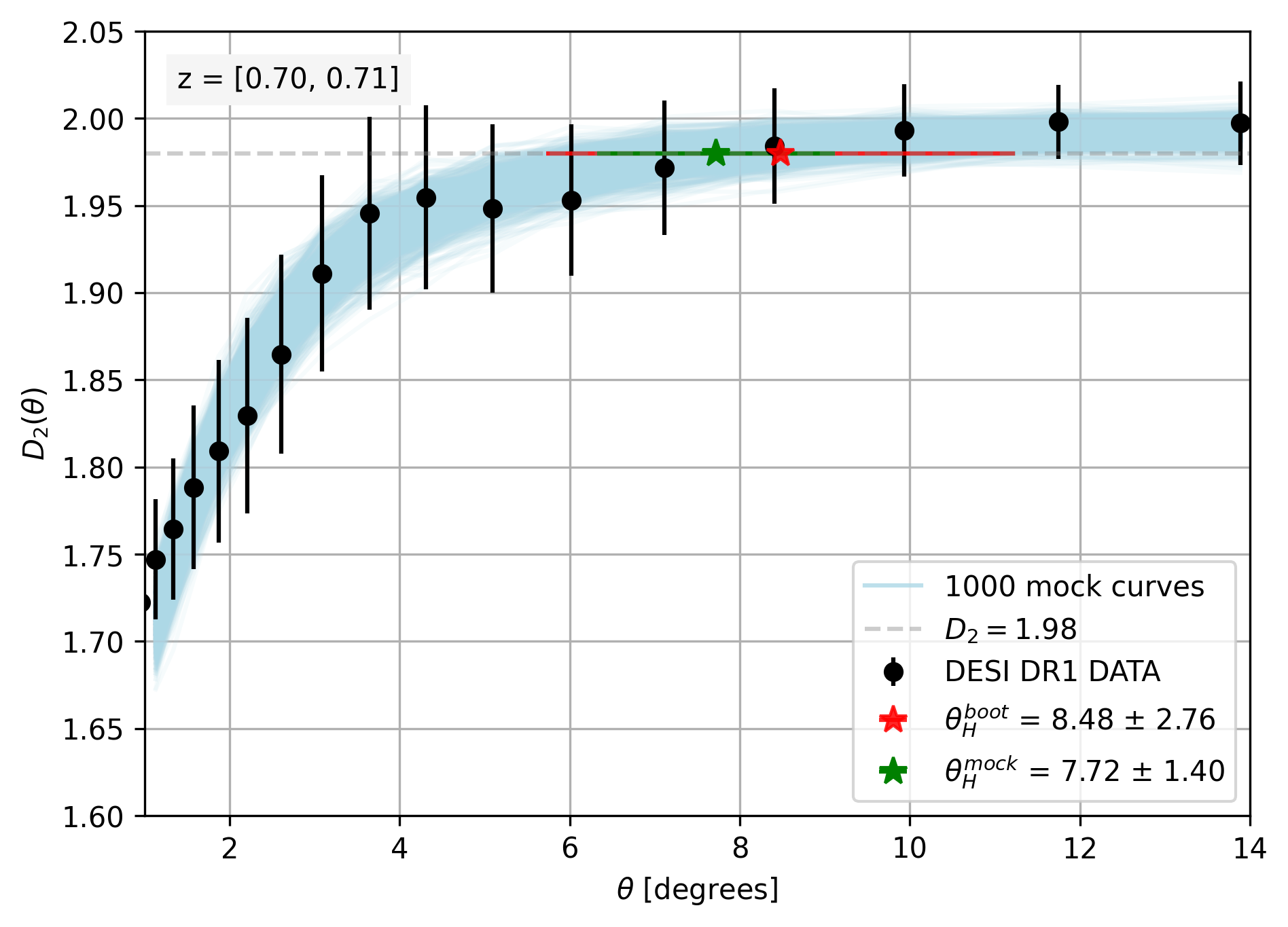}
    \caption{Measurements of the correlation dimension, $D_2(\theta)$, in the same redshift range $0.70 < z < 0.71$. The left panel shows measurements using the eBOSS DR16 catalog, while the right panel shows measurements using the DESI DR1 catalog, both shows the North Galactic Cap (NGC). }
    \label{fig:thetah_d2curves_NGC_070z071}
\end{figure}

\begin{figure}[h!]
    \centering
    \includegraphics[width=0.45\textwidth]{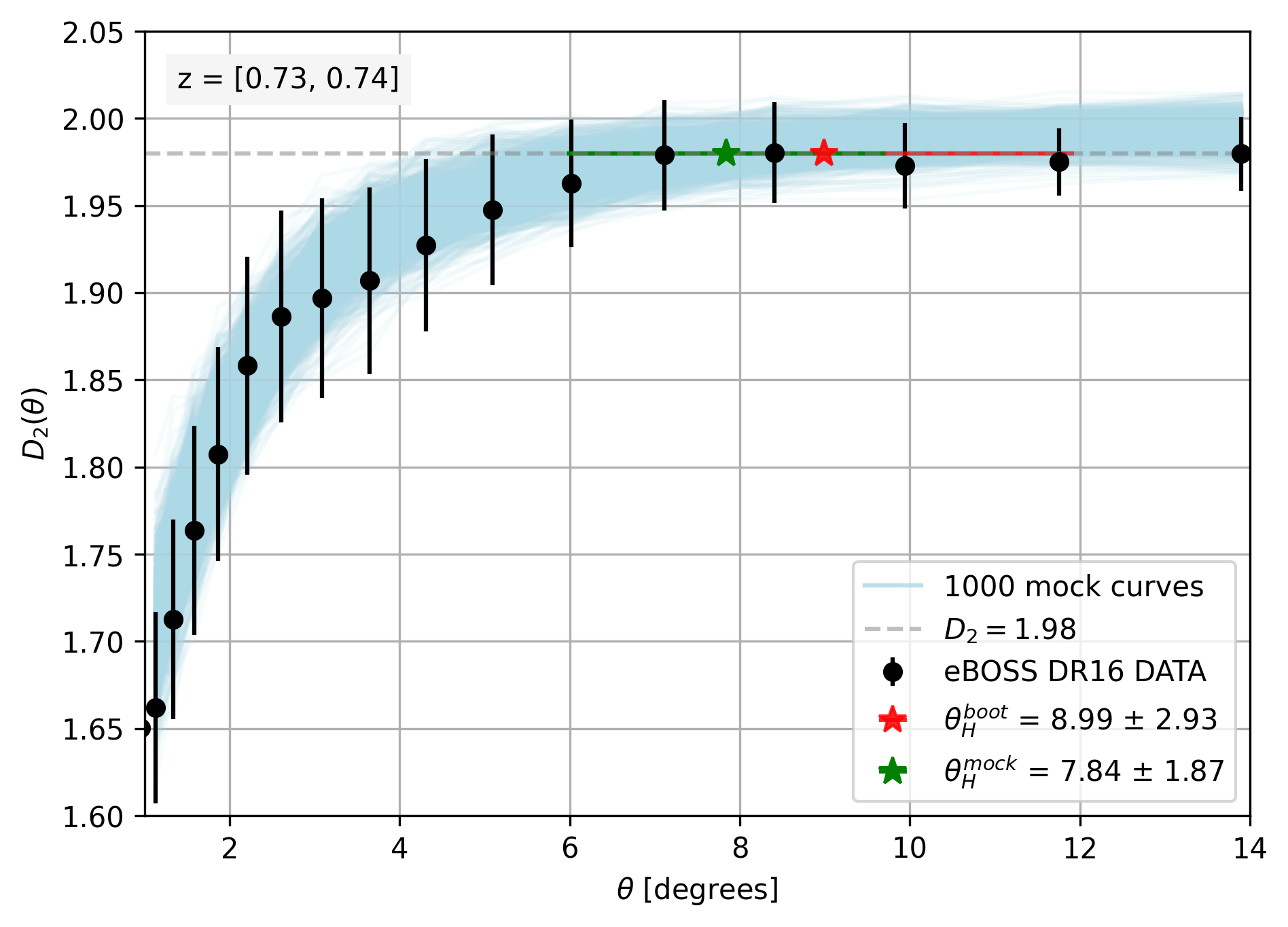}
    \hfill
    \includegraphics[width=0.45\textwidth]{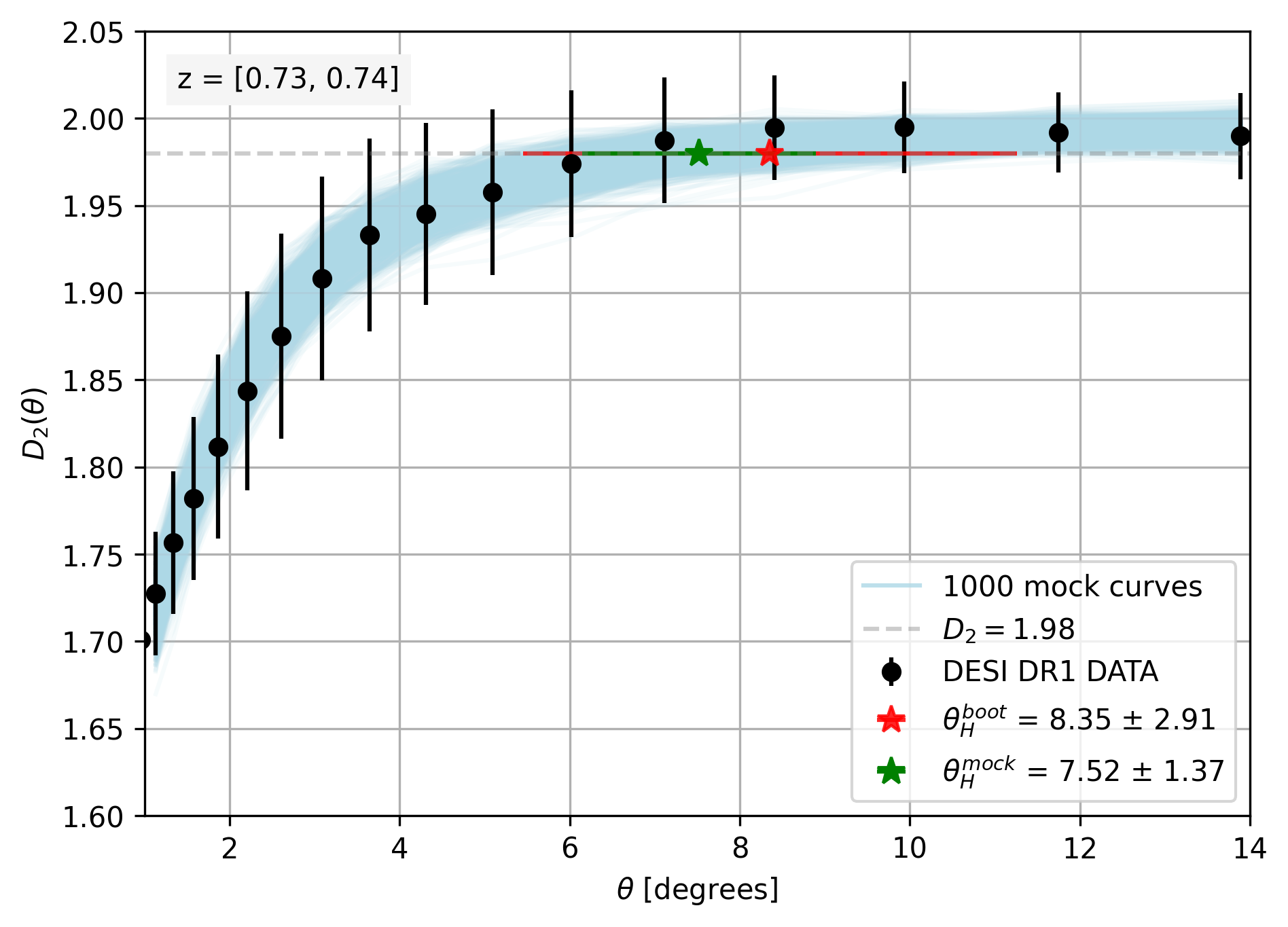}
    \caption{Measurements of the correlation dimension, $D_2(\theta)$, in the same redshift range $0.73 < z < 0.74$. The left panel shows measurements using the eBOSS DR16 catalog, while the right panel shows measurements using the DESI DR1 catalog, both shows the North Galactic Cap (NGC). }
    \label{fig:thetah_d2curves_NGC_073z074}
\end{figure}

Figures \ref{fig:thetah_d2curves_NGC_067z068}, \ref{fig:thetah_d2curves_NGC_070z071}, and \ref{fig:thetah_d2curves_NGC_073z074} show this comparative result, where the values of $\theta_H^{boot}$ and $\theta_H^{mock}$ are provided. For each redshift range, the $\theta_H^{boot}$ values from each catalog agree with each other within their respective uncertainties. Moreover, when comparing $\theta_H^{mock}$ with the corresponding $\theta_H^{boot}$ values at each redshift range, we also find agreement within the respective uncertainties. The increased galaxy density of DESI DR1 relative to eBOSS DR16 translates directly into reduced uncertainties on $\theta_H$, as seen in the narrower confidence bands in Figures \ref{fig:thetah_d2curves_NGC_067z068}, \ref{fig:thetah_d2curves_NGC_070z071}, and \ref{fig:thetah_d2curves_NGC_073z074}, highlighting one of the key advantages of the new dataset.

\begin{figure}[h!]
    \centering
    \includegraphics[width=0.60\textwidth]{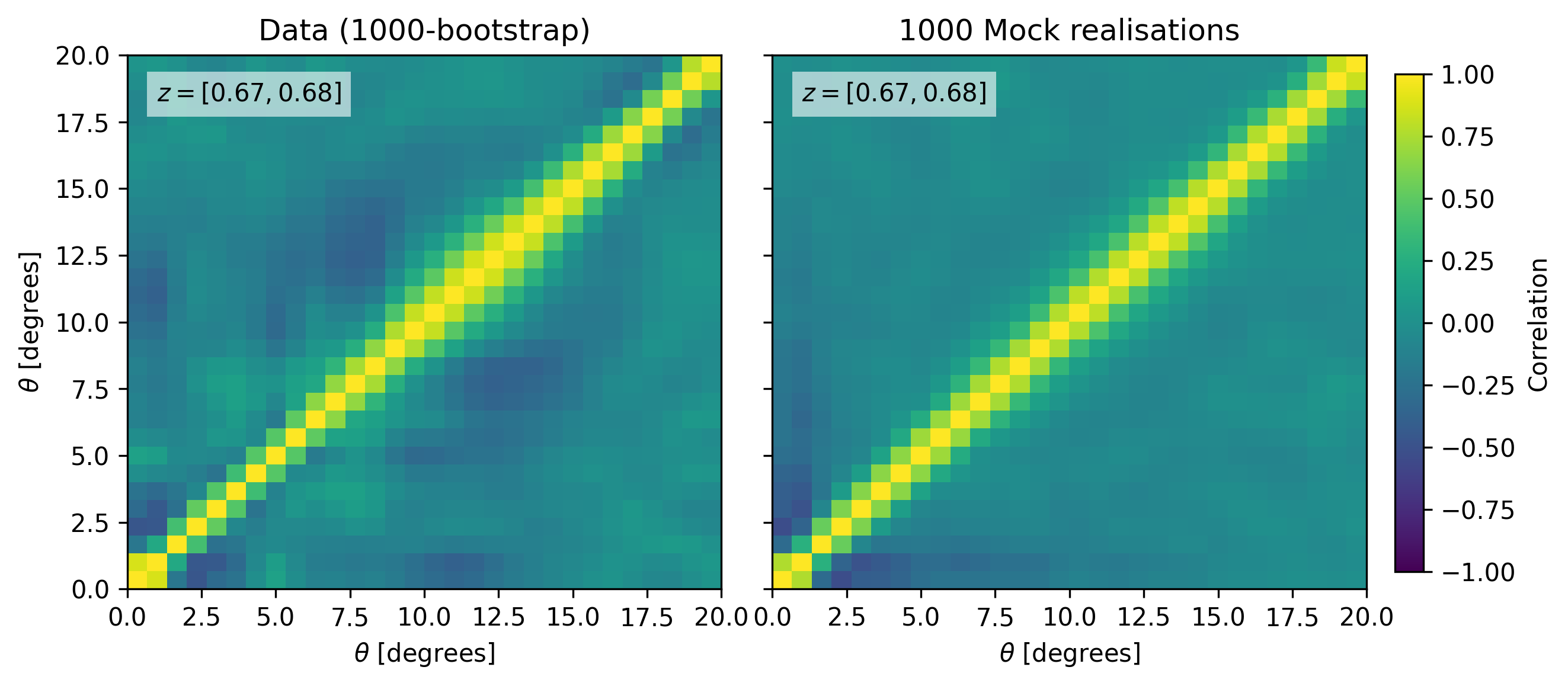}
    \vspace{0.2cm}
    \includegraphics[width=0.60\textwidth]{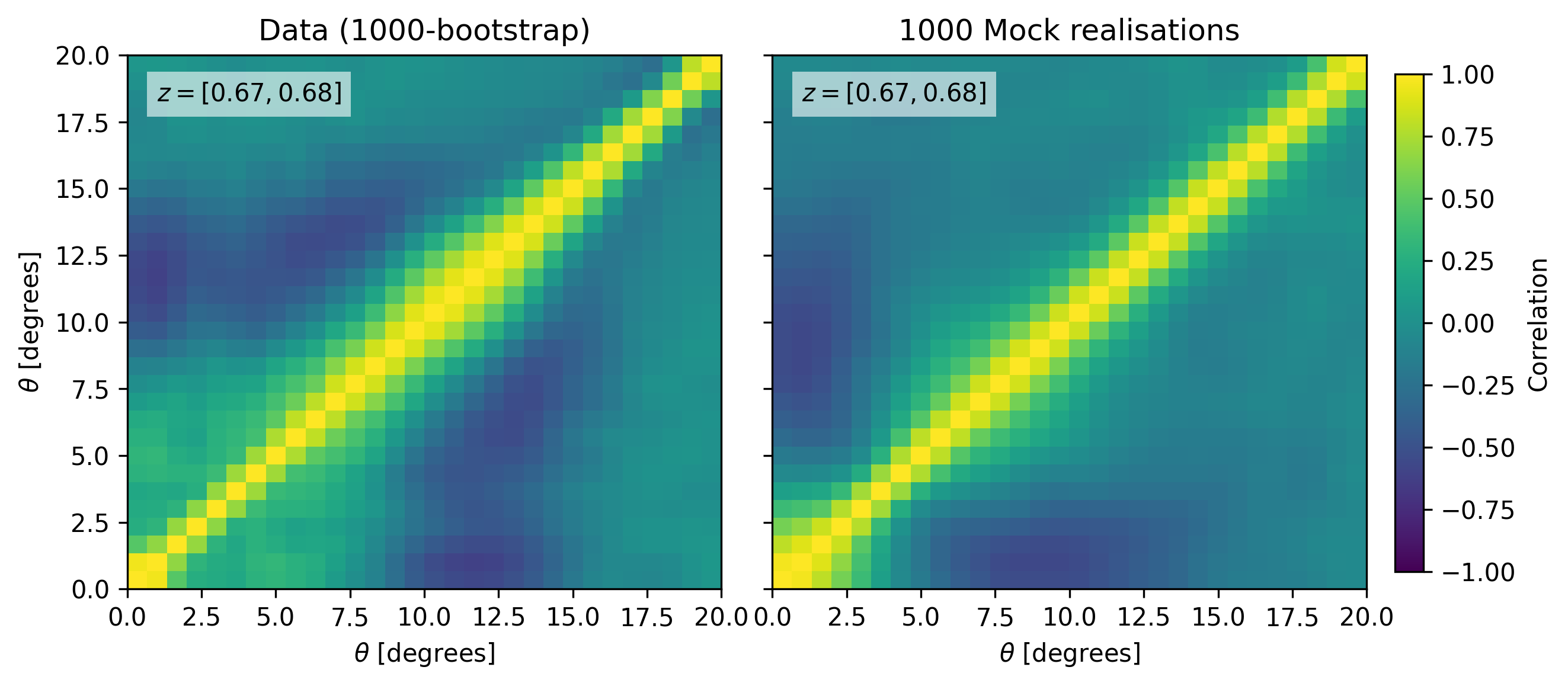}
    \caption{Correlation matrices for correlation dimension, $D_2(\theta)$, in the same redshift range $0.67< z < 0.68$. The left panels show the results obtained from 1000-bootstrap resamplings, and the right panels correspond to the results obtained from 1000 mock catalogs. The upper panel shows the matrices using the eBOSS DR16 catalog, while the bottom panel shows the matrices using the DESI DR1 catalog, both shows the North Galactic Cap (NGC).}
    \label{fig:correlation:matrix_NGC_067z068}
\end{figure}

\begin{figure}[h!]
    \centering
    \includegraphics[width=0.60\textwidth]{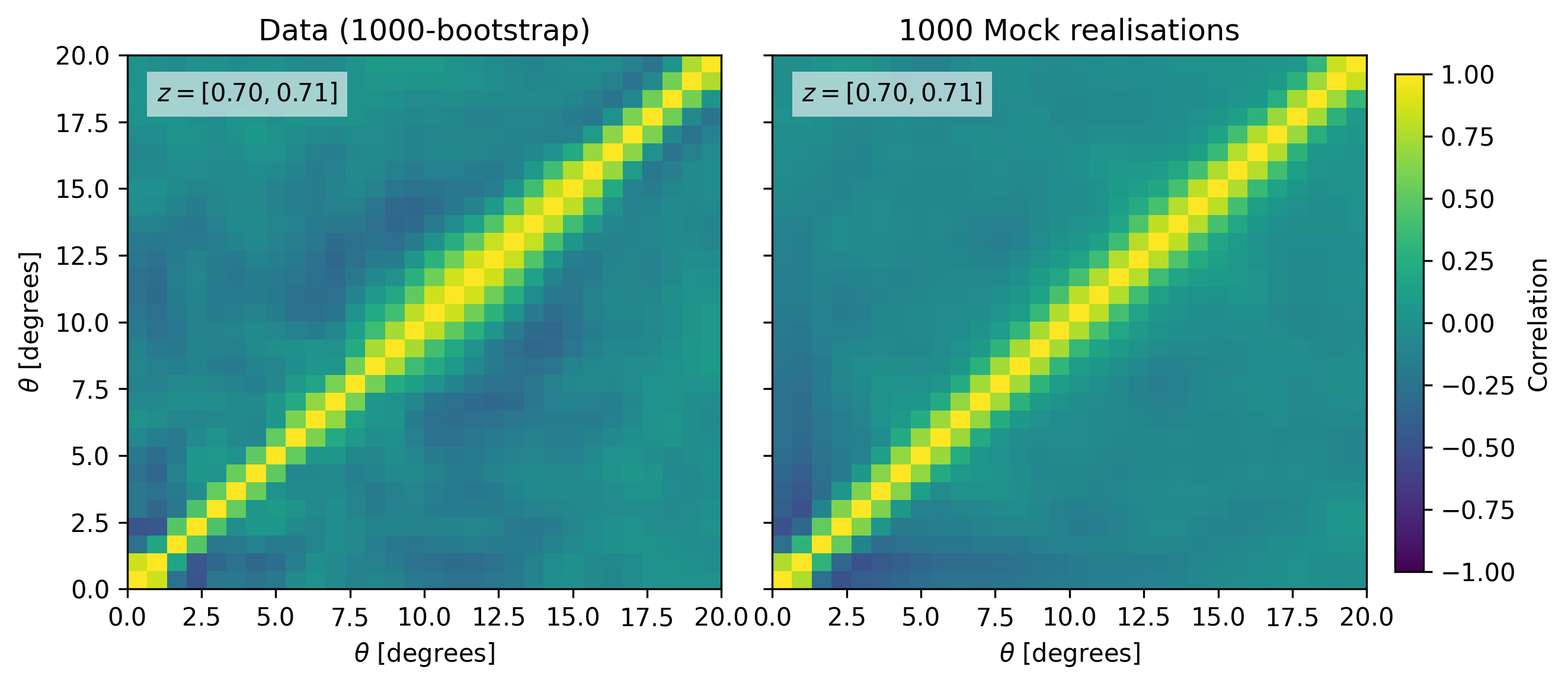}
    \vspace{0.2cm}
    \includegraphics[width=0.60\textwidth]{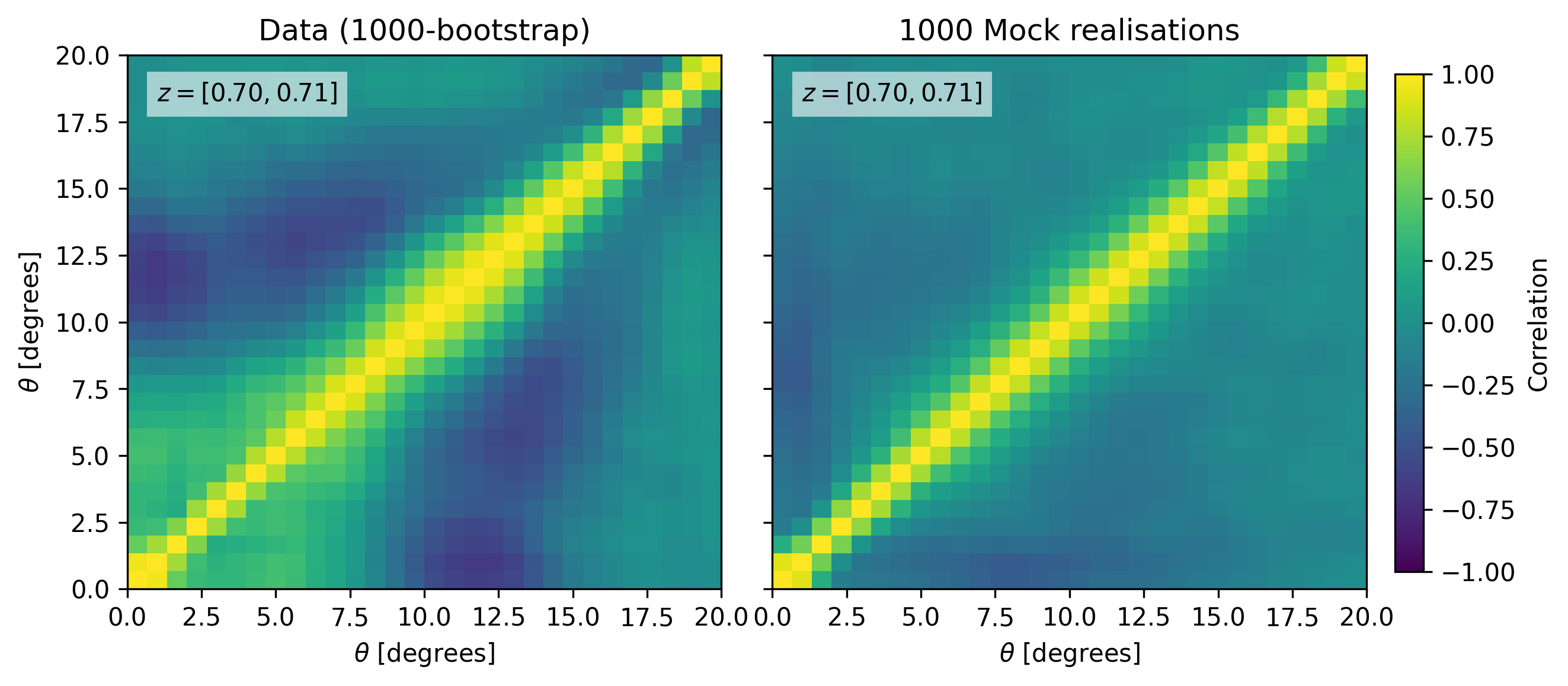}
    \caption{ Correlation matrices for correlation dimension, $D_2(\theta)$, in the same redshift range $0.70< z < 0.71$. The left panels show the results obtained from 1000-bootstrap resamplings, and the right panels correspond to the results obtained from 1000 mock catalogs. The upper panel shows the matrices using the eBOSS DR16 catalog, while the bottom panel shows the matrices using the DESI DR1 catalog, both shows the North Galactic Cap (NGC).}
    \label{fig:correlationmatrix_NGC_070z071}
\end{figure}

\begin{figure}[h!]
    \centering
    \includegraphics[width=0.60\textwidth]{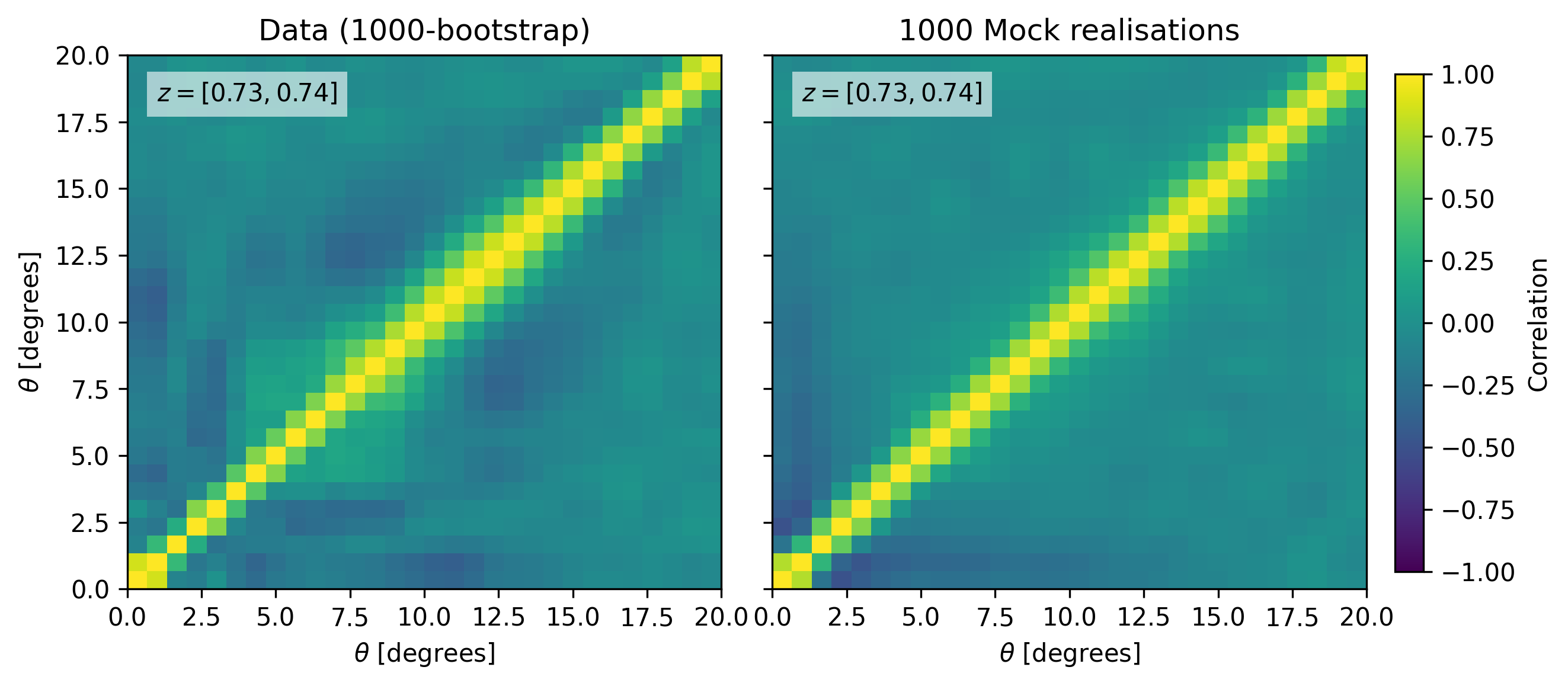}
    \vspace{0.2cm}
    \includegraphics[width=0.60\textwidth]{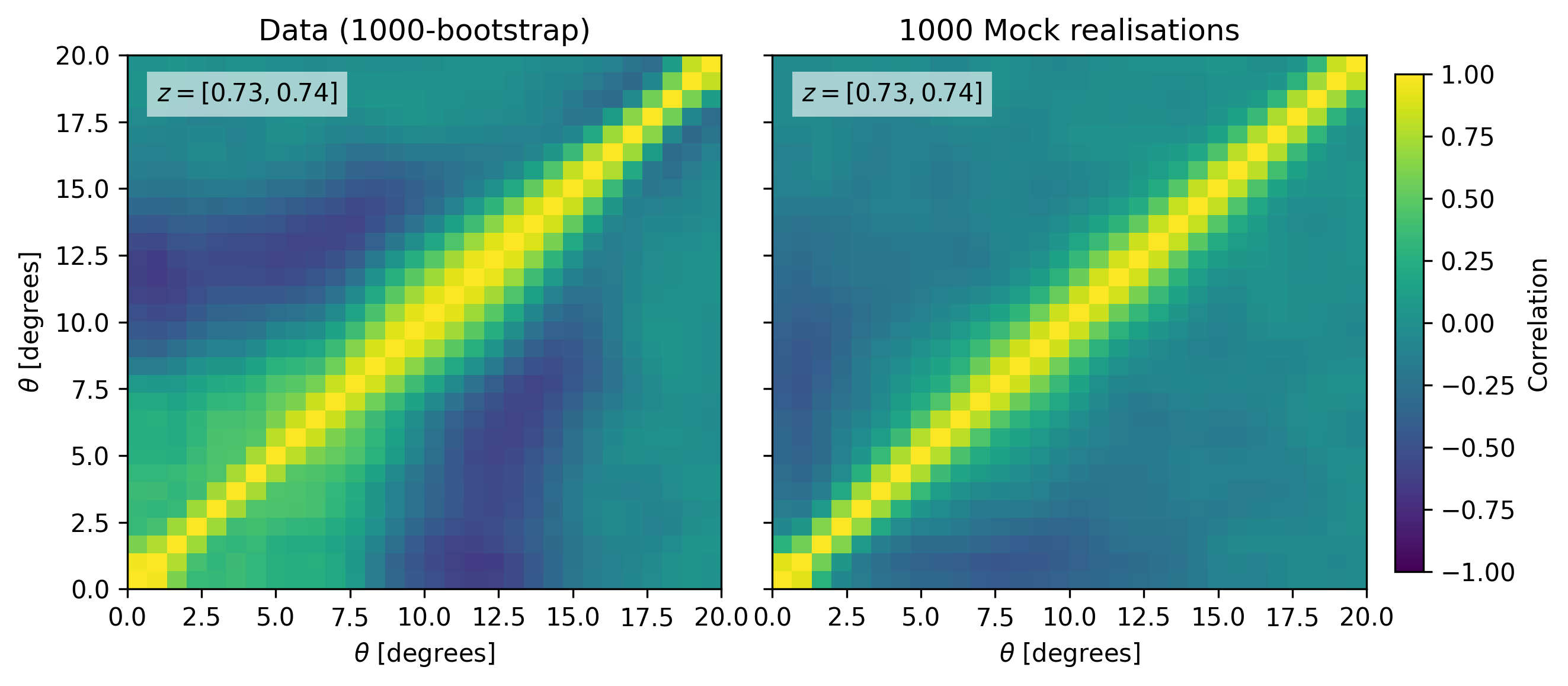}
    \caption{Correlation matrices for correlation dimension, $D_2(\theta)$, in the same redshift range $0.73< z < 0.74$. The left panels show the results obtained from 1000-bootstrap resamplings, and the right panels correspond to the results obtained from 1000 mock catalogs. The upper panel shows the matrices using the eBOSS DR16 catalog, while the bottom panel shows the matrices using the DESI DR1 catalog, both shows the North Galactic Cap (NGC).}
    \label{fig:correlationmatrix_NGC_073z074}
\end{figure}

Figures \ref{fig:correlation:matrix_NGC_067z068}, \ref{fig:correlationmatrix_NGC_070z071} and \ref{fig:correlationmatrix_NGC_073z074} show a comparative analysis of the $D_2(\theta)$ correlation matrices between the surveys for each redshift range. We can observe that the DESI DR1 data exhibit slightly stronger correlations than the eBOSS DR16 data, particularly at smaller angular scales. This pattern is consistently observed across all selected redshift ranges. We also observe good agreement between the correlation matrices obtained from the mocks and those derived from the real data.

\begin{table}[h!]
\centering
\resizebox{0.8\textwidth}{!}{
\begin{tabular}{lcl|clc|l|clc|}
\cline{4-6} \cline{8-10}
\multicolumn{2}{c}{\multirow{2}{*}{}}                              & \multirow{6}{*}{} & \multicolumn{3}{c|}{NGC}                                                                & \multirow{6}{*}{} & \multicolumn{3}{c|}{SGC}                                                                \\ \cline{4-6} \cline{8-10} 
\multicolumn{2}{c}{}                                               &                   & \multicolumn{1}{c|}{eBOSS DR16}  & \multicolumn{1}{l|}{\multirow{5}{*}{}} & DESI DR1    &                   & \multicolumn{1}{c|}{eBOSS DR16}  & \multicolumn{1}{l|}{\multirow{5}{*}{}} & DESI DR1    \\ \cline{1-2} \cline{4-4} \cline{6-6} \cline{8-8} \cline{10-10} 
\multicolumn{1}{|c|}{$z$}           & \multicolumn{1}{c|}{$\bar{z}$} &                   & \multicolumn{1}{c|}{\# galaxies} & \multicolumn{1}{l|}{}                  & \# galaxies &                   & \multicolumn{1}{c|}{\# galaxies} & \multicolumn{1}{l|}{}                  & \# galaxies \\ \cline{1-2} \cline{4-4} \cline{6-6} \cline{8-8} \cline{10-10} 
\multicolumn{1}{|l|}{0.67 - 0.68} & \multicolumn{1}{c|}{0.675}     &                   & \multicolumn{1}{c|}{4 209}       & \multicolumn{1}{l|}{}                  & 25 002      &                   & \multicolumn{1}{c|}{2 632}       & \multicolumn{1}{l|}{}                  & 10 990      \\ \cline{1-2} \cline{4-4} \cline{6-6} \cline{8-8} \cline{10-10} 
\multicolumn{1}{|l|}{0.70 - 0.71} & \multicolumn{1}{c|}{0.705}     &                   & \multicolumn{1}{c|}{4 073}       & \multicolumn{1}{l|}{}                  & 25 828      &                   & \multicolumn{1}{c|}{2 651}       & \multicolumn{1}{l|}{}                  & 11 555      \\ \cline{1-2} \cline{4-4} \cline{6-6} \cline{8-8} \cline{10-10} 
\multicolumn{1}{|l|}{0.73 - 0.74} & \multicolumn{1}{c|}{0.735}     &                   & \multicolumn{1}{c|}{4 210}       & \multicolumn{1}{l|}{}                  & 27 568      &                   & \multicolumn{1}{c|}{2 656}       & \multicolumn{1}{l|}{}                  & 11 773      \\ \cline{1-2} \cline{4-6} \cline{8-10} 
\end{tabular}
}
\caption{Number of galaxies per redshift range for each catalog and their respective hemispheres.}
\label{tab:num_galaxies}
\end{table}

\FloatBarrier
\subsection{DESI DR1 NGC x SGC}
In addition, an extended comparison between the NGC and SGC was performed for the DESI DR1 data. For this analysis, the redshift ranges $0.41 < z < 0.42$, $0.89 < z < 0.90$ and $1.08 < z < 1.09$ were selected, in order to represent the lower, middle and upper end of the catalog coverage (see Figure \ref{fig:histogram_NGCandSGC}). We note that the redshift coverage of the DESI DR1 catalog is significantly larger than that of the SDSS eBOSS DR16 catalog, for this reason the intervals chosen for this comparison are more widely spaced.

\begin{figure}[h!]
    \centering
    \includegraphics[width=0.45\textwidth]{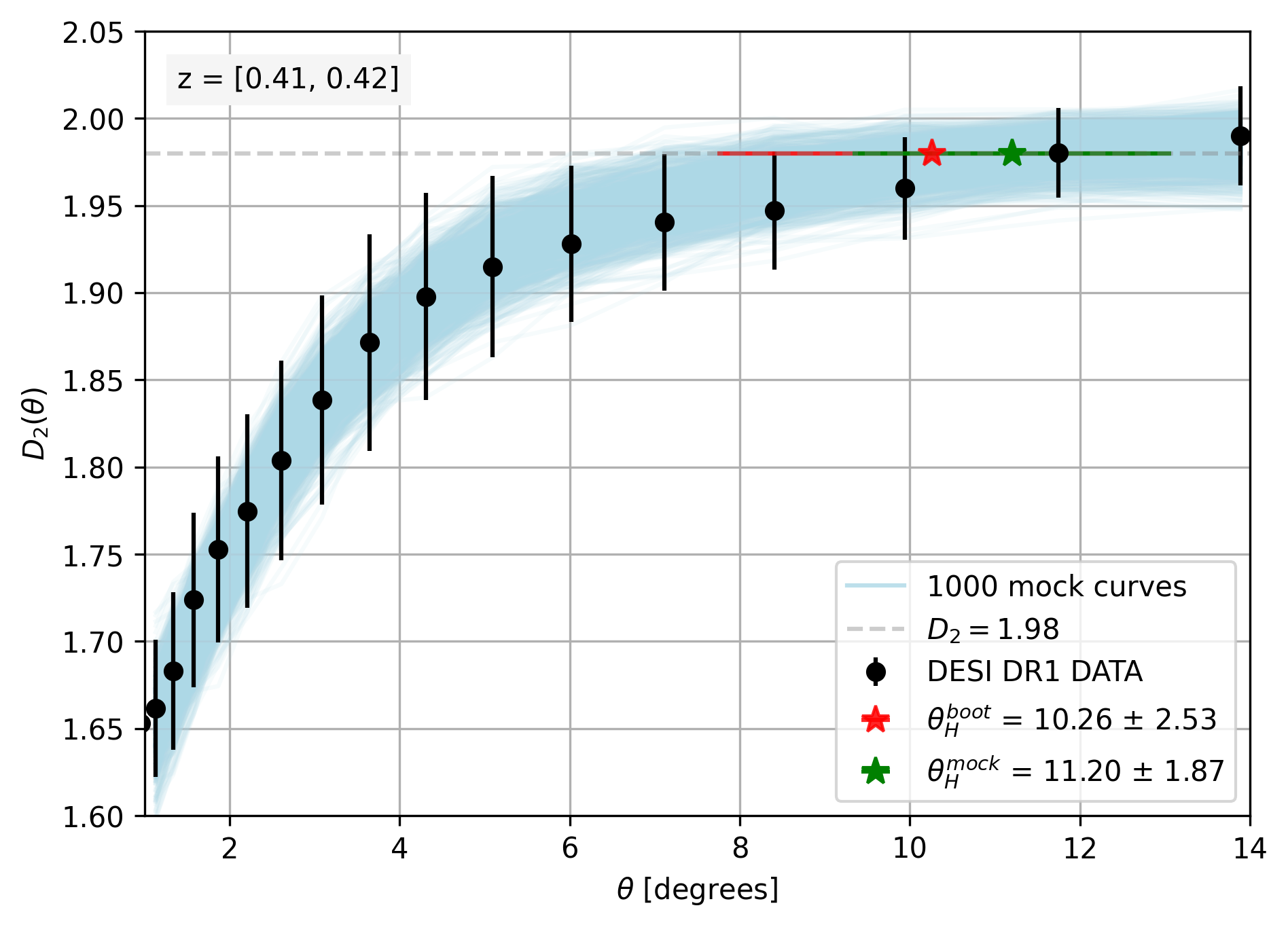}
    \hfill
    \includegraphics[width=0.45\textwidth]{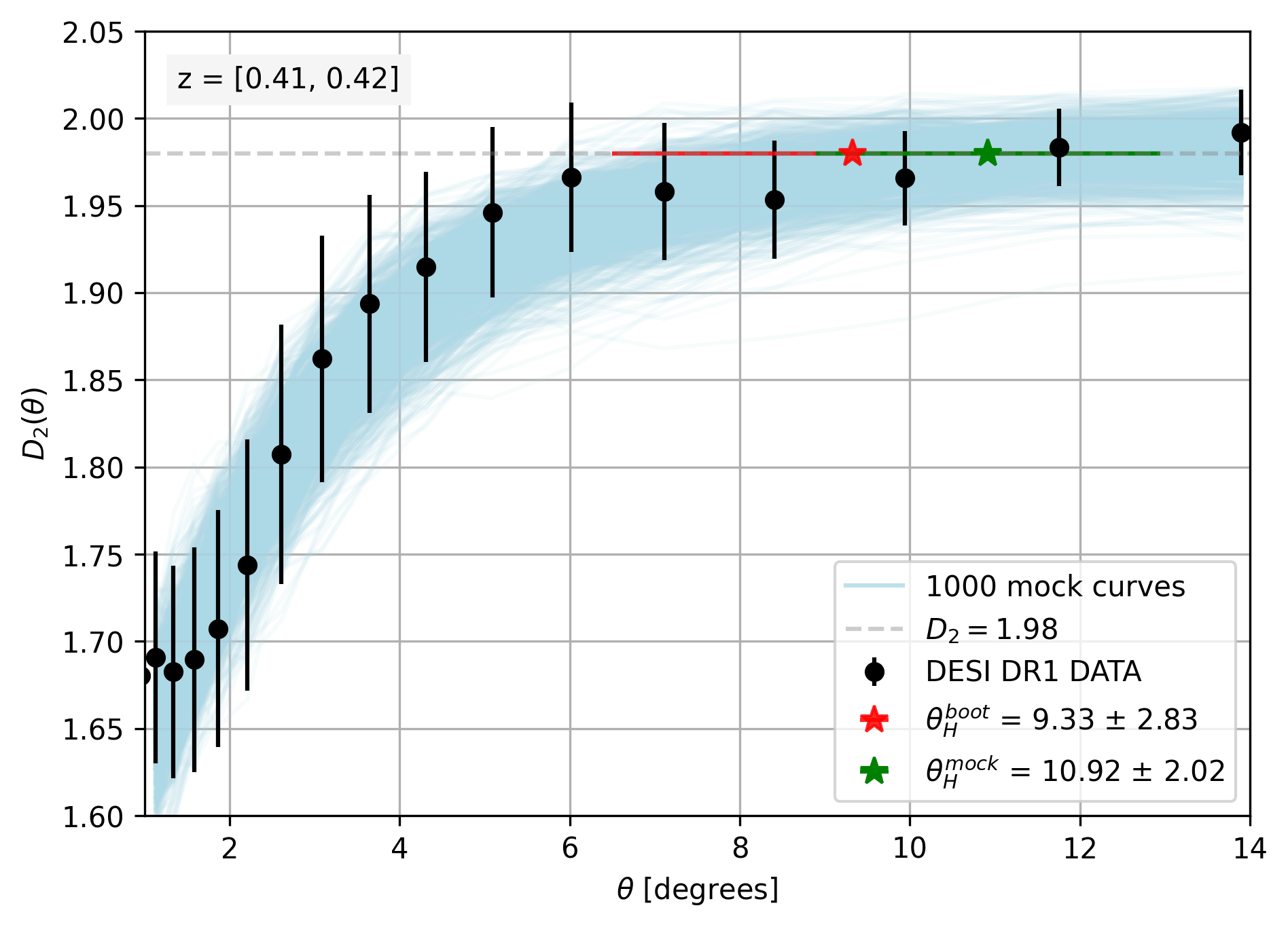}
    \caption{Measurements of the correlation dimension, $D_2(\theta)$, in the same redshift range $0.41 < z < 0.42$. The left panel shows measurements using North Galactic Cap (NGC), while the right panel shows measurements using the South Galactic Cap (SGC), both shows the DESI DR1 catalog.  }
    \label{fig:thetah_d2curves_NGCxSGC_041z042}
\end{figure}

\begin{figure}[h!]
    \centering
    \includegraphics[width=0.45\textwidth]{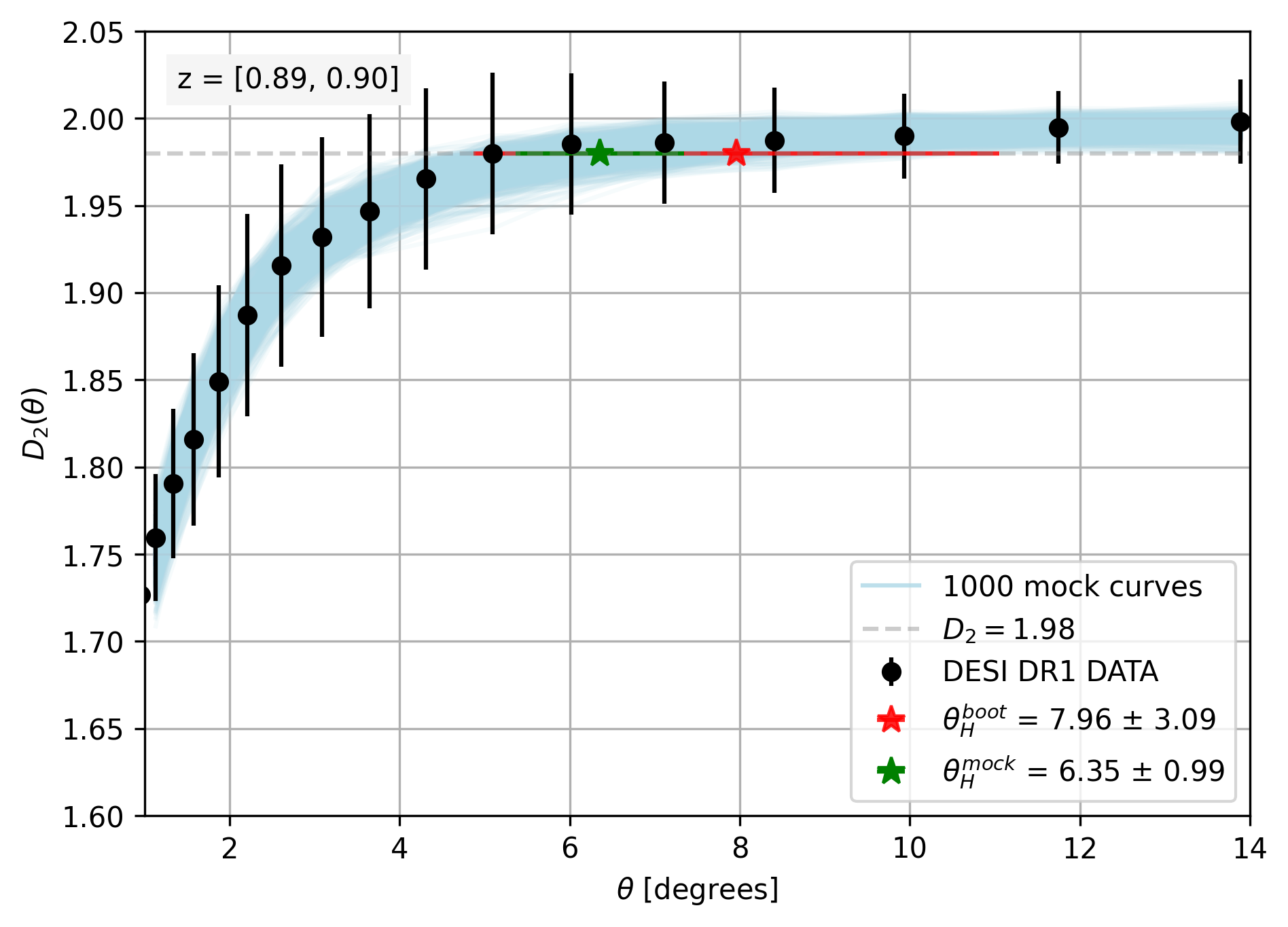}
    \hfill
    \includegraphics[width=0.45\textwidth]{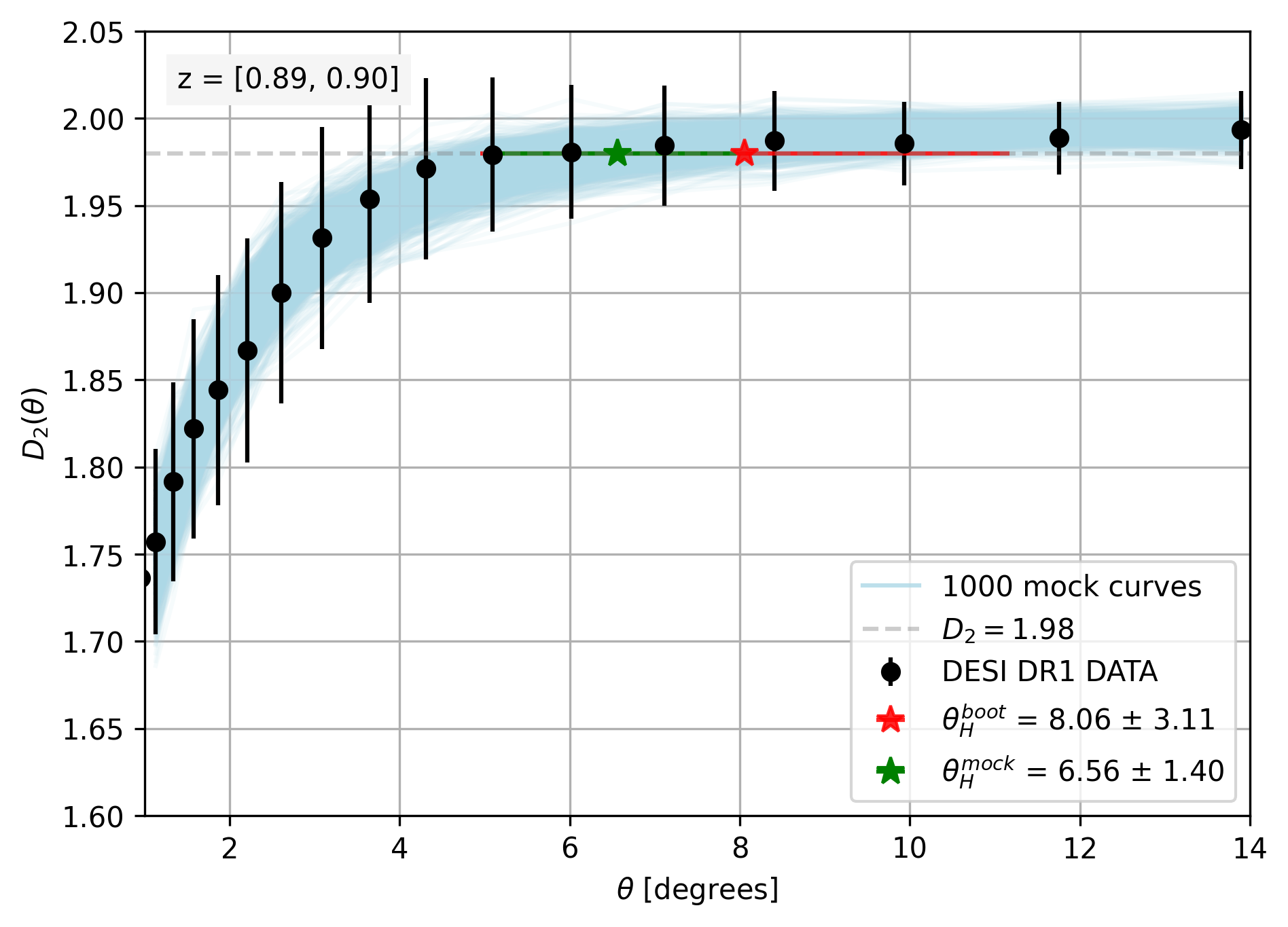}
    \caption{Measurements of the correlation dimension, $D_2(\theta)$, in the same redshift range $0.89 < z < 0.90$. The left panel shows measurements using North Galactic Cap (NGC), while the right panel shows measurements using the South Galactic Cap (SGC), both shows the DESI DR1 catalog. }
    \label{fig:thetah_d2curves_NGCxSGC_089z090}
\end{figure}

\begin{figure}[h!]
    \centering
    \includegraphics[width=0.45\textwidth]{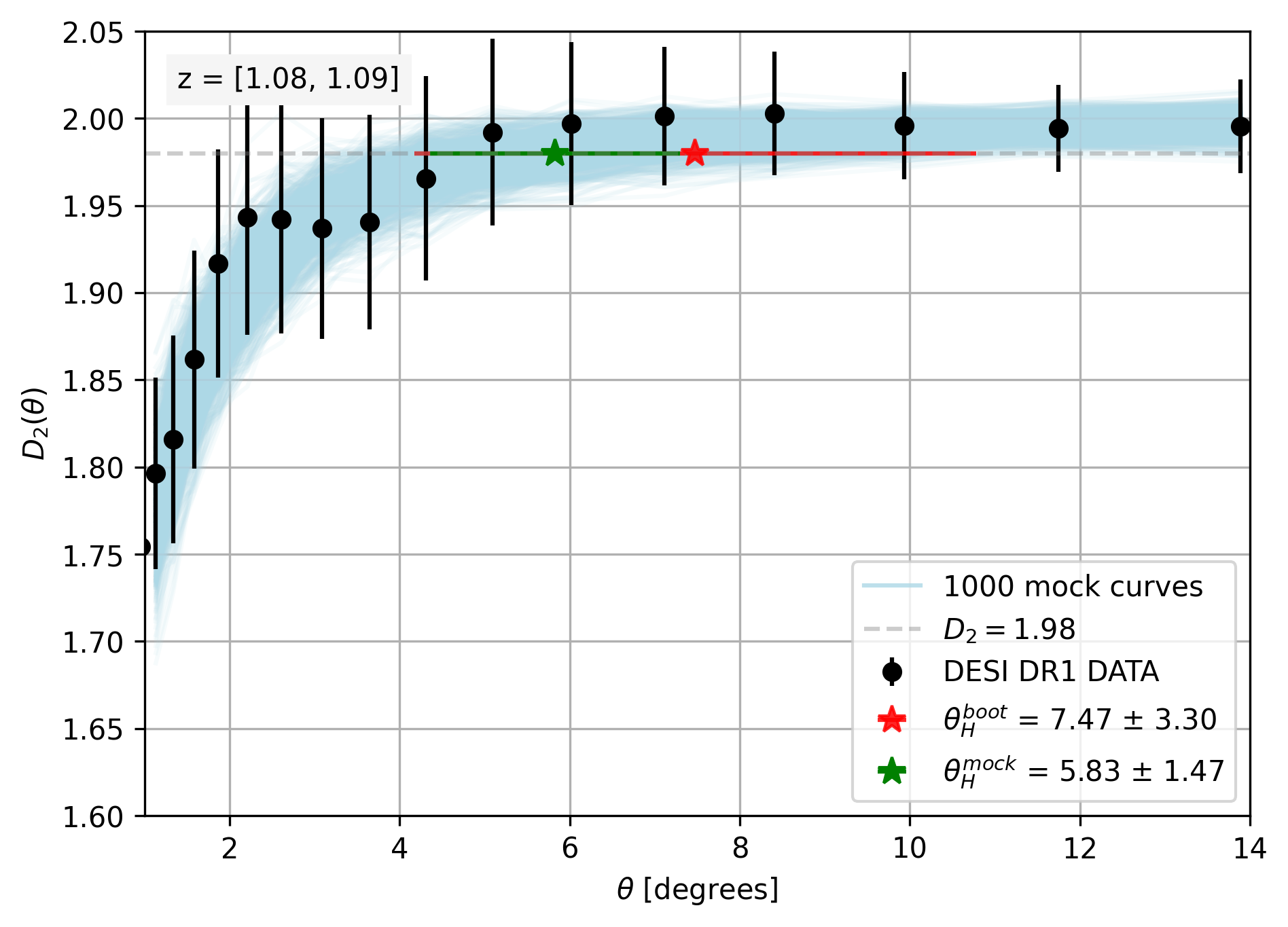}
    \hfill
    \includegraphics[width=0.45\textwidth]{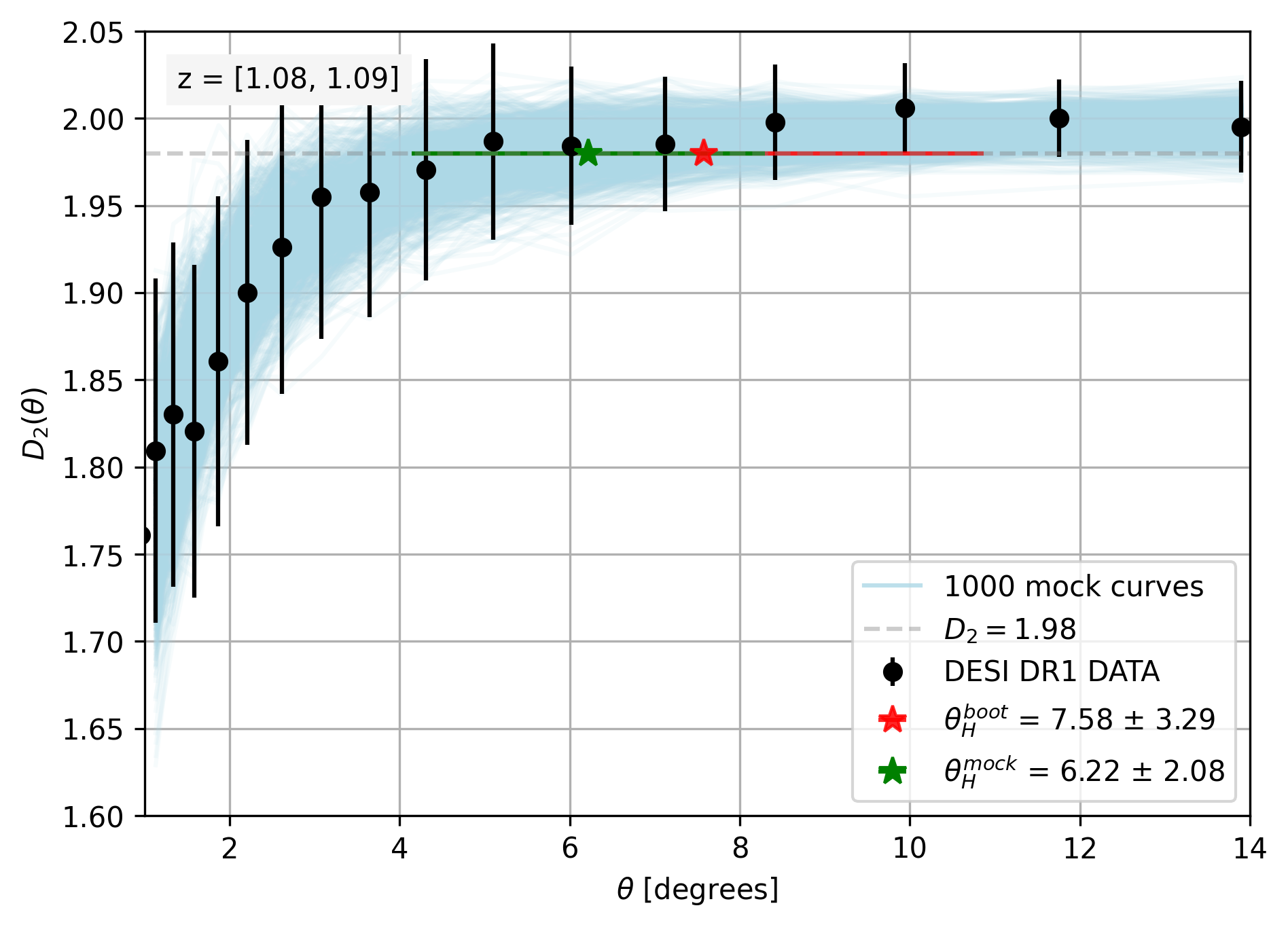}
    \caption{Measurements of the correlation dimension, $D_2(\theta)$, in the same redshift range $1.08 < z < 1.09$. The left panel shows measurements using North Galactic Cap (NGC), while the right panel shows measurements using the South Galactic Cap (SGC), both shows the DESI DR1 catalog.  }
    \label{fig:thetah_d2curves_NGCxSGC_108z109}
\end{figure}

The number density of LRG galaxies is presented at Table \ref{tab:num_galaxies_onlyDESI}, while the galaxy distributions per these redshift ranges are shown in the Appendix \ref{appendix_footprints} in Figure \ref{fig:onlyDESI_footprint_byz}. 
In the Figures \ref{fig:thetah_d2curves_NGCxSGC_041z042}, \ref{fig:thetah_d2curves_NGCxSGC_089z090} and \ref{fig:thetah_d2curves_NGCxSGC_108z109} we present the comparative analysis, displaying the values of $\theta_H^{boot}$ and $\theta_H^{mock}$. For all redshift ranges considered, the $\theta_H^{boot}$ meeasurements obtained from each sky portion are consistent with one another within their respective uncertainties. Furthermore, the $\theta_H^{mock}$ measurements are consistent with the corresponding $\theta_H^{boot}$ values within the associated uncertainties.

\begin{figure}[h!]
    \centering
    \includegraphics[width=0.60\textwidth]{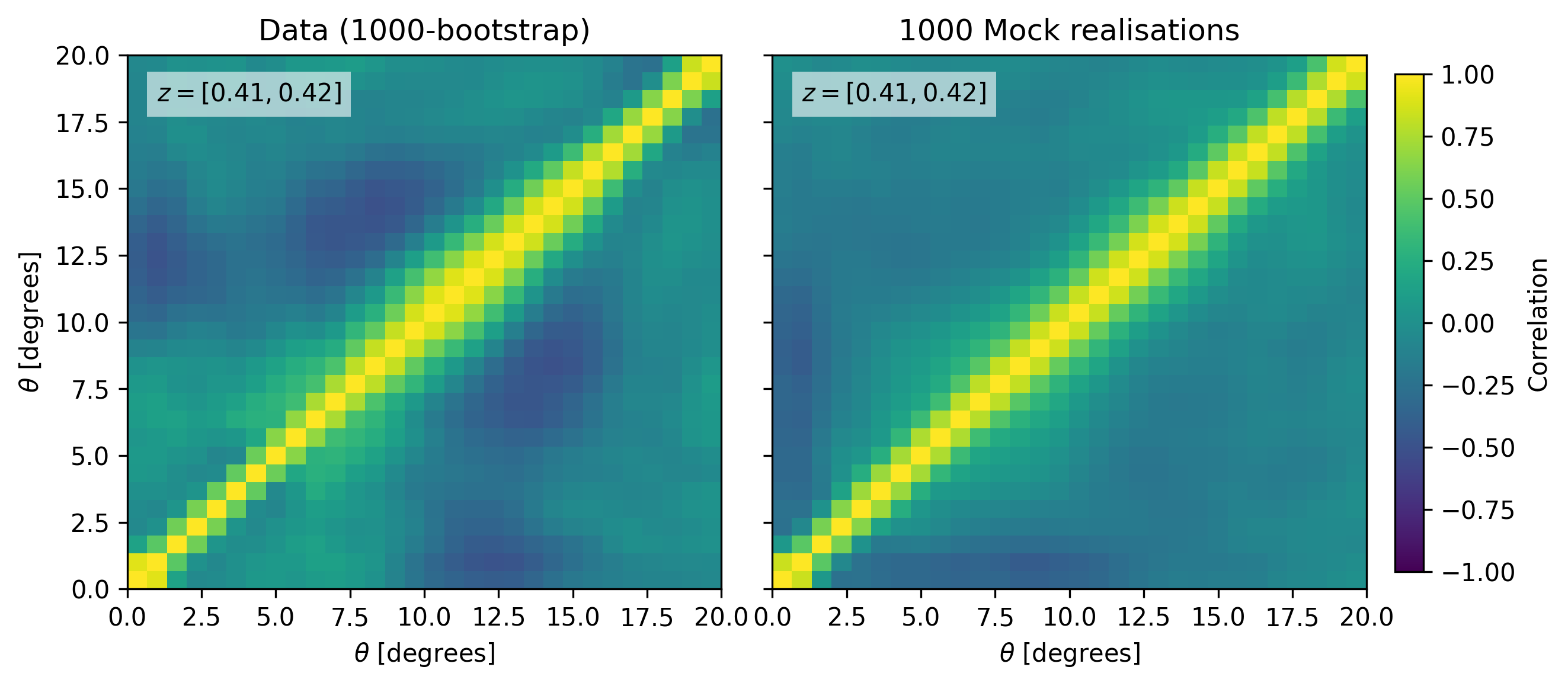}
    \vfill
    \includegraphics[width=0.60\textwidth]{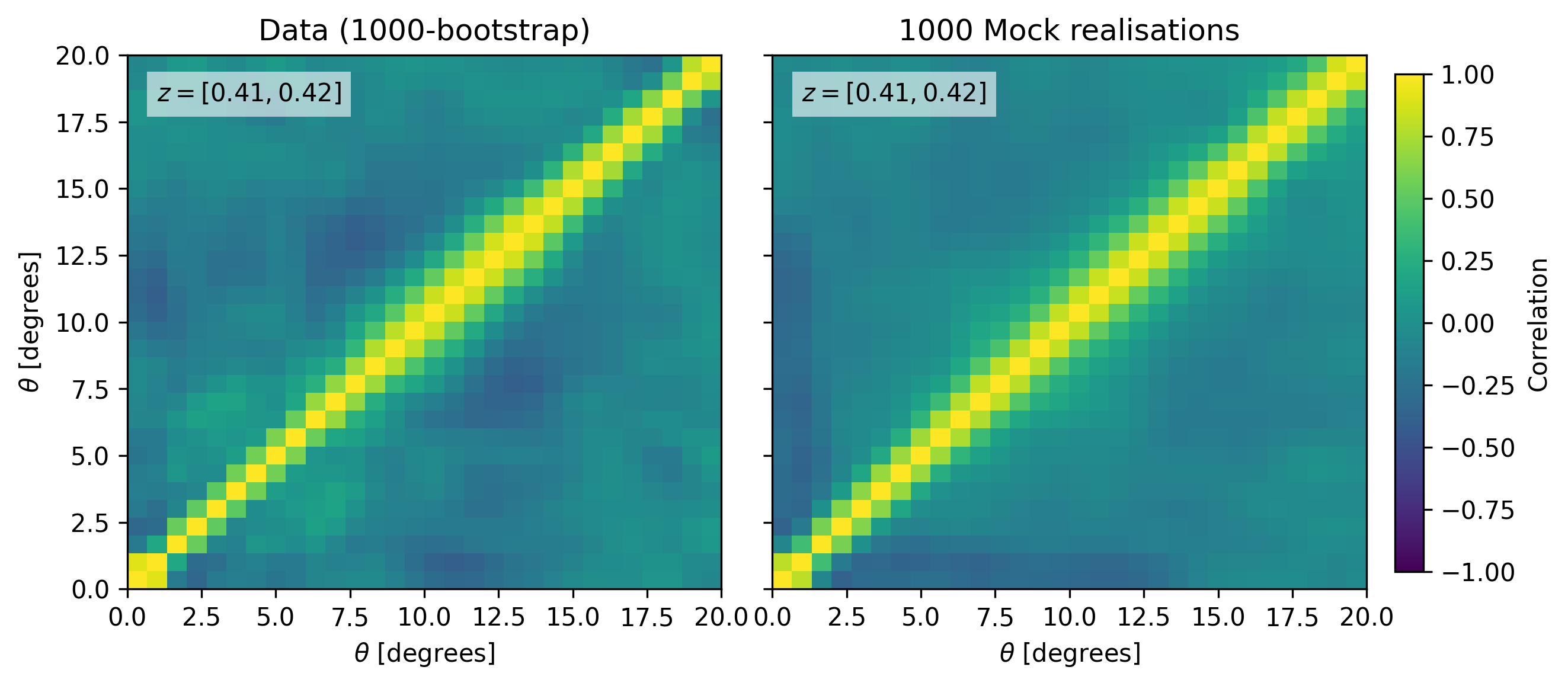}
    \caption{Correlation matrices for correlation dimension, $D_2(\theta)$, in the same redshift range $0.41< z < 0.42$. The left panels show the results obtained from 1000-bootstrap resamplings, and the right panels correspond to the results obtained from 1000 mock catalogs. The upper panel shows the matrices using the North Galactic Cap, while the bottom panel shows the matrices using the South Galactic Cap, both shows the DESI DR1 data.}
    \label{fig:correlation:matrix_NGCvsSGC_041z042}
\end{figure}

\begin{figure}[h!]
    \centering
    \includegraphics[width=0.60\textwidth]{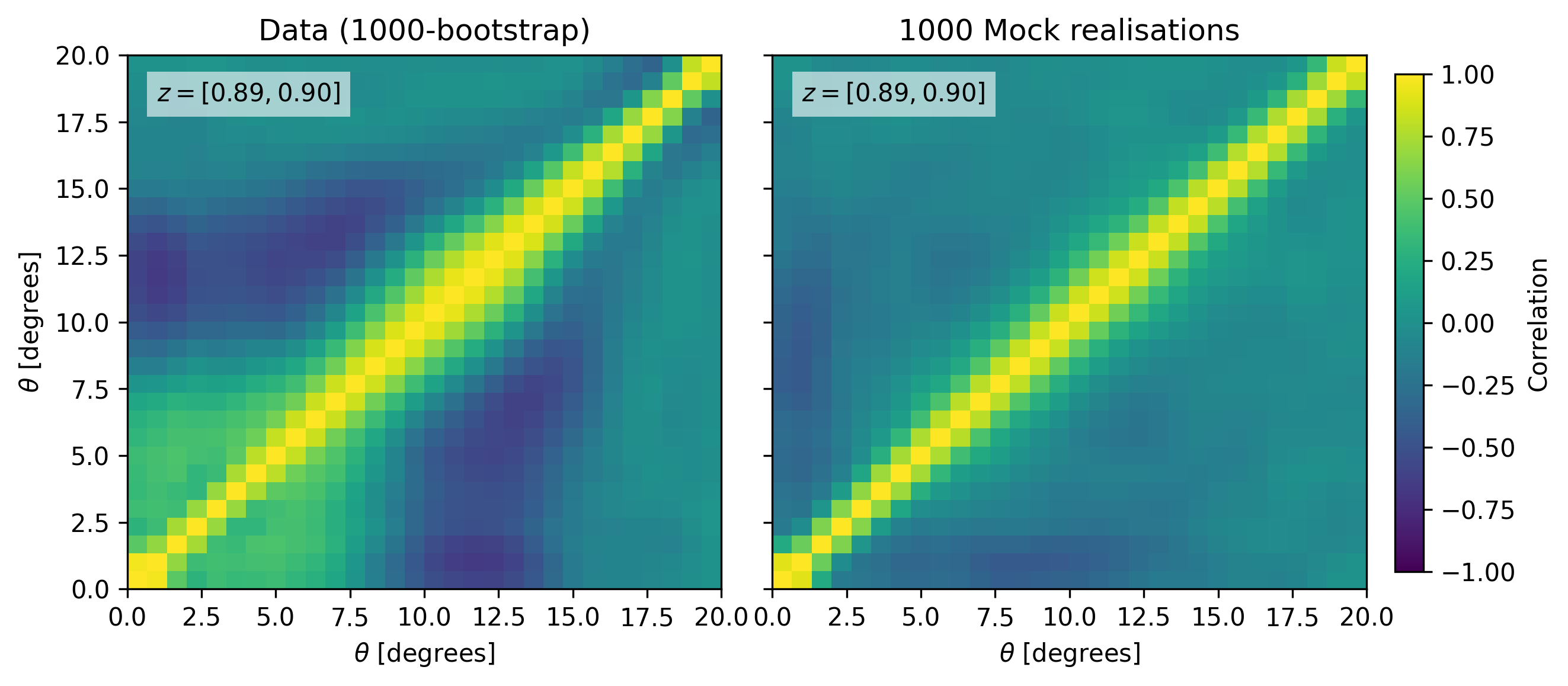}
    \vfill
    \includegraphics[width=0.60\textwidth]{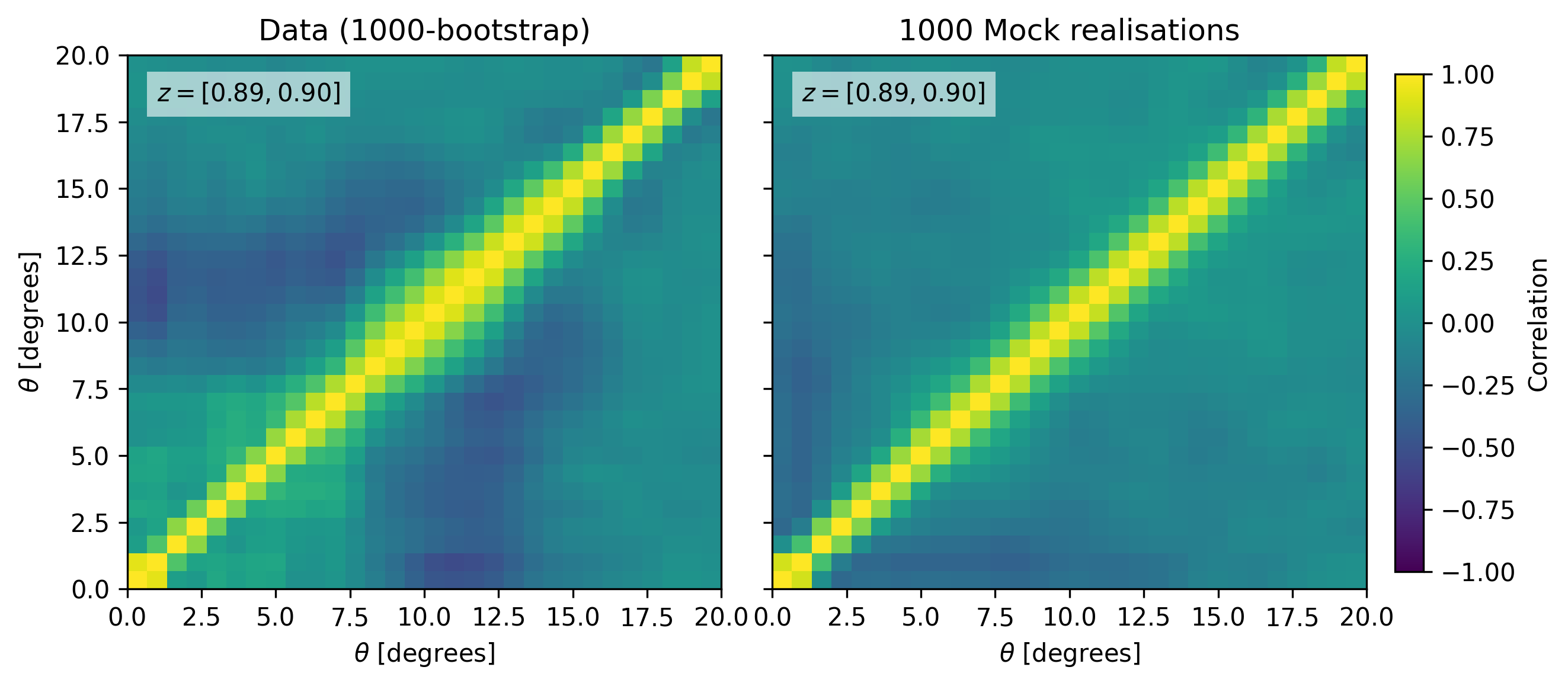}
    \caption{ Correlation matrices for correlation dimension, $D_2(\theta)$, in the same redshift range $0.89< z < 0.90$.The left panels show the results obtained from 1000-bootstrap resamplings, and the right panels correspond to the results obtained from 1000 mock catalogs. The upper panel shows the matrices using the North Galactic Cap, while the bottom panel shows the matrices using the South Galactic Cap, both shows the DESI DR1 data.}
    \label{fig:correlationmatrix_NGCvsSGC_089z090}
\end{figure}

\begin{figure}[h!]
    \centering
    \includegraphics[width=0.60\textwidth]{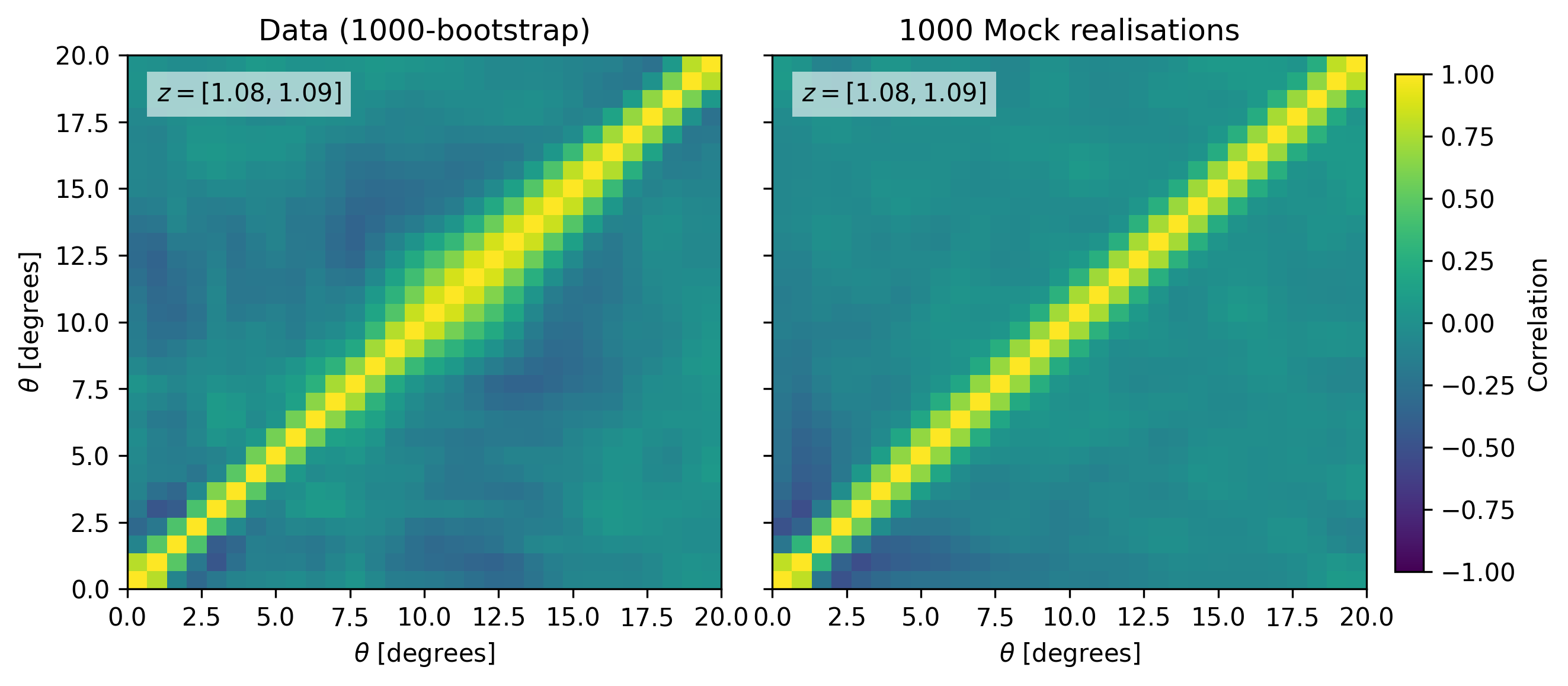}
    \vfill
    \includegraphics[width=0.60\textwidth]{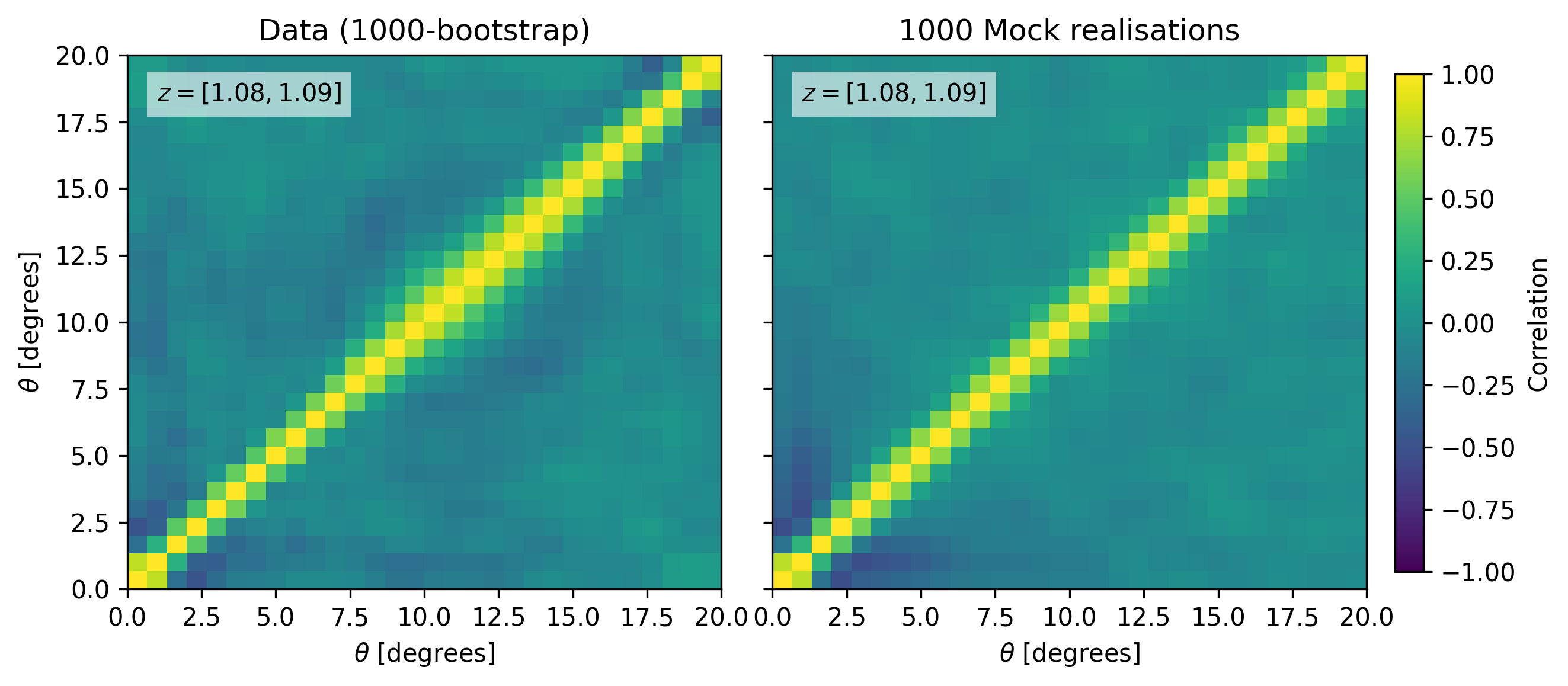}
    \caption{Correlation matrices for correlation dimension, $D_2(\theta)$, in the same redshift range $1.08< z < 1.09$. The left panels show the results obtained from 1000-bootstrap resamplings, and the right panels correspond to the results obtained from 1000 mock catalogs. The upper panel shows the matrices using the North Galactic Cap, while the bottom panel shows the matrices using the South Galactic Cap, both shows the DESI DR1 data.}
    \label{fig:correlationmatrix_NGCvsSGC_108z109}
\end{figure}

By the same token of the DESI versus eBOSS comparison, Figures \ref{fig:correlation:matrix_NGCvsSGC_041z042}, \ref{fig:correlationmatrix_NGCvsSGC_089z090} and \ref{fig:correlationmatrix_NGCvsSGC_108z109} show a comparative analysis of the $D_2(\theta)$ correlation matrices between the NGC and SGC regions of the DESI DR1 survey for each redshift range. 
\begin{table}[h!]
\resizebox{0.4\textwidth}{!}{
\begin{tabular}{lcl|clllcll|}
\cline{4-10}
                                  & \multicolumn{1}{l}{}           &                       & \multicolumn{7}{c|}{DESI DR1}                                                     \\ \cline{1-2} \cline{4-10} 
\multicolumn{1}{|c|}{$z$}         & \multicolumn{1}{c|}{$\bar{z}$} &                       & \multicolumn{3}{c|}{NGC}    & \multicolumn{1}{l|}{} & \multicolumn{3}{c|}{SGC}    \\ \cline{1-2} \cline{4-6} \cline{8-10} 
\multicolumn{1}{|l|}{0.41 - 0.42} & \multicolumn{1}{c|}{0.455}     &                       & \multicolumn{3}{c|}{13 152} & \multicolumn{1}{l|}{} & \multicolumn{3}{c|}{6 335}  \\ \cline{1-2} \cline{4-6} \cline{8-10} 
\multicolumn{1}{|l|}{0.89 - 0.90} & \multicolumn{1}{c|}{0.895}     &                       & \multicolumn{3}{c|}{26 626} & \multicolumn{1}{l|}{} & \multicolumn{3}{c|}{11 884} \\ \cline{1-2} \cline{4-6} \cline{8-10} 
\multicolumn{1}{|c|}{1.08 - 1.09} & \multicolumn{1}{c|}{1.085}     & \multicolumn{1}{c|}{} & \multicolumn{3}{c|}{6 017}  & \multicolumn{1}{c|}{} & \multicolumn{3}{c|}{2 672}  \\ \cline{1-2} \cline{4-6} \cline{8-10} 
\end{tabular}
}
\caption{Number of galaxies per redshift range for each hemisphere of DESI DR1 catalog.}
\label{tab:num_galaxies_onlyDESI}
\end{table}

\FloatBarrier
\section{Conclusions}
In this work, we perform a test of the Cosmological Principle by measuring the angular homogeneity scale, $\theta_H$, using the most recent available Luminous Red Galaxy catalog from DESI DR1 survey. Our methodology relies exclusively on a two-dimensional (angular) approach, carried out in narrow redshift ranges, thereby minimizing the dependence on cosmological models.

Our results show that the homogeneity scale is consistently identified across all redshift intervals analyzed, both in the North and South Galactic Caps of DESI DR1 survey. In addition, the estimates of $\theta_H$ (and, for completeness, the estimates of the spatial homogeneity scale, $R_H$) exhibit agreement between different sky regions, as well as between the observational data and simulations based on the $\Lambda$CDM model. Furthermore, we find agreement with measurements previously obtained from the SDSS-IV eBOSS DR16 survey, highlighting the stability of inferences drawn from different observational datasets. Therefore, this work contributes to supporting the physical validity of the Cosmological Principle, one of the fundamental pillars of the current standard cosmological model, $\Lambda$CDM.

\section{Acknowledgements}

Funding for the Sloan Digital Sky Survey IV has been provided by the Alfred P. Sloan Foundation, the U.S. Department of Energy Office of Science, and the Participating Institutions. SDSS acknowledges support and resources from the Center for High-Performance Computing at the University of Utah. The SDSS website is www.sdss.org.

SDSS is managed by the Astrophysical Research Consortium for the Participating Institutions of the SDSS Collaboration including the Brazilian Participation Group, the Carnegie Institution for Science, Carnegie Mellon University, Center for Astrophysics | Harvard \& Smithsonian (CfA), the Chilean Participation Group, the French Participation Group, Instituto de Astrofísica de Canarias, The Johns Hopkins University, Kavli Institute for the Physics and Mathematics of the Universe (IPMU) / University of Tokyo, the Korean Participation Group, Lawrence Berkeley National Laboratory, Leibniz Institut für Astrophysik Potsdam (AIP), Max-Planck-Institut für Astronomie (MPIA Heidelberg), Max-Planck-Institut für Astrophysik (MPA Garching), Max-Planck-Institut für Extraterrestrische Physik (MPE), National Astronomical Observatories of China, New Mexico State University, New York University, University of Notre Dame, Observatório Nacional / MCTI, The Ohio State University, Pennsylvania State University, Shanghai Astronomical Observatory, United Kingdom Participation Group, Universidad Nacional Autónoma de México, University of Arizona, University of Colorado Boulder, University of Oxford, University of Portsmouth, University of Utah, University of Virginia, University of Washington, University of Wisconsin, Vanderbilt University, and Yale University.

This research used data obtained with the Dark Energy Spectroscopic Instrument (DESI). DESI construction and operations is managed by the Lawrence Berkeley National Laboratory. This material is based upon work supported by the U.S. Department of Energy, Office of Science, Office of High-Energy Physics, under Contract No. DE–AC02–05CH11231, and by the National Energy Research Scientific Computing Center, a DOE Office of Science User Facility under the same contract. Additional support for DESI was provided by the U.S. National Science Foundation (NSF), Division of Astronomical Sciences under Contract No. AST-0950945 to the NSF’s National Optical-Infrared Astronomy Research Laboratory; the Science and Technology Facilities Council of the United Kingdom; the Gordon and Betty Moore Foundation; the Heising-Simons Foundation; the French Alternative Energies and Atomic Energy Commission (CEA); the National Council of Humanities, Science and Technology of Mexico (CONAHCYT); the Ministry of Science and Innovation of Spain (MICINN), and by the DESI Member Institutions: www.desi.lbl.gov/collaborating-institutions. The DESI collaboration is honored to be permitted to conduct scientific research on I’oligam Du’ag (Kitt Peak), a mountain with particular significance to the Tohono O’odham Nation. Any opinions, findings, and conclusions or recommendations expressed in this material are those of the author(s) and do not necessarily reflect the views of the U.S. National Science Foundation, the U.S. Department of Energy, or any of the listed funding agencies. 

This study was financed in part by the Coordenação de Aperfeiçoamento de Pessoal de Nível Superior - Brasil (CAPES) - finance code 001. MLD acknowledges CAPES for financial support. CAPB acknowledges financial support from the CNPq grant 306630/2025-7. PSF acknowledges the support from the start-up funding of Zhejiang University, Zhejiang provincial top level research support program.

\newpage
\appendix 

\section{Measurements of the angular homogeneity scale}\label{appendix_values_thetah_rh}

In this section, we present the measurements of $\theta_H$ and $R_H$ obtained for each of the 70 $\Delta z$'s intervals in both the NGC and SGC hemispheres of the DESI DR1 survey, in Table \ref{tab:values_thetaH_rh}.
In the case of the spatial homogeneity scale, we adopt the standard $\Lambda$CDM cosmological model, using the best-fit values from the Planck collaboration \cite{Planck:2018vyg} ($H_0 = 67.4 \pm 0.5$ km/s/Mpc and $\Omega_m = 0.315 \pm 0.007$).We can see they are in good agreement with previous analyses~\cite{Goncalves:2017dzs}, and the maximum estimate of $260$ Mpc h$^{-1}$ reported in~\cite{Yadav:2010cc} within uncertainties.
\begin{equation}
    R_H = d_A(z) \theta_H, 
\end{equation}
\begin{equation}
    d_A(z) = \frac{c}{1+z} \int_0^z \frac{dz^\prime}{H(z^\prime)}, \text{ where } H(z) = H_0 \sqrt{\Omega_m (1 + z)^3 + (1 - \Omega_m)}.
\end{equation}
\begin{equation}
    \sigma_{R_H}^2 =
\left(\frac{\partial R_H}{\partial \theta_H}\sigma_{\theta_H}\right)^2 
\end{equation}

\begin{center}
\small
\setlength{\tabcolsep}{4pt}

\begin{longtable}{|c|c|c|c|c|}
\caption{Measurements of $\theta_H$ and $R_H$ from each redshift range of width $\Delta z = 0.01$, using DESI DR1 NGC and SGC data.}
\label{tab:values_thetaH_rh} \\

\hline
$z_{\rm mean}$ &
$\theta_H^{\rm NGC}$ [degrees] &
$R_H^{\rm NGC}$ [Mpc] &
$\theta_H^{\rm SGC}$ [degrees] &
$R_H^{\rm SGC}$ [Mpc] \\
\hline
\endfirsthead

\hline
$z_{\rm mean}$ &
$\theta_H^{\rm NGC}$ [degrees] &
$R_H^{\rm NGC}$ [Mpc] &
$\theta_H^{\rm SGC}$ [degrees] &
$R_H^{\rm SGC}$ [Mpc] \\
\hline
\endhead

\hline
\endfoot

\hline
\endlastfoot
0.405          & 10.45 $\pm$ 2.59     & 210.62 $\pm$ 52.15  & 9.85 $\pm$ 2.47      & 198.49 $\pm$ 49.73 \\ \hline
0.415          & 10.26 $\pm$ 2.53     & 209.85 $\pm$ 51.68  & 9.33 $\pm$ 2.83      & 190.76 $\pm$ 57.79 \\ \hline
0.425          & 9.55 $\pm$ 2.68      & 198.06 $\pm$ 55.63  & 10.17 $\pm$ 2.80     & 210.97 $\pm$ 58.00 \\ \hline
0.435          & 9.38 $\pm$ 2.43      & 197.19 $\pm$ 51.12  & 9.04 $\pm$ 2.76      & 190.06 $\pm$ 58.01 \\ \hline
0.445          & 9.00 $\pm$ 2.83      & 191.57 $\pm$ 60.22  & 9.44 $\pm$ 2.79      & 201.13 $\pm$ 59.45 \\ \hline
0.455          & 8.97 $\pm$ 2.75      & 193.45 $\pm$ 59.20  & 9.86 $\pm$ 2.42      & 212.63 $\pm$ 52.16 \\ \hline
0.465          & 9.75 $\pm$ 2.58      & 212.92 $\pm$ 56.26  & 8.49 $\pm$ 2.65      & 185.37 $\pm$ 57.96 \\ \hline
0.475          & 9.14 $\pm$ 2.60      & 201.83 $\pm$ 57.40  & 10.46 $\pm$ 2.48     & 230.94 $\pm$ 54.84 \\ \hline
0.485          & 8.85 $\pm$ 2.52      & 197.67 $\pm$ 56.39  & 9.62 $\pm$ 2.44      & 214.94 $\pm$ 54.54 \\ \hline
0.495          & 9.43 $\pm$ 2.64      & 212.96 $\pm$ 59.62  & 9.08 $\pm$ 2.81      & 205.13 $\pm$ 63.39 \\ \hline
0.505          & 9.14 $\pm$ 2.72      & 208.56 $\pm$ 62.15  & 7.94 $\pm$ 2.67      & 181.26 $\pm$ 60.85 \\ \hline
0.515          & 8.67 $\pm$ 2.60      & 200.01 $\pm$ 60.06  & 9.06 $\pm$ 2.64      & 209.03 $\pm$ 60.79 \\ \hline
0.525          & 9.28 $\pm$ 2.79      & 216.09 $\pm$ 64.99  & 8.66 $\pm$ 2.63      & 201.69 $\pm$ 61.15 \\ \hline
0.535          & 9.63 $\pm$ 2.55      & 226.44 $\pm$ 59.98  & 8.56 $\pm$ 2.92      & 201.25 $\pm$ 68.76 \\ \hline
0.545          & 9.19 $\pm$ 2.70      & 218.01 $\pm$ 64.07  & 9.41 $\pm$ 2.36      & 223.39 $\pm$ 55.98 \\ \hline
0.555          & 8.36 $\pm$ 2.46      & 200.20 $\pm$ 58.88  & 8.69 $\pm$ 2.88      & 208.00 $\pm$ 68.93 \\ \hline
0.565          & 9.31 $\pm$ 2.57      & 224.75 $\pm$ 62.08  & 8.86 $\pm$ 2.99      & 214.02 $\pm$ 72.12 \\ \hline
0.575          & 9.22 $\pm$ 2.71      & 224.69 $\pm$ 65.93  & 9.18 $\pm$ 2.95      & 223.60 $\pm$ 71.89 \\ \hline
0.585          & 8.29 $\pm$ 2.75      & 203.56 $\pm$ 67.62  & 8.83 $\pm$ 2.18      & 216.77 $\pm$ 53.44 \\ \hline
0.595          & 8.28 $\pm$ 2.65      & 204.87 $\pm$ 65.53  & 8.53 $\pm$ 2.79      & 211.24 $\pm$ 69.05 \\ \hline
0.605          & 8.58 $\pm$ 2.77      & 213.91 $\pm$ 69.15  & 9.07 $\pm$ 2.96      & 226.14 $\pm$ 73.91 \\ \hline
0.615          & 8.74 $\pm$ 2.84      & 219.70 $\pm$ 71.37  & 8.38 $\pm$ 2.75      & 210.66 $\pm$ 69.13 \\ \hline
0.625          & 8.55 $\pm$ 2.82      & 216.47 $\pm$ 71.45  & 8.38 $\pm$ 2.91      & 211.98 $\pm$ 73.73 \\ \hline
0.635          & 8.42 $\pm$ 2.98      & 214.69 $\pm$ 76.01  & 8.85 $\pm$ 2.69      & 225.63 $\pm$ 68.46 \\ \hline
0.645          & 8.54 $\pm$ 2.82      & 219.17 $\pm$ 72.46  & 8.62 $\pm$ 2.87      & 221.20 $\pm$ 73.67 \\ \hline
0.655          & 8.69 $\pm$ 2.62      & 224.36 $\pm$ 67.75  & 8.32 $\pm$ 2.89      & 214.93 $\pm$ 74.53 \\ \hline
0.665          & 8.71 $\pm$ 2.91      & 226.42 $\pm$ 75.66  & 8.41 $\pm$ 2.96      & 218.48 $\pm$ 77.03 \\ \hline
0.675          & 9.07 $\pm$ 2.63      & 237.11 $\pm$ 68.79  & 8.24 $\pm$ 2.92      & 215.50 $\pm$ 76.27 \\ \hline
0.685          & 8.61 $\pm$ 2.77      & 226.48 $\pm$ 72.84  & 8.68 $\pm$ 2.91      & 228.34 $\pm$ 76.61 \\ \hline
0.695          & 8.65 $\pm$ 2.83      & 228.96 $\pm$ 74.80  & 8.67 $\pm$ 2.75      & 229.42 $\pm$ 72.79 \\ \hline
0.705          & 8.48 $\pm$ 2.76      & 225.67 $\pm$ 73.35  & 7.81 $\pm$ 3.06      & 207.76 $\pm$ 81.50 \\ \hline
0.715          & 8.15 $\pm$ 2.88      & 218.01 $\pm$ 77.17  & 8.31 $\pm$ 2.85      & 222.29 $\pm$ 76.21 \\ \hline
0.725          & 7.62 $\pm$ 2.34      & 204.84 $\pm$ 62.83  & 8.97 $\pm$ 2.95      & 241.29 $\pm$ 79.36 \\ \hline
0.735          & 8.35 $\pm$ 2.91      & 225.89 $\pm$ 78.56  & 8.35 $\pm$ 2.89      & 225.80 $\pm$ 78.11 \\ \hline
0.745          & 8.86 $\pm$ 2.78      & 240.88 $\pm$ 75.47  & 8.52 $\pm$ 2.99      & 231.40 $\pm$ 81.23 \\ \hline
0.755          & 8.85 $\pm$ 2.75      & 241.66 $\pm$ 75.09  & 8.72 $\pm$ 3.03      & 238.09 $\pm$ 82.62 \\ \hline
0.765          & 8.55 $\pm$ 2.71      & 234.49 $\pm$ 74.38  & 8.05 $\pm$ 3.17      & 220.72 $\pm$ 87.04 \\ \hline
0.775          & 8.07 $\pm$ 2.74      & 222.50 $\pm$ 75.49  & 8.64 $\pm$ 2.85      & 238.14 $\pm$ 78.68 \\ \hline
0.785          & 7.84 $\pm$ 3.15      & 217.13 $\pm$ 87.21  & 7.90 $\pm$ 2.74      & 218.78 $\pm$ 75.96 \\ \hline
0.795          & 8.62 $\pm$ 2.88      & 239.53 $\pm$ 80.17  & 7.65 $\pm$ 2.93      & 212.74 $\pm$ 81.38 \\ \hline
0.805          & 7.52 $\pm$ 2.87      & 210.06 $\pm$ 80.00  & 8.23 $\pm$ 3.05      & 229.91 $\pm$ 85.29 \\ \hline
0.815          & 8.26 $\pm$ 2.73      & 231.56 $\pm$ 76.47  & 8.57 $\pm$ 2.99      & 240.27 $\pm$ 83.92 \\ \hline
0.825          & 8.02 $\pm$ 3.04      & 225.65 $\pm$ 85.46  & 8.37 $\pm$ 3.11      & 235.65 $\pm$ 87.53 \\ \hline
0.835          & 7.67 $\pm$ 2.92      & 216.62 $\pm$ 82.58  & 7.80 $\pm$ 3.11      & 220.25 $\pm$ 87.84 \\ \hline
0.845          & 8.08 $\pm$ 2.90      & 229.03 $\pm$ 82.10  & 8.25 $\pm$ 2.99      & 234.00 $\pm$ 84.70 \\ \hline
0.855          & 7.99 $\pm$ 3.01      & 227.43 $\pm$ 85.78  & 8.35 $\pm$ 2.85      & 237.70 $\pm$ 81.23 \\ \hline
0.865          & 7.89 $\pm$ 2.94      & 225.22 $\pm$ 83.99  & 7.41 $\pm$ 3.05      & 211.59 $\pm$ 87.10 \\ \hline
0.875          & 7.82 $\pm$ 2.98      & 223.98 $\pm$ 85.37  & 8.24 $\pm$ 3.10      & 236.17 $\pm$ 88.89 \\ \hline
0.885          & 7.72 $\pm$ 3.13      & 222.10 $\pm$ 89.89  & 7.90 $\pm$ 3.02      & 227.24 $\pm$ 86.96 \\ \hline
0.895          & 7.96 $\pm$ 3.09      & 229.63 $\pm$ 89.14  & 8.06 $\pm$ 3.11      & 232.41 $\pm$ 89.75 \\ \hline
0.905          & 7.62 $\pm$ 2.69      & 220.42 $\pm$ 77.85  & 7.28 $\pm$ 3.00      & 210.74 $\pm$ 86.95 \\ \hline
0.915          & 7.61 $\pm$ 3.11      & 220.97 $\pm$ 90.24  & 7.49 $\pm$ 3.02      & 217.44 $\pm$ 87.52 \\ \hline
0.925          & 7.34 $\pm$ 2.75      & 213.58 $\pm$ 79.98  & 7.20 $\pm$ 2.97      & 209.60 $\pm$ 86.57 \\ \hline
0.935          & 8.14 $\pm$ 3.19      & 237.54 $\pm$ 92.99  & 6.80 $\pm$ 2.97      & 198.65 $\pm$ 86.79 \\ \hline
0.945          & 7.95 $\pm$ 3.09      & 232.69 $\pm$ 90.57  & 7.41 $\pm$ 2.99      & 217.07 $\pm$ 87.59 \\ \hline
0.955          & 7.97 $\pm$ 3.23      & 233.87 $\pm$ 94.69  & 7.94 $\pm$ 3.28      & 233.06 $\pm$ 96.17 \\ \hline
0.965          & 7.97 $\pm$ 3.12      & 234.50 $\pm$ 91.74  & 7.78 $\pm$ 2.94      & 229.00 $\pm$ 86.45 \\ \hline
0.975          & 8.05 $\pm$ 3.21      & 237.58 $\pm$ 94.78  & 7.23 $\pm$ 3.09      & 213.36 $\pm$ 91.21 \\ \hline
0.985          & 7.59 $\pm$ 3.18      & 224.63 $\pm$ 94.03  & 8.03 $\pm$ 2.94      & 237.48 $\pm$ 86.86 \\ \hline
0.995          & 7.37 $\pm$ 3.15      & 218.56 $\pm$ 93.48  & 7.39 $\pm$ 3.14      & 219.10 $\pm$ 93.23 \\ \hline
1.005          & 7.54 $\pm$ 2.92      & 224.09 $\pm$ 86.68  & 7.17 $\pm$ 3.05      & 213.00 $\pm$ 90.65 \\ \hline
1.015          & 7.68 $\pm$ 3.04      & 228.68 $\pm$ 90.44  & 6.92 $\pm$ 2.57      & 205.96 $\pm$ 76.49 \\ \hline
1.025          & 8.17 $\pm$ 2.99      & 243.76 $\pm$ 89.14  & 7.33 $\pm$ 2.91      & 218.73 $\pm$ 86.92 \\ \hline
1.035          & 7.39 $\pm$ 3.25      & 221.05 $\pm$ 97.34  & 8.27 $\pm$ 3.06      & 247.46 $\pm$ 91.64 \\ \hline
1.045          & 7.87 $\pm$ 3.23      & 235.99 $\pm$ 96.66  & 6.55 $\pm$ 3.18      & 196.17 $\pm$ 95.44 \\ \hline
1.055          & 7.41 $\pm$ 3.18      & 222.61 $\pm$ 95.64  & 7.03 $\pm$ 3.18      & 211.23 $\pm$ 95.46 \\ \hline
1.065          & 7.95 $\pm$ 3.29      & 239.31 $\pm$ 98.98  & 6.93 $\pm$ 3.04      & 208.35 $\pm$ 91.38 \\ \hline
1.075          & 7.93 $\pm$ 3.18      & 239.04 $\pm$ 95.83  & 7.27 $\pm$ 3.12      & 219.01 $\pm$ 94.18 \\ \hline
1.085          & 7.47 $\pm$ 3.30      & 225.70 $\pm$ 99.75  & 7.58 $\pm$ 3.29      & 228.75 $\pm$ 99.41 \\ \hline
1.095          & 7.09 $\pm$ 3.37      & 214.32 $\pm$ 101.95 & 7.69 $\pm$ 3.18      & 232.49 $\pm$ 96.26 \\ \hline
\end{longtable}
\end{center}

\section{Footprints by redshifts}\label{appendix_footprints}

\begin{figure}[h!]
    \centering
    \includegraphics[width=0.98\textwidth]{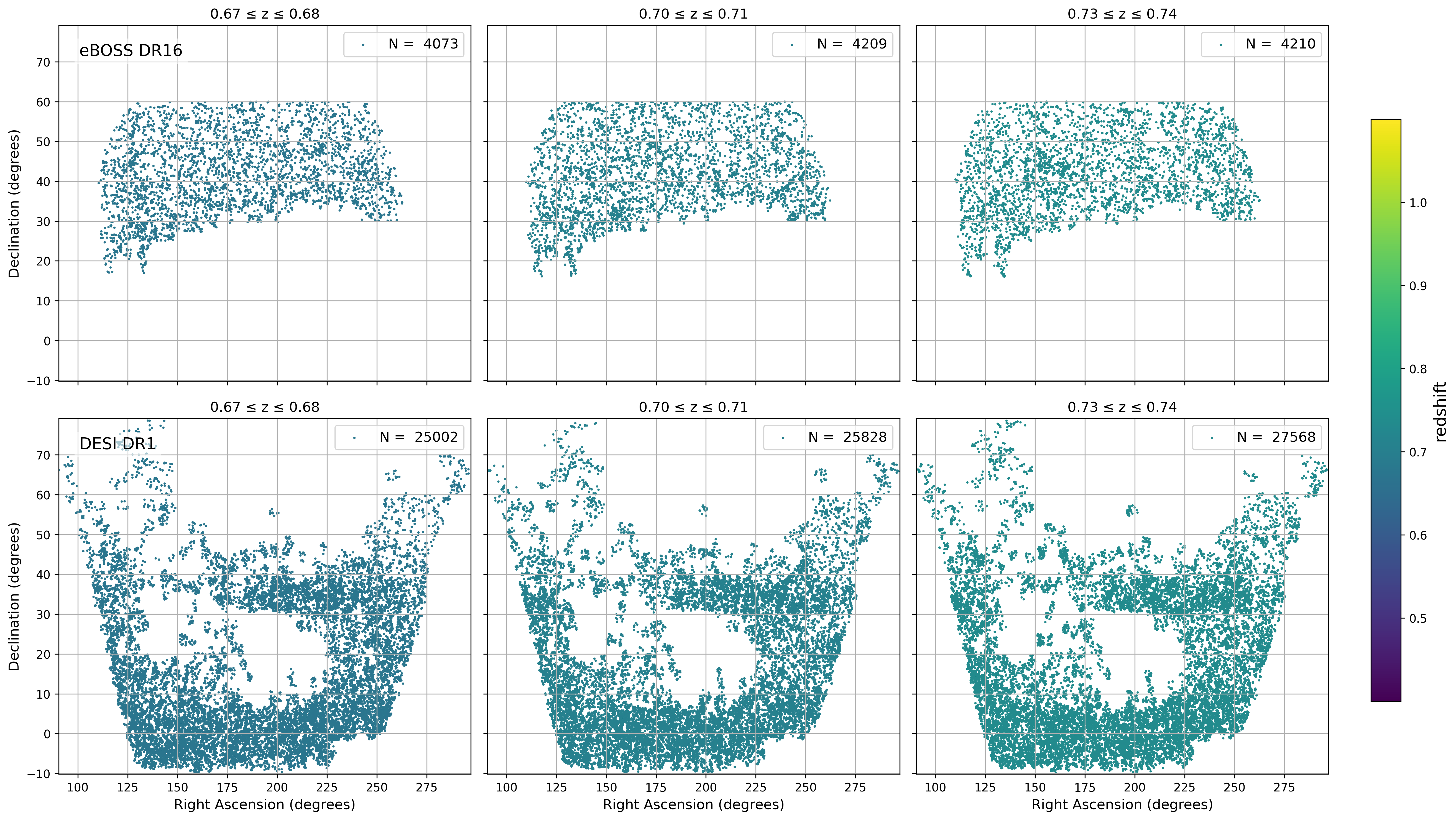}
    \caption{Footprint of North Galactic Cap distribution for each redshift range. Upper panel shows the eBOSS DR16 data and the bottom panel shows the DESI DR1 data.}
     \label{fig:NGC_footprint_byz}
\end{figure}

\begin{figure}[h!]
    \centering
    \includegraphics[width=0.98\textwidth]{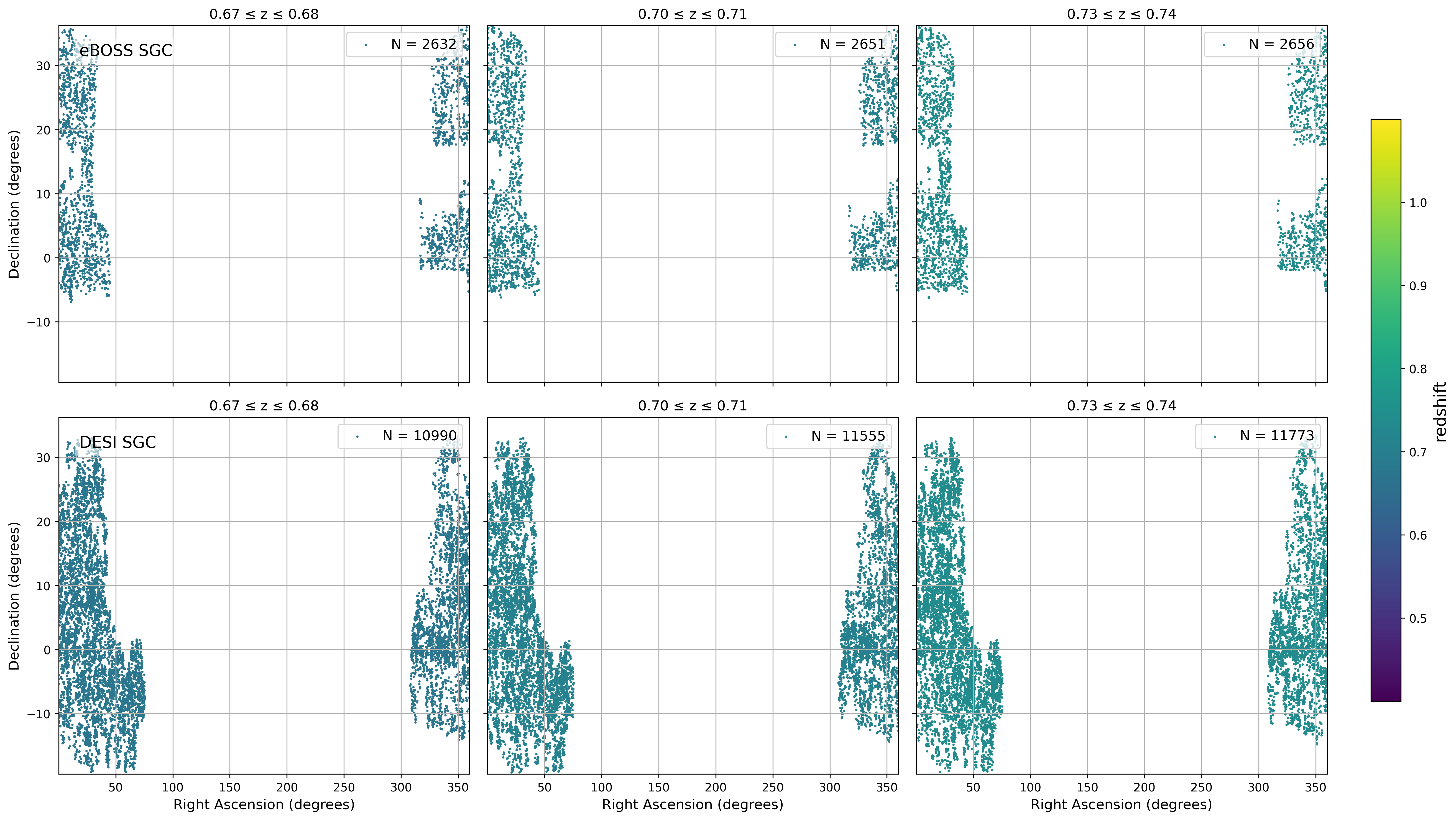}
    \caption{Footprint of South Galactic Cap distribution for each redshift range. Upper panel shows the eBOSS DR16 data and the bottom panel shows the DESI DR1 data.}
    \label{fig:SGC_footprint_byz}
\end{figure}

\begin{figure}[h!]
    \centering
    \includegraphics[width=0.98\textwidth]{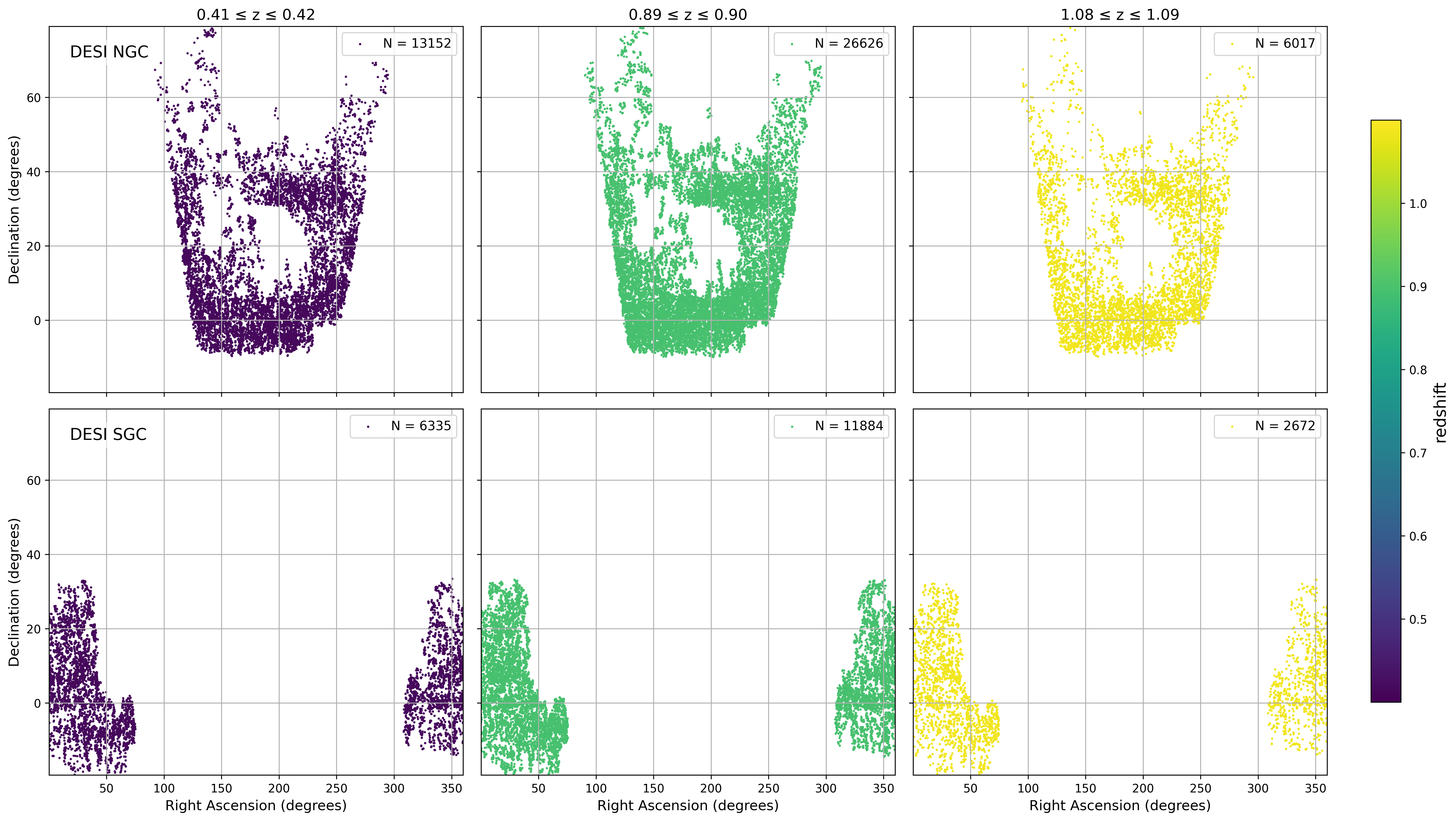}
    \caption{Footprint of DESI DR1 catalog distribution for each redshift range. Upper panel shows the North Galactic Cap data and the bottom panel shows the South Galactic Cap data. }
     \label{fig:onlyDESI_footprint_byz}
\end{figure}

In this section we present the distribution of LRGs by redshift. The comparison between eBOSS DR16 and DESI DR1 are shown in Figure \ref{fig:NGC_footprint_byz} (NGC) and Figure \ref{fig:SGC_footprint_byz} (SGC). While the comparison between the both hemisphere of the DESI data are shown in Figure \ref{fig:onlyDESI_footprint_byz}.


\clearpage
\section{Analysis with eBOSS DR16 SGC}\label{appendix_ebossSGC}

In this section we show some analysis using the South Galactic Cap from eBOSS DR16 data. 
\begin{figure}[h!]
    \centering
    \includegraphics[width=0.5\textwidth]{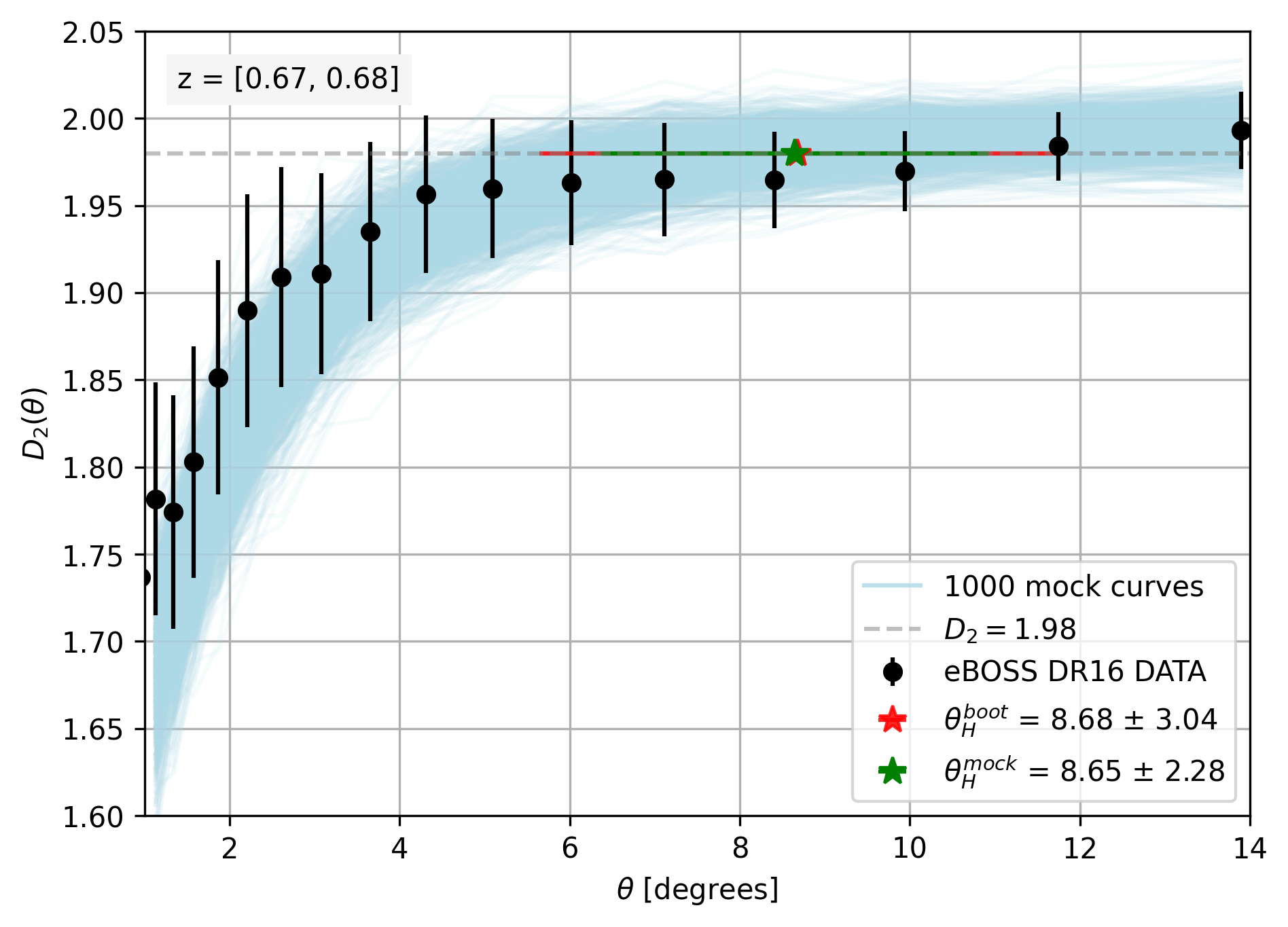}
    \vfill
    \includegraphics[width=0.5\textwidth]{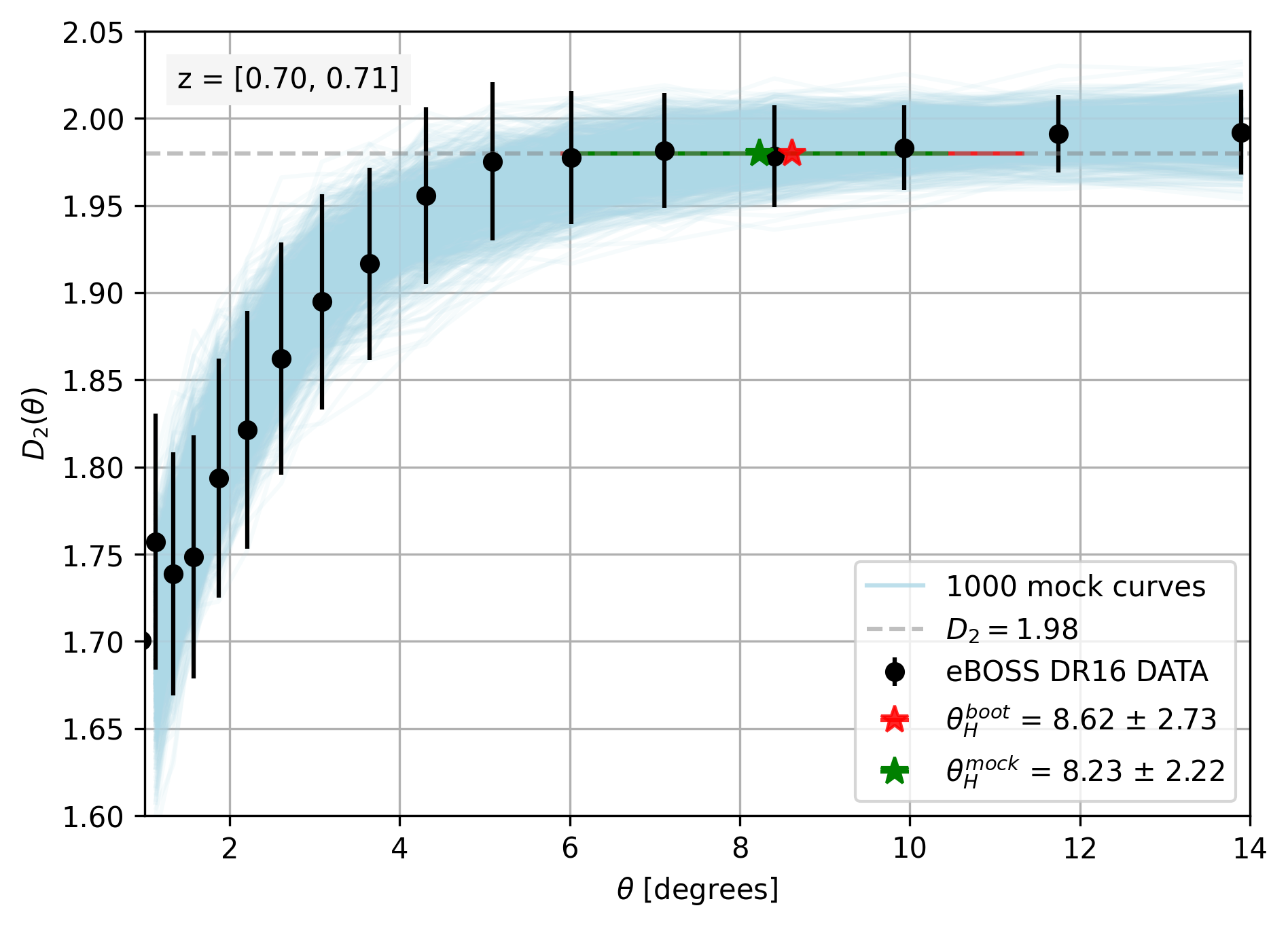}
    \vfill
    \includegraphics[width=0.5\textwidth]{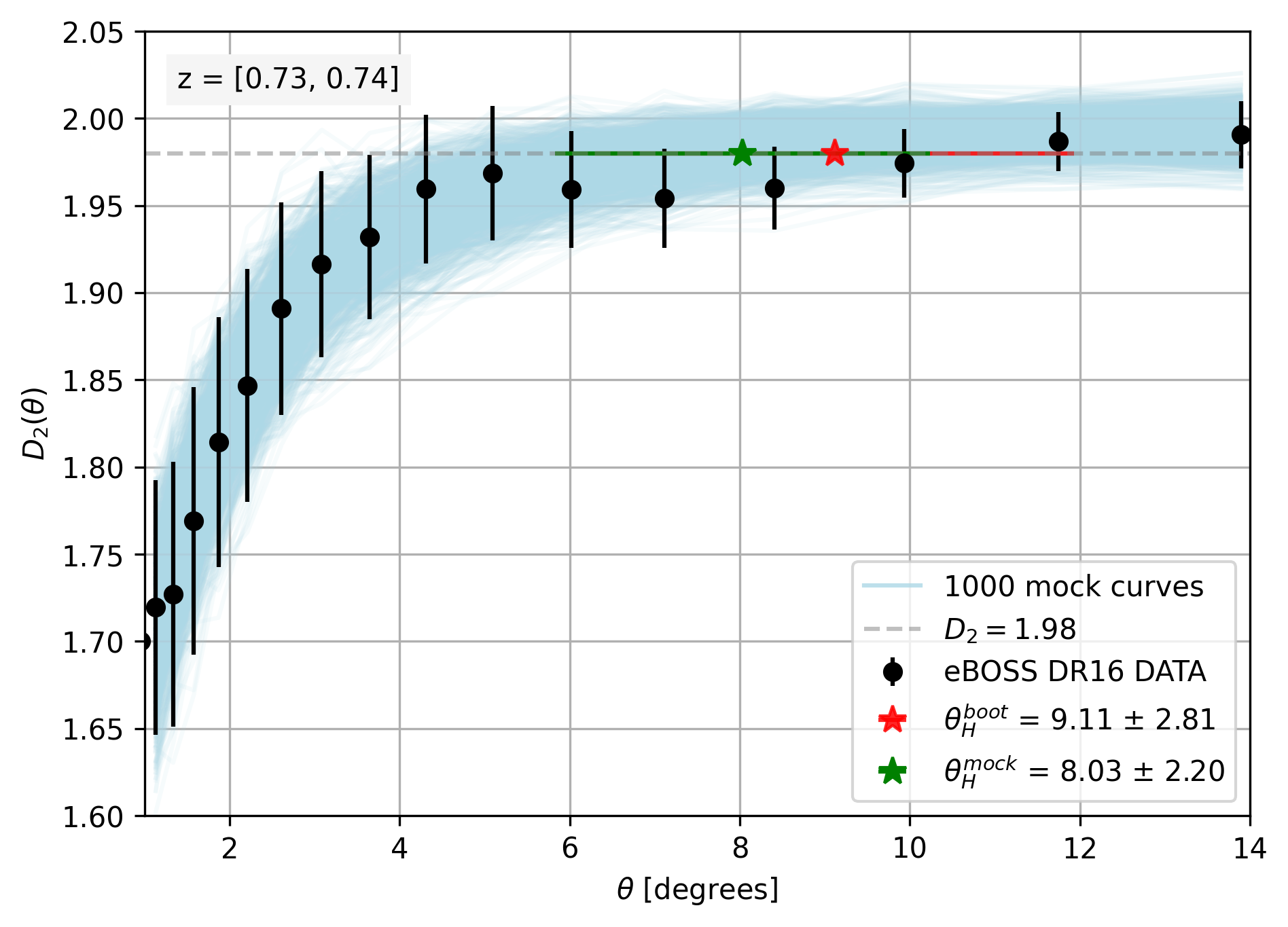}
    \caption{Measurements of the correlation dimension, $D_2(\theta)$, for three different redshift ranges. All panels shows measurements using the South Galactic Cap (SGC) from eBOSS DR16 catalog. }
    \label{fig:D2_ebossSGC}
\end{figure}

\begin{figure}[h!]
    \centering
    \includegraphics[width=0.60\textwidth]{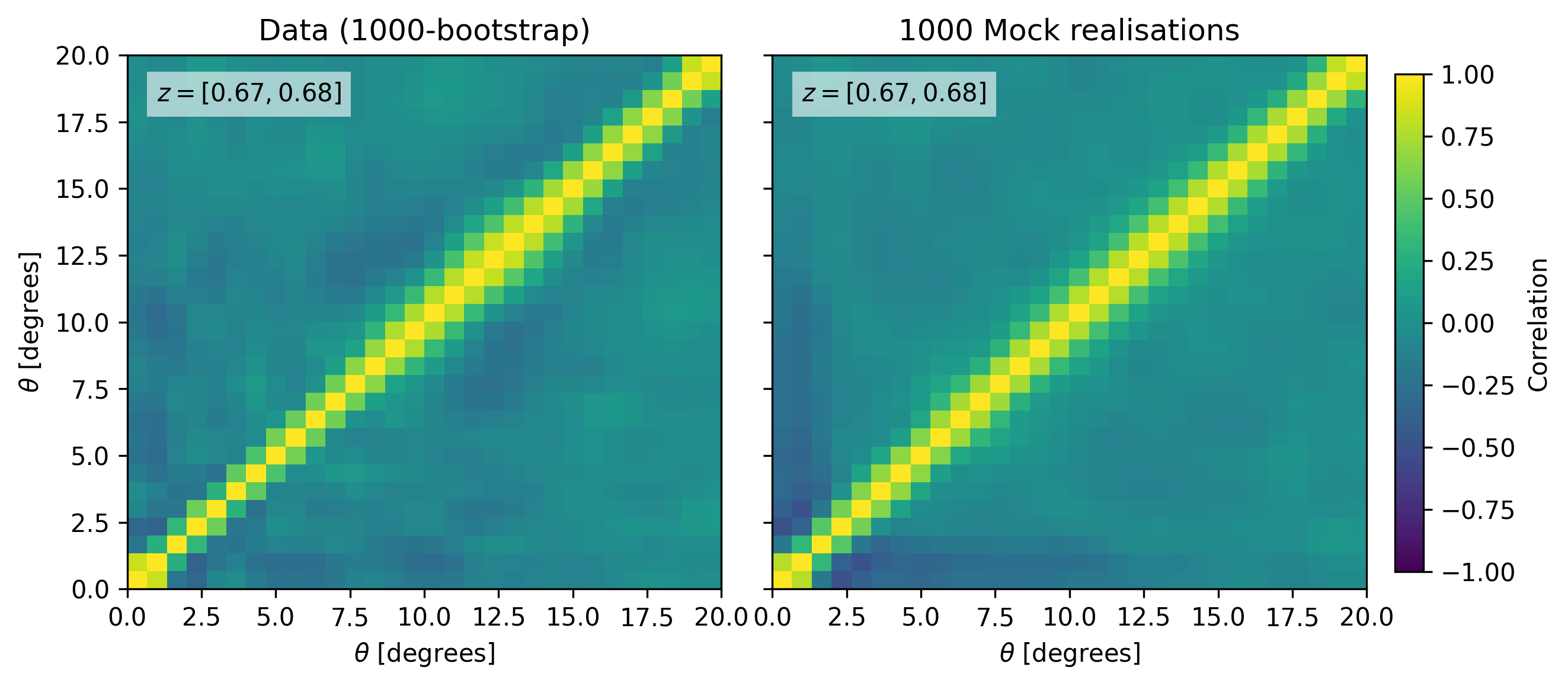}
    \vfill
    \includegraphics[width=0.60\textwidth]{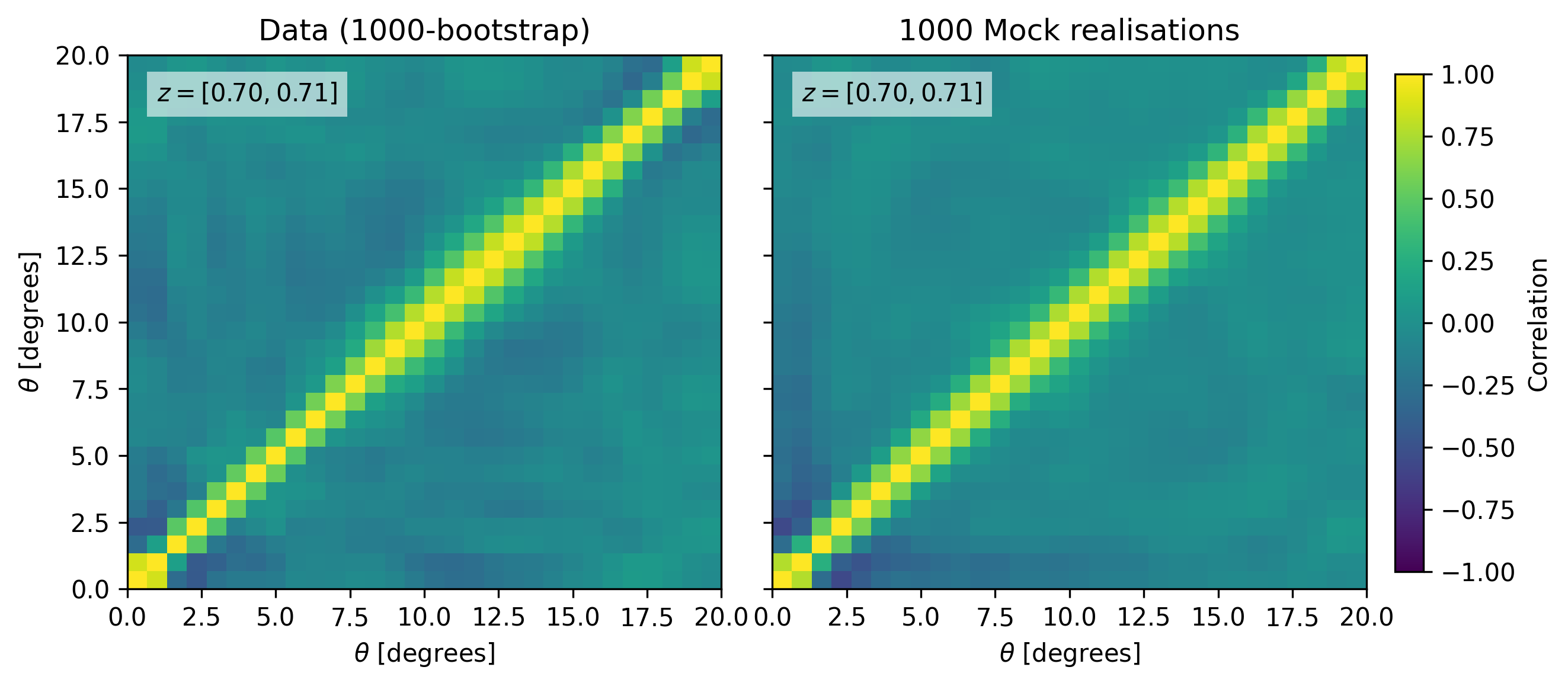}
    \vfill
    \includegraphics[width=0.60\textwidth]{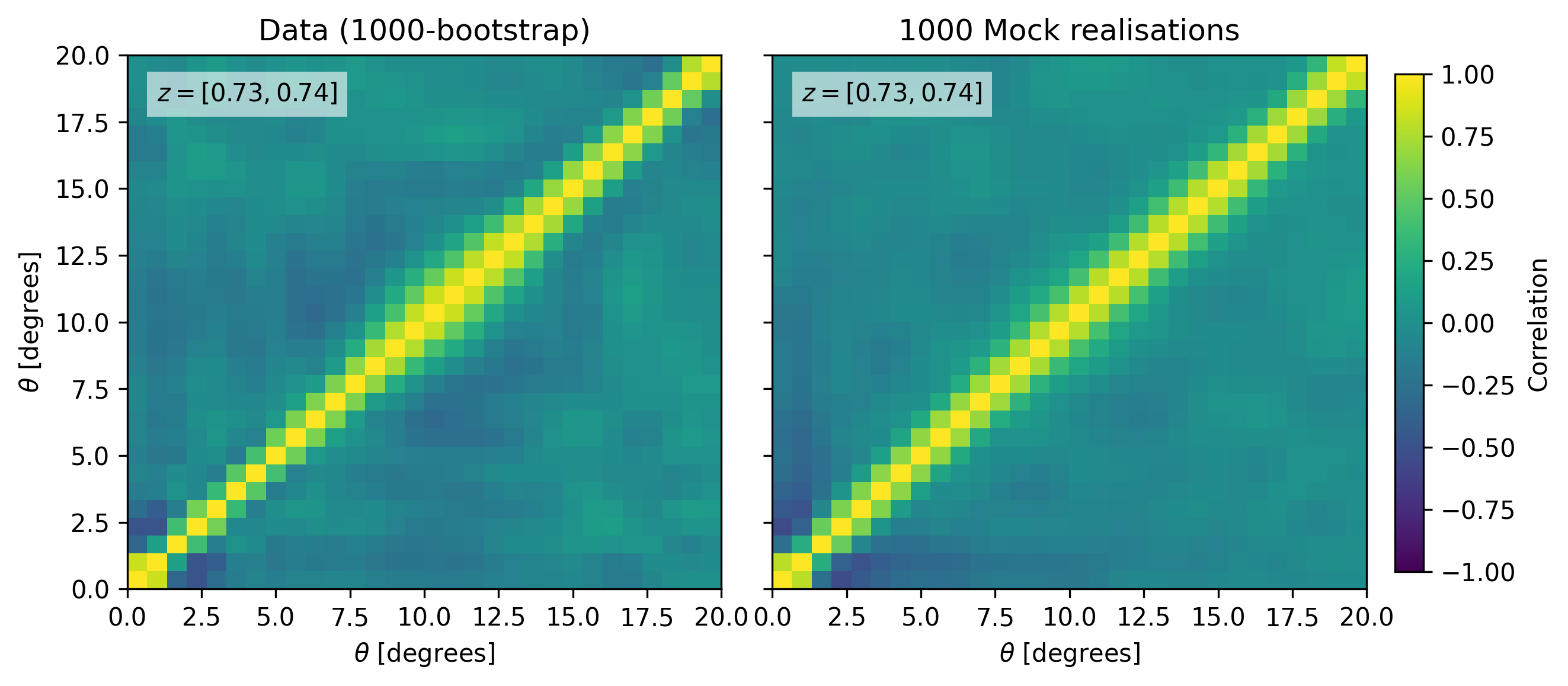}
    \caption{Comparison of correlation matrices for correlation dimension, $D_2(\theta)$, for three different redshift ranges. All panels shows measurements using the South Galactic Cap (SGC) from eBOSS DR16 catalog.  The left panels show the results obtained from 1000-bootstrap resamplings, and the right panels correspond to the results obtained from 1000 mock catalogs.}
    \label{fig:correlation_matrix_ebossSGC}
\end{figure}

\newpage 

\newpage
\section{24 bins x 30 bins}\label{appendix_bins}

\begin{figure}[h!]
    \centering
    \includegraphics[width=0.60\textwidth]{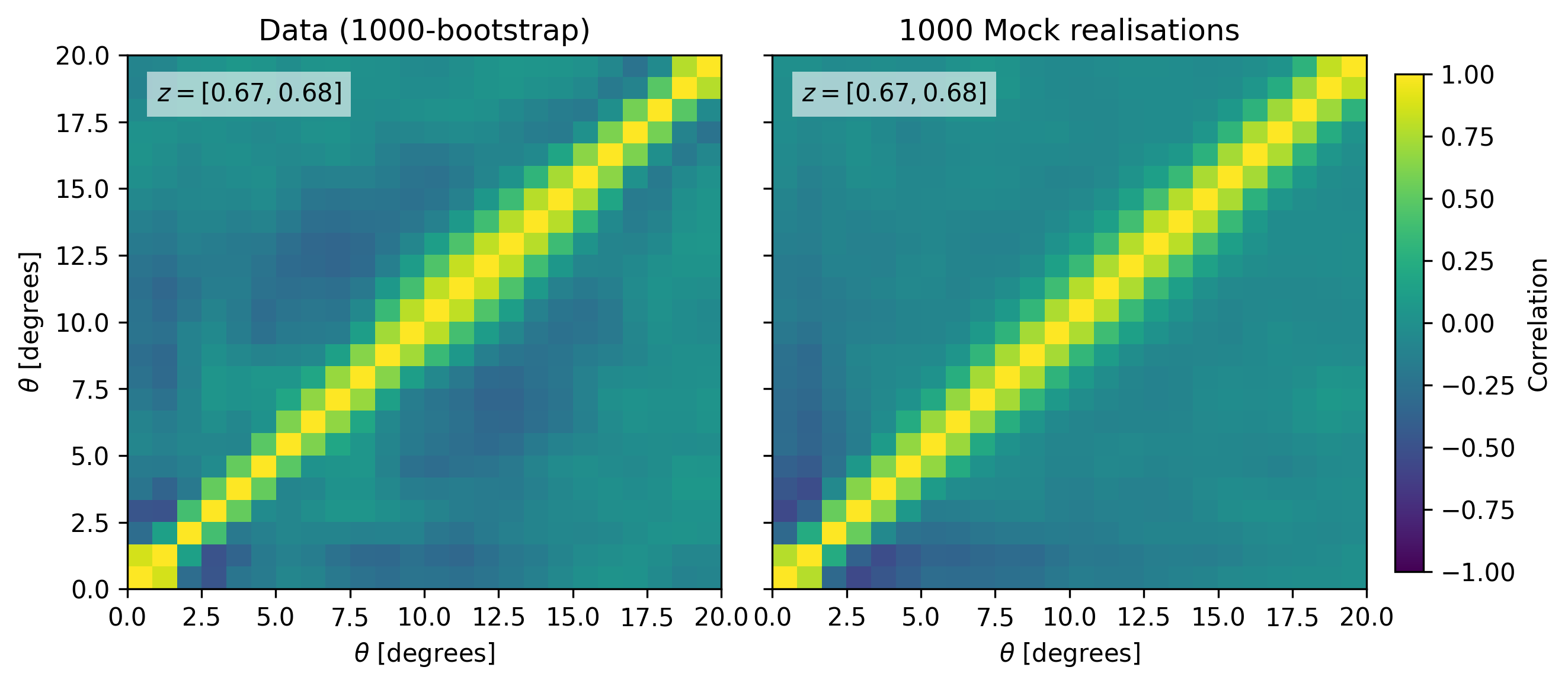}
    \vfill
    \includegraphics[width=0.60\textwidth]{plots/correlation_matrix_ebossNGC_nbins30_0.67z0.68.png}
    \caption{Comparison of correlation matrices for correlation dimension, $D_2(\theta)$, for different number os bins in the same redshift range $0.67 < z < 0.68$. The left panels show the results obtained from 1000-bootstrap resamplings, and the right panels correspond to the results obtained from 1000 mock catalogs. The upper panels shows the analysis with 24 bins while the bottom panel shows with 30 bins. Both shows the matrices using the North Galactic Cap from eBOSS DR16 data. }
    \label{fig:correlation_matrix_24x30bins_eboss}
\end{figure}

\begin{figure}[h!]
    \centering
    \includegraphics[width=0.60\textwidth]{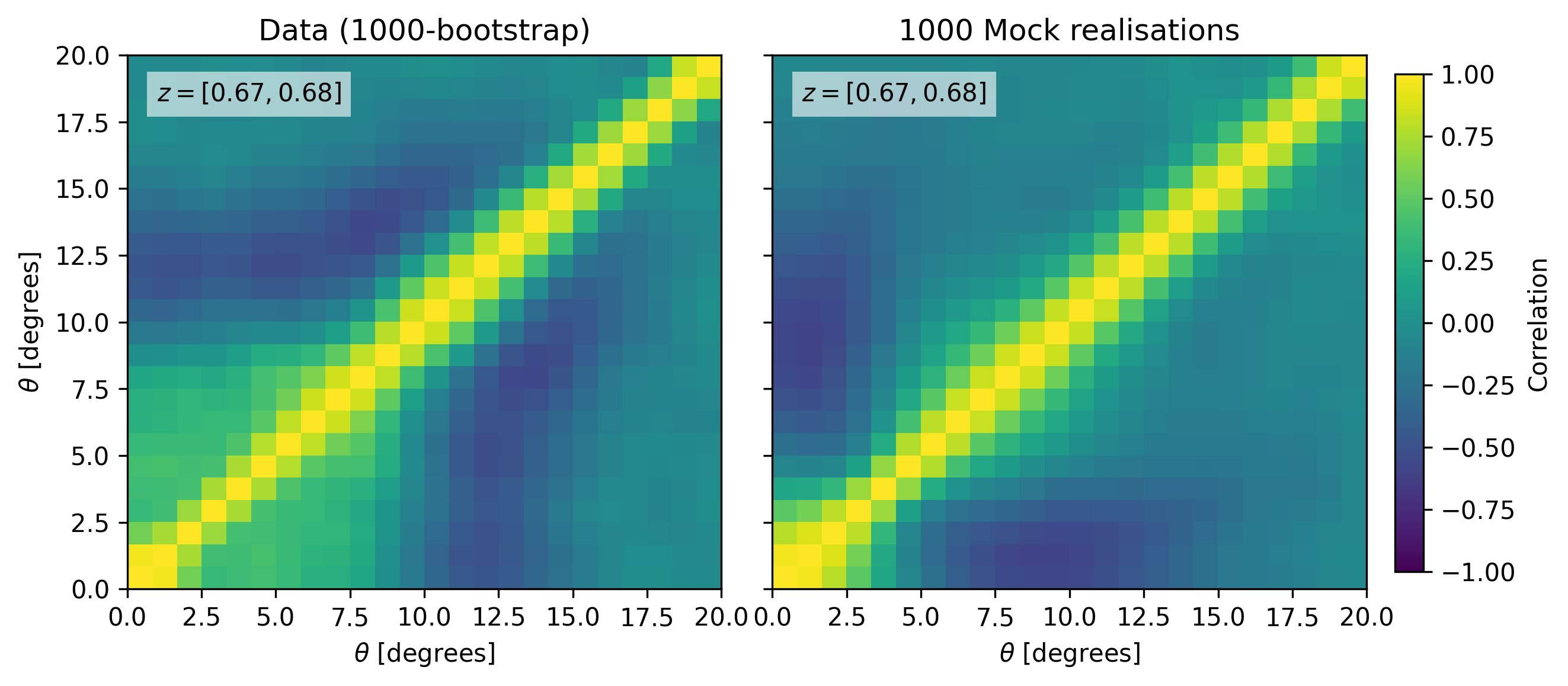}
    \vfill
    \includegraphics[width=0.60\textwidth]{plots/correlation_matrix_desi_nbins30_0.67z0.68.png}
    \caption{Comparison of correlation matrices for correlation dimension, $D_2(\theta)$, for different number os bins in the same redshift range $0.67 < z < 0.68$. The left panels show the results obtained from 1000-bootstrap resamplings, and the right panels correspond to the results obtained from 1000 mock catalogs. The upper panels shows the analysis with 24 bins while the bottom panel shows with 30 bins. Both shows the matrices using the North Galactic Cap from DESI DR1 data. }
    \label{fig:correlation_matrix_24x30bins_desi}
\end{figure}

In this section, we present a comparison of the $D_2(\theta)$ correlation matrices obtained using the same dataset within the same redshift range, varying only the number of bins adopted in the pair-counting procedure of the {\sc TreeCorr} package. By using 30 bins, we observe stronger correlations and greater consistency in the data. Figure \ref{fig:correlation_matrix_24x30bins_eboss} shows the comparison for the eBOSS DR16 data, while Figure \ref{fig:correlation_matrix_24x30bins_desi} presents the corresponding comparison for the DESI DR1 data.


\clearpage
\bibliography{bib}

@article{eBOSS:2020mzp,
    author = "Ross, Ashley J. and others",
    collaboration = "eBOSS",
    title = "{The Completed SDSS-IV extended Baryon Oscillation Spectroscopic Survey: Large-scale structure catalogues for cosmological analysis}",
    eprint = "2007.09000",
    archivePrefix = "arXiv",
    primaryClass = "astro-ph.CO",
    doi = "10.1093/mnras/staa2416",
    journal = "Mon. Not. Roy. Astron. Soc.",
    volume = "498",
    number = "2",
    pages = "2354--2371",
    year = "2020"
}

@article{DESI:2025fxa,
    author = "Abdul Karim, M. and others",
    collaboration = "DESI",
    title = "{Data Release 1 of the Dark Energy Spectroscopic Instrument}",
    eprint = "2503.14745",
    archivePrefix = "arXiv",
    primaryClass = "astro-ph.CO",
    reportNumber = "FERMILAB-PUB-25-0163-PPD",
    doi = "10.3847/1538-3881/ae4c43",
    journal = "Astron. J.",
    volume = "171",
    number = "5",
    pages = "285",
    year = "2026"
}

@article{Alonso:2013boa,
    author = "Alonso, D. and Bueno belloso, A. and S{\'a}nchez, F. J. and Garc{\'\i}a-Bellido, J. and S{\'a}nchez, E.",
    title = "{Measuring the transition to homogeneity with photometric redshift surveys}",
    eprint = "1312.0861",
    archivePrefix = "arXiv",
    primaryClass = "astro-ph.CO",
    doi = "10.1093/mnras/stu255",
    journal = "Mon. Not. Roy. Astron. Soc.",
    volume = "440",
    number = "1",
    pages = "10--23",
    year = "2014"
}

@article{Andrade:2022imy,
    author = "Andrade, Uendert and Gon{\c{c}}alves, Rodrigo S. and Carvalho, Gabriela C. and Bengaly, Carlos A. P. and Carvalho, Joel C. and Alcaniz, Jailson",
    title = "{The angular scale of homogeneity with SDSS-IV DR16 luminous red galaxies}",
    eprint = "2205.07819",
    archivePrefix = "arXiv",
    primaryClass = "astro-ph.CO",
    doi = "10.1088/1475-7516/2022/10/088",
    journal = "JCAP",
    volume = "10",
    pages = "088",
    year = "2022"
}

@article{Scrimgeour:2012wt,
    author = "Scrimgeour, Morag and others",
    title = "{The WiggleZ Dark Energy Survey: the transition to large-scale cosmic homogeneity}",
    eprint = "1205.6812",
    archivePrefix = "arXiv",
    primaryClass = "astro-ph.CO",
    doi = "10.1111/j.1365-2966.2012.21402.x",
    journal = "Mon. Not. Roy. Astron. Soc.",
    volume = "425",
    pages = "116--134",
    year = "2012"
}

@article{Laurent:2016eqo,
    author = "Laurent, Pierre and others",
    title = "{A 14 $h^{-3}$ Gpc$^3$ study of cosmic homogeneity using BOSS DR12 quasar sample}",
    eprint = "1602.09010",
    archivePrefix = "arXiv",
    primaryClass = "astro-ph.CO",
    doi = "10.1088/1475-7516/2016/11/060",
    journal = "JCAP",
    volume = "11",
    pages = "060",
    year = "2016"
}

@article{Ntelis:2017nrj,
    author = "Ntelis, Pierros and others",
    title = "{Exploring cosmic homogeneity with the BOSS DR12 galaxy sample}",
    eprint = "1702.02159",
    archivePrefix = "arXiv",
    primaryClass = "astro-ph.CO",
    doi = "10.1088/1475-7516/2017/06/019",
    journal = "JCAP",
    volume = "06",
    pages = "019",
    year = "2017"
}

@article{Goncalves:2020erb,
    author = "Gon{\c{c}}alves, Rodrigo S. and Carvalho, Gabriela C. and Andrade, Uendert and Bengaly, Carlos A. P. and Carvalho, Joel C. and Alcaniz, Jailson",
    title = "{Measuring the cosmic homogeneity scale with SDSS-IV DR16 Quasars}",
    eprint = "2010.06635",
    archivePrefix = "arXiv",
    primaryClass = "astro-ph.CO",
    doi = "10.1088/1475-7516/2021/03/029",
    journal = "JCAP",
    volume = "03",
    pages = "029",
    year = "2021"
}

@article{Goncalves:2017dzs,
    author = "Gon{\c{c}}alves, R. S. and Carvalho, G. C. and Bengaly, C. A. P. and Carvalho, J. C. and Bernui, A. and Alcaniz, J. S. and Maartens, R.",
    title = "{Cosmic homogeneity: a spectroscopic and model-independent measurement}",
    eprint = "1710.02496",
    archivePrefix = "arXiv",
    primaryClass = "astro-ph.CO",
    doi = "10.1093/mnrasl/slx202",
    journal = "Mon. Not. Roy. Astron. Soc.",
    volume = "475",
    number = "1",
    pages = "L20--L24",
    year = "2018"
}

@article{Landy:1993yu,
    author = "Landy, Stephen D. and Szalay, Alexander S.",
    title = "{Bias and variance of angular correlation functions}",
    doi = "10.1086/172900",
    journal = "Astrophys. J.",
    volume = "412",
    pages = "64",
    year = "1993"
}

@article{DESI:2024aax,
    author = "Adame, A. G. and others",
    collaboration = "DESI",
    title = "{DESI 2024 II: sample definitions, characteristics, and two-point clustering statistics}",
    eprint = "2411.12020",
    archivePrefix = "arXiv",
    primaryClass = "astro-ph.CO",
    reportNumber = "FERMILAB-PUB-24-0850-PPD",
    doi = "10.1088/1475-7516/2025/07/017",
    journal = "JCAP",
    volume = "07",
    pages = "017",
    year = "2025"
}

@article{ezmocks,
    author = "Zhao, Cheng and others",
    collaboration = "eBOSS",
    title = "{The completed SDSS-IV extended Baryon Oscillation Spectroscopic Survey: 1000 multi-tracer mock catalogues with redshift evolution and systematics for galaxies and quasars of the final data release}",
    eprint = "2007.08997",
    archivePrefix = "arXiv",
    primaryClass = "astro-ph.CO",
    doi = "10.1093/mnras/stab510",
    journal = "Mon. Not. Roy. Astron. Soc.",
    volume = "503",
    number = "1",
    pages = "1149--1173",
    year = "2021"
}

@article{uchuu,
    author = "Fern{\'a}ndez-Garc{\'\i}a, E. and others",
    title = "{DESI DR2 reference mocks: clustering results from Uchuu-BGS and LRG}",
    eprint = "2507.01593",
    archivePrefix = "arXiv",
    primaryClass = "astro-ph.GA",
    reportNumber = "FERMILAB-PUB-25-0441-PPD",
    doi = "10.1088/1475-7516/2026/05/002",
    journal = "JCAP",
    volume = "05",
    pages = "002",
    year = "2026"
}

@article{Feldman:1993ky,
    author = "Feldman, Hume A. and Kaiser, Nick and Peacock, John A.",
    title = "{Power spectrum analysis of three-dimensional redshift surveys}",
    eprint = "astro-ph/9304022",
    archivePrefix = "arXiv",
    reportNumber = "UM-AC-93-5",
    doi = "10.1086/174036",
    journal = "Astrophys. J.",
    volume = "426",
    pages = "23--37",
    year = "1994"
}

@article{Planck:2018vyg,
    author = "Aghanim, N. and others",
    collaboration = "Planck",
    title = "{Planck 2018 results. VI. Cosmological parameters}",
    eprint = "1807.06209",
    archivePrefix = "arXiv",
    primaryClass = "astro-ph.CO",
    doi = "10.1051/0004-6361/201833910",
    journal = "Astron. Astrophys.",
    volume = "641",
    pages = "A6",
    year = "2020",
    note = "[Erratum: Astron.Astrophys. 652, C4 (2021)]"
}

@article{SupernovaCosmologyProject:1998vns,
    author = "Perlmutter, S. and others",
    collaboration = "Supernova Cosmology Project",
    title = "{Measurements of $\Omega$ and $\Lambda$ from 42 High Redshift Supernovae}",
    eprint = "astro-ph/9812133",
    archivePrefix = "arXiv",
    reportNumber = "LBNL-41801, LBL-41801",
    doi = "10.1086/307221",
    journal = "Astrophys. J.",
    volume = "517",
    pages = "565--586",
    year = "1999"
}

@article{SupernovaSearchTeam:1998fmf,
    author = "Riess, Adam G. and others",
    collaboration = "Supernova Search Team",
    title = "{Observational evidence from supernovae for an accelerating universe and a cosmological constant}",
    eprint = "astro-ph/9805201",
    archivePrefix = "arXiv",
    doi = "10.1086/300499",
    journal = "Astron. J.",
    volume = "116",
    pages = "1009--1038",
    year = "1998"
}

@article{Planck:2018yye,
    author = "Akrami, Y. and others",
    collaboration = "Planck",
    title = "{Planck 2018 results. IV. Diffuse component separation}",
    eprint = "1807.06208",
    archivePrefix = "arXiv",
    primaryClass = "astro-ph.CO",
    doi = "10.1051/0004-6361/201833881",
    journal = "Astron. Astrophys.",
    volume = "641",
    pages = "A4",
    year = "2020"
}

@article{DiValentino:2021izs,
    author = "Di Valentino, Eleonora and Mena, Olga and Pan, Supriya and Visinelli, Luca and Yang, Weiqiang and Melchiorri, Alessandro and Mota, David F. and Riess, Adam G. and Silk, Joseph",
    title = "{In the realm of the Hubble tension{\textemdash}a review of solutions}",
    eprint = "2103.01183",
    archivePrefix = "arXiv",
    primaryClass = "astro-ph.CO",
    reportNumber = "IPPP/20/108",
    doi = "10.1088/1361-6382/ac086d",
    journal = "Class. Quant. Grav.",
    volume = "38",
    number = "15",
    pages = "153001",
    year = "2021"
}

@article{Shah:2021onj,
    author = "Shah, Paul and Lemos, Pablo and Lahav, Ofer",
    title = "{A buyer{\textquoteright}s guide to the Hubble constant}",
    eprint = "2109.01161",
    archivePrefix = "arXiv",
    primaryClass = "astro-ph.CO",
    doi = "10.1007/s00159-021-00137-4",
    journal = "Astron. Astrophys. Rev.",
    volume = "29",
    number = "1",
    pages = "9",
    year = "2021"
}

@article{Riess:2021jrx,
    author = "Riess, Adam G. and others",
    title = "{A Comprehensive Measurement of the Local Value of the Hubble Constant with 1 km s$^{-1}$ Mpc$^{-1}$ Uncertainty from the Hubble Space Telescope and the SH0ES Team}",
    eprint = "2112.04510",
    archivePrefix = "arXiv",
    primaryClass = "astro-ph.CO",
    doi = "10.3847/2041-8213/ac5c5b",
    journal = "Astrophys. J. Lett.",
    volume = "934",
    number = "1",
    pages = "L7",
    year = "2022"
}

@article{Hogg:2004vw,
    author = "Hogg, David W. and Eisenstein, Daniel J. and Blanton, Michael R. and Bahcall, Neta A. and Brinkmann, J. and Gunn, James E. and Schneider, Donald P.",
    title = "{Cosmic homogeneity demonstrated with luminous red galaxies}",
    eprint = "astro-ph/0411197",
    archivePrefix = "arXiv",
    doi = "10.1086/429084",
    journal = "Astrophys. J.",
    volume = "624",
    pages = "54--58",
    year = "2005"
}

@article{Sarkar:2009iga,
    author = "Sarkar, Prakash and Yadav, Jaswant and Pandey, Biswajit and Bharadwaj, Somnath",
    title = "{The scale of homogeneity of the galaxy distribution in SDSS DR6}",
    eprint = "0906.3431",
    archivePrefix = "arXiv",
    primaryClass = "astro-ph.CO",
    doi = "10.1111/j.1745-3933.2009.00738.x",
    journal = "Mon. Not. Roy. Astron. Soc.",
    volume = "399",
    pages = "L128--L131",
    year = "2009"
}

@article{Pandey:2013xz,
    author = "Pandey, Biswajit",
    title = "{A method for testing the cosmic homogeneity with Shannon entropy}",
    eprint = "1301.4961",
    archivePrefix = "arXiv",
    primaryClass = "astro-ph.CO",
    doi = "10.1093/mnras/stt134",
    journal = "Mon. Not. Roy. Astron. Soc.",
    volume = "430",
    pages = "3376",
    year = "2013"
}

@article{Pandey:2015xea,
    author = "Pandey, Biswajit and Sarkar, Suman",
    title = "{Testing homogeneity in the Sloan Digital Sky Survey Data Release Twelve with Shannon entropy}",
    eprint = "1507.03124",
    archivePrefix = "arXiv",
    primaryClass = "astro-ph.CO",
    doi = "10.1093/mnras/stv2166",
    journal = "Mon. Not. Roy. Astron. Soc.",
    volume = "454",
    number = "3",
    pages = "2647--2656",
    year = "2015"
}

@article{Sarkar:2016fir,
    author = "Sarkar, Suman and Pandey, Biswajit",
    title = "{An information theory based search for homogeneity on the largest accessible scale}",
    eprint = "1607.06194",
    archivePrefix = "arXiv",
    primaryClass = "astro-ph.CO",
    doi = "10.1093/mnrasl/slw145",
    journal = "Mon. Not. Roy. Astron. Soc.",
    volume = "463",
    number = "1",
    pages = "L12--L16",
    year = "2016"
}

@article{Goncalves:2018sxa,
    author = "Gon{\c{c}}alves, R. S. and Carvalho, G. C. and Bengaly, C. A. P. and Carvalho, J. C. and Alcaniz, J. S.",
    title = "{Measuring the scale of cosmic homogeneity with SDSS-IV DR14 quasars}",
    eprint = "1809.11125",
    archivePrefix = "arXiv",
    primaryClass = "astro-ph.CO",
    doi = "10.1093/mnras/sty2670",
    journal = "Mon. Not. Roy. Astron. Soc.",
    volume = "481",
    number = "4",
    pages = "5270--5274",
    year = "2018"
}

@article{Shao:2024qrd,
    author = "Shao, Xiaoyun and Bengaly, Carlos A. P. and Gon{\c{c}}alves, Rodrigo S. and Carvalho, Gabriela C. and Alcaniz, Jailson",
    title = "{Cosmological constraints from angular homogeneity scale measurements}",
    eprint = "2409.06009",
    archivePrefix = "arXiv",
    primaryClass = "astro-ph.CO",
    doi = "10.1140/epjc/s10052-025-13987-4",
    journal = "Eur. Phys. J. C",
    volume = "85",
    number = "3",
    pages = "225",
    year = "2025"
}

@article{Avila:2021mbj,
    author = "Avila, Felipe and Bernui, Armando and Nunes, Rafael C. and de Carvalho, Edilson and Novaes, Camila P.",
    title = "{The homogeneity scale and the growth rate of cosmic structures}",
    eprint = "2111.08541",
    archivePrefix = "arXiv",
    primaryClass = "astro-ph.CO",
    doi = "10.1093/mnras/stab3122",
    journal = "Mon. Not. Roy. Astron. Soc.",
    volume = "509",
    number = "2",
    pages = "2994--3003",
    year = "2021"
}

@article{Kim:2021osl,
    author = "Kim, Yigon and Park, Chan-Gyung and Noh, Hyerim and Hwang, Jai-chan",
    title = "{CMASS galaxy sample and the ontological status of the cosmological principle}",
    eprint = "2112.04134",
    archivePrefix = "arXiv",
    primaryClass = "astro-ph.CO",
    doi = "10.1051/0004-6361/202141909",
    journal = "Astron. Astrophys.",
    volume = "660",
    pages = "A139",
    year = "2022"
}

@article{Yadav:2010cc,
    author = "Yadav, Jaswant K. and Bagla, J. S. and Khandai, Nishikanta",
    title = "{Fractal Dimension as a measure of the scale of Homogeneity}",
    eprint = "1001.0617",
    archivePrefix = "arXiv",
    primaryClass = "astro-ph.CO",
    doi = "10.1111/j.1365-2966.2010.16612.x",
    journal = "Mon. Not. Roy. Astron. Soc.",
    volume = "405",
    pages = "2009",
    year = "2010"
}

@article{Zhang:2010fa,
    author = "Zhang, Pengjie and Stebbins, Albert",
    title = "{Confirmation of the Copernican Principle at Gpc Radial Scale and above from the Kinetic Sunyaev Zel'dovich Effect Power Spectrum}",
    eprint = "1009.3967",
    archivePrefix = "arXiv",
    primaryClass = "astro-ph.CO",
    reportNumber = "FERMILAB-PUB-10-373-A",
    doi = "10.1103/PhysRevLett.107.041301",
    journal = "Phys. Rev. Lett.",
    volume = "107",
    pages = "041301",
    year = "2011"
}

@article{Valkenburg:2012td,
    author = "Valkenburg, Wessel and Marra, Valerio and Clarkson, Chris",
    title = "{Testing the Copernican principle by constraining spatial homogeneity}",
    eprint = "1209.4078",
    archivePrefix = "arXiv",
    primaryClass = "astro-ph.CO",
    doi = "10.1093/mnrasl/slt140",
    journal = "Mon. Not. Roy. Astron. Soc.",
    volume = "438",
    pages = "L6--L10",
    year = "2014"
}

@article{Hoyle:2012pb,
    author = "Hoyle, Ben and Tojeiro, Rita and Jimenez, Raul and Heavens, Alan and Clarkson, Chris and Maartens, Roy",
    title = "{Testing homogeneity with galaxy star formation history}",
    eprint = "1209.6181",
    archivePrefix = "arXiv",
    primaryClass = "astro-ph.CO",
    doi = "10.1088/2041-8205/762/1/L9",
    journal = "Astrophys. J. Lett.",
    volume = "762",
    pages = "L9",
    year = "2013"
}

@article{Kraljic:2014aea,
    author = "Kraljic, David",
    title = "{Characterizing cosmic inhomogeneity with anomalous diffusion}",
    eprint = "1410.4107",
    archivePrefix = "arXiv",
    primaryClass = "astro-ph.CO",
    doi = "10.1093/mnras/stv1199",
    journal = "Mon. Not. Roy. Astron. Soc.",
    volume = "451",
    number = "4",
    pages = "3393--3399",
    year = "2015"
}

@article{Jimenez:2019cll,
    author = "Jimenez, Raul and Maartens, Roy and Khalifeh, Ali Rida and Caldwell, Robert R. and Heavens, Alan F. and Verde, Licia",
    title = "{Measuring the Homogeneity of the Universe Using Polarization Drift}",
    eprint = "1902.11298",
    archivePrefix = "arXiv",
    primaryClass = "astro-ph.CO",
    doi = "10.1088/1475-7516/2019/05/048",
    journal = "JCAP",
    volume = "05",
    pages = "048",
    year = "2019"
}

@article{Camarena:2021mjr,
    author = "Camarena, David and Marra, Valerio and Sakr, Ziad and Clarkson, Chris",
    title = "{The Copernican principle in light of the latest cosmological~data}",
    eprint = "2107.02296",
    archivePrefix = "arXiv",
    primaryClass = "astro-ph.CO",
    doi = "10.1093/mnras/stab3077",
    journal = "Mon. Not. Roy. Astron. Soc.",
    volume = "509",
    number = "1",
    pages = "1291--1302",
    year = "2021"
}

@article{Alonso:2014xca,
    author = "Alonso, David and Salvador, Ana Isabel and S{\'a}nchez, Francisco Javier and Bilicki, Maciej and Garc{\'\i}a-Bellido, Juan and S{\'a}nchez, Eusebio",
    title = "{Homogeneity and isotropy in the Two Micron All Sky Survey Photometric Redshift catalogue}",
    eprint = "1412.5151",
    archivePrefix = "arXiv",
    primaryClass = "astro-ph.CO",
    doi = "10.1093/mnras/stv309",
    journal = "Mon. Not. Roy. Astron. Soc.",
    volume = "449",
    number = "1",
    pages = "670--684",
    year = "2015"
}

@article{Avila:2019gdb,
    author = "Avila, F. and Novaes, C. P. and Bernui, A. and de Carvalho, E. and Nogueira-Cavalcante, J. P.",
    title = "{The angular scale of homogeneity in the Local Universe with the SDSS blue galaxies}",
    eprint = "1906.10744",
    archivePrefix = "arXiv",
    primaryClass = "astro-ph.CO",
    doi = "10.1093/mnras/stz1765",
    journal = "Mon. Not. Roy. Astron. Soc.",
    volume = "488",
    number = "1",
    pages = "1481--1487",
    year = "2019"
}

@article{Camacho-Quevedo:2021bvt,
    author = "Camacho-Quevedo, Benjamin and Gazta{\~n}aga, Enrique",
    title = "{A measurement of the scale of homogeneity in the early Universe}",
    eprint = "2106.14303",
    archivePrefix = "arXiv",
    primaryClass = "astro-ph.CO",
    doi = "10.1088/1475-7516/2022/04/044",
    journal = "JCAP",
    volume = "04",
    number = "04",
    pages = "044",
    year = "2022"
}

@article{Jarvis:2003wq,
    author = "Jarvis, Micael and Bernstein, G. and Jain, B.",
    title = "{The skewness of the aperture mass statistic}",
    eprint = "astro-ph/0307393",
    archivePrefix = "arXiv",
    doi = "10.1111/j.1365-2966.2004.07926.x",
    journal = "Mon. Not. Roy. Astron. Soc.",
    volume = "352",
    pages = "338--352",
    year = "2004"
}

@ARTICLE{Eisenstein2001,
       author = {Eisenstein, Daniel J. and others},
        title = "{Spectroscopic Target Selection for the Sloan Digital Sky Survey: The Luminous Red Galaxy Sample}",
      journal = {The Astronomical Journal},
         year = 2001,
        month = nov,
       volume = {122},
       number = {5},
        pages = {2267-2280},
          doi = {10.1086/323717},
archivePrefix = {arXiv},
       eprint = {astro-ph/0108153},
 primaryClass = {astro-ph},
       adsurl = {https://ui.adsabs.harvard.edu/abs/2001AJ....122.2267E}
}

@ARTICLE{Hogg2003,
       author = {Hogg, David W. and others},
        title = "{The Overdensities of Galaxy Environments as a Function of Luminosity and Color}",
      journal = {The Astrophysical Journal Letters},
         year = 2003,
        month = mar,
       volume = {585},
       number = {1},
        pages = {L5-L9},
          doi = {10.1086/374238},
archivePrefix = {arXiv},
       eprint = {astro-ph/0212085},
 primaryClass = {astro-ph},
       adsurl = {https://ui.adsabs.harvard.edu/abs/2003ApJ...585L...5H}
}

@ARTICLE{Zehavi2005,
       author = {Zehavi, Idit and others},
        title = "{The Luminosity and Color Dependence of the Galaxy Correlation Function}",
      journal = {The Astrophysical Journal},
         year = 2005,
        month = sep,
       volume = {630},
       number = {1},
        pages = {1-27},
          doi = {10.1086/431891},
archivePrefix = {arXiv},
       eprint = {astro-ph/0408569},
 primaryClass = {astro-ph},
       adsurl = {https://ui.adsabs.harvard.edu/abs/2005ApJ...630....1Z}
}

@article{Brout:2022vxf,
    author = "Brout, Dillon and others",
    title = "{The Pantheon+ Analysis: Cosmological Constraints}",
    eprint = "2202.04077",
    archivePrefix = "arXiv",
    primaryClass = "astro-ph.CO",
    doi = "10.3847/1538-4357/ac8e04",
    journal = "Astrophys. J.",
    volume = "938",
    number = "2",
    pages = "110",
    year = "2022"
}

@misc{Rubin:2023jdq,
    author = "Rubin, David and others",
    title = "{Union Through UNITY: Cosmology with 2,000 SNe Using a Unified Bayesian Framework}",
    eprint = "2311.12098",
    archivePrefix = "arXiv",
    primaryClass = "astro-ph.CO",
    month = "11",
    year = "2023"
}

@article{DES:2024jxu,
    author = "Abbott, T. M. C. and others",
    collaboration = "DES",
    title = "{The Dark Energy Survey: Cosmology Results with {\ensuremath{\sim}}1500 New High-redshift Type Ia Supernovae Using the Full 5 yr Data Set}",
    eprint = "2401.02929",
    archivePrefix = "arXiv",
    primaryClass = "astro-ph.CO",
    reportNumber = "FERMILAB-PUB-23-0821-PPD, DES-2023-805",
    doi = "10.3847/2041-8213/ad6f9f",
    journal = "Astrophys. J. Lett.",
    volume = "973",
    number = "1",
    pages = "L14",
    year = "2024"
}
\end{document}